\documentclass[10pt,twocolumn]{article} 

\usepackage{times}

\usepackage{geometry}              
\usepackage{microtype}
\usepackage{graphicx}
\usepackage{subfigure}
\usepackage{booktabs}
\usepackage{float}

\usepackage{hyperref}
\usepackage{xcolor} 

\definecolor{darkblue}{rgb}{0,0.08,0.50}
\hypersetup{
    colorlinks=true,        
    linkcolor=darkblue,         
    citecolor=darkblue,         
    filecolor=darkblue,      
    urlcolor=darkblue           
}

\usepackage{natbib}                
\setcitestyle{authoryear,round,citesep={;},aysep={,},yysep={;}}
\newcommand{\yrcite}[1]{\citeyearpar{#1}}
\renewcommand{\cite}[1]{\citep{#1}}

\usepackage{fancyhdr} 

\usepackage{caption}

\newcommand{\documenttitle}{Adversarial Robustness of Partitioned Quantum Classifiers}

\fancypagestyle{plain}{%
  \fancyhf{} 
  \fancyfoot[C]{\thepage} 
}

\renewenvironment{abstract}{
	\centerline{\large \bf Abstract}
     \vspace{-0.1in}\begin{quote}
}
{\par\end{quote}\vskip 0.15in}

\usepackage{enumitem}
\usepackage{bm}
\usepackage{braket}
\usepackage{bbold}
\usepackage{qcircuit}

\usepackage{ifthen}
\newboolean{twocolumnmode}
\setboolean{twocolumnmode}{true}
\newboolean{icml-citation-format}
\setboolean{icml-citation-format}{true}

\newcommand{\reffig}[1]{Fig. \ref{#1}}

\usepackage{amsthm}
\usepackage{amsmath}
\usepackage{mathtools}
\usepackage{amssymb}

\usepackage{titlesec}
\theoremstyle{plain}
\newtheorem{theorem}{Theorem}[section]

\newtheorem{lemma}[theorem]{Lemma}
\newtheorem{corollary}[theorem]{Corollary}
\theoremstyle{definition}

\theoremstyle{remark}

\DeclareMathOperator{\argmax}{argmax}
\DeclareMathOperator{\argmin}{argmin}

\begin{document}

\title{\documenttitle}
\author{Pouya Kananian and Hans-Arno Jacobsen\\[0.5em]
Department of Electrical and Computer Engineering,  University of Toronto, Toronto, Canada\\[0.5em]
\texttt{pouya.kananian@mail.utoronto.ca, jacobsen@eecg.toronto.edu}}
\date{} 
\maketitle
\thispagestyle{plain}

\begin{abstract}
Adversarial robustness in quantum classifiers is a critical area of study, providing insights into their performance compared to classical models and uncovering potential advantages inherent to quantum machine learning. 
In the NISQ era of quantum computing, circuit cutting is a notable technique for simulating circuits that exceed the qubit limitations of current devices,  enabling the distribution of a quantum circuit's execution across multiple quantum processing units through classical communication.  In contrast, when quantum communication is available,  teleportation-based methods can be used to support the distribution of the quantum circuit.
We study the robustness of partitioned quantum classifiers to adversarial perturbations targeting wire cutting or quantum state teleportation and show a link between such perturbations and  implementing adversarial gates within intermediate layers of a quantum classifier. 
We then proceed to study the latter problem from both a theoretical and experimental perspective.
\end{abstract}

\section{Introduction}
\label{sec:intro}

Quantum machine learning (QML) has emerged as a thriving and rapidly evolving field of study \cite{biamonte2017quantum, abbas2021power,liu2021rigorous,cerezo2022challenges, larocca2024review}.
Especially,  variational quantum classifiers hold particular significance due to their practical applications in the Noisy Intermediate-Scale Quantum (NISQ) era of quantum computing \cite{cerezo2021variational, bharti2022noisy}.  
An increasingly important domain in QML that has captured attention recently is adversarial robustness of quantum classifiers \cite{lu2020quantum,liu2020vulnerability,du2021quantum,weber2021optimal,liao2021robust, gong2022enhancing,gong2022universal,anil2024generating,dowling2024adversarial}. 
This area is vital for comparing the performance of quantum classifiers against classical models and exploring the potential advantages within QML \cite{west2023towards}.

Current NISQ era quantum devices feature a small number of noisy qubits.
To address the limited qubit count in these devices, 
numerous methods \cite{bravyi2016trading, peng2020simulating,yuan2021quantum,mitarai2021constructing,mitarai2021overhead,fujii2022deep,eddins2022doubling}
 have been suggested to expand the size of quantum systems through the use of classical processing.
A notable category of such approaches, known as circuit cutting \cite{lowe2023fast} or circuit knitting \cite{piveteau2023circuit}, involves partitioning quantum circuits into smaller fragments that can be executed on devices with fewer qubits than required by the original circuit.
After execution, classical post-processing can be employed to combine the measurement outcomes 
and simulate quantum circuits that surpass the qubit limitations of a given device.  
Most circuit cutting approaches are primarily based on
quasiprobability simulation,  
a commonly employed technique in quantum error mitigation \cite{temme2017error,endo2018practical,piveteau2022quasiprobability} and classical simulation of quantum systems \cite{pashayan2015estimating, howard2017application,seddon2019quantifying,seddon2021quantifying}.
Quantum channels describe the evolution of quantum states,  with the identity channel mapping a quantum state to itself. 
Quasiprobability decomposition allows us to decompose the identity channel into a linear combination of measurement and state-preparation channels. Similarly,  it can break down a non-local channel into a sum of tensor products of local channels. This process is known as wire cutting 
\cite{peng2020simulating,uchehara2022rotation,lowe2023fast,pednault2023alternative,brenner2023optimal,harada2023doubly,harrow2024optimal}
when applied to the identity channel
and gate cutting \cite{mitarai2021constructing,mitarai2021overhead,
piveteau2023circuit, schmitt2023cutting, ufrecht2023cutting, ufrecht2024optimal,harrow2024optimal}
when applied to non-local channels.

Quantum teleportation \cite{bennett1993teleporting,ren2017ground} is a key method in quantum communication that allows the transmission of an unknown quantum state using shared entanglement and classical communication \cite{cuomo2020towards,cacciapuoti2020entanglement}.
In the absence of quantum communication,  circuit cutting can be employed to distribute the execution of a quantum circuit across multiple quantum processing units,  combining quantum and classical computational resources \cite{barral2024review}. 
Conversely,  when quantum communication is available,  teleportation-based methods \cite{bennett1993teleporting,bouwmeester1997experimental,boschi1998experimental,narottama2023federated} can be employed to distribute quantum circuits across nodes.  While circuit cutting leads to an exponential sampling overhead, quantum teleportation does not; however, it consumes entangled qubit pairs shared between the parties.

This paper focuses on how distributing a quantum classifier's circuit through circuit cutting or teleportation-based methods can inadvertently increase its exposure to adversarial attacks. 
When a model is distributed across multiple processors, each device is exposed to the risk of tampering by malicious inputs or compromised participants.
If an adversary manipulates the subcircuits created through circuit cutting in any way,    
combining their outcomes may no longer reproduce the output of the original circuit that was meant to be simulated.
This also holds true when teleportation-based methods are used for distributing the circuit. 
However,  adversarial robustness of partitioned quantum classifiers has not, to the best of our knowledge, been previously studied, highlighting a crucial gap in the distributed QML literature.

\begin{figure*}[tb]
  \centering
  \includegraphics[width=\textwidth]{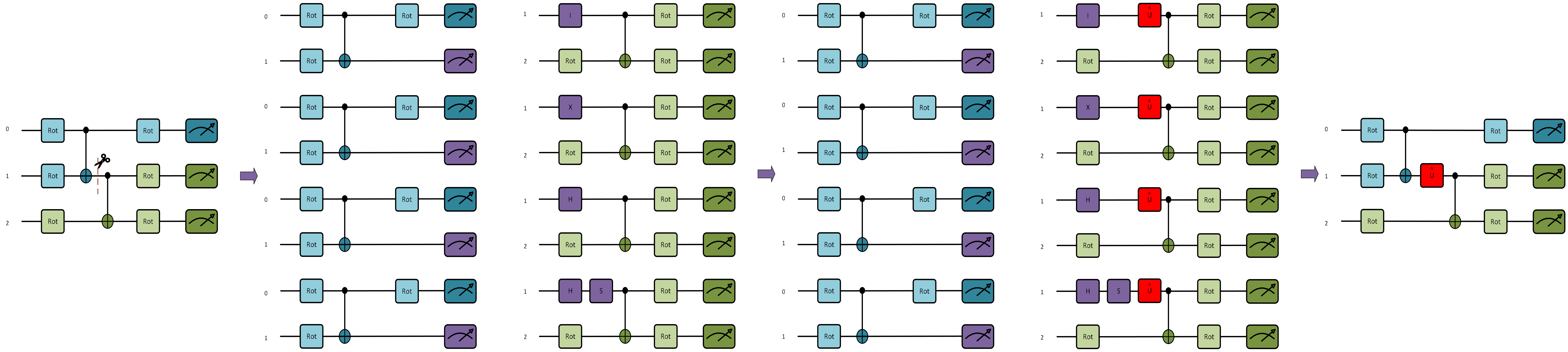}
  \caption{Wire cutting applied to a simple quantum circuit, decomposing it into multiple sub-circuits.  In the resulting sub-circuits, the cut wire from the original circuit is replaced by measurement and state preparation channels, shown in purple.  The input states of the green sub-circuits are adversarially modified through the application of a perturbation gate $\hat{U}$.  When the results of the sub-circuits are recombined to reconstruct the original circuit, this adversarial attack leads to the insertion of $\hat{U}$ in the intermediate layers of the reconstructed circuit.}
  \label{fig:adv-cut-intro}
\end{figure*}

One possible method of manipulating the subcircuits is through adversarial perturbations,  which denote small alterations to the input states of classifiers,  designed to deceive the models into making inaccurate predictions \cite{szegedy2013intriguing}.  
Unlike gate cutting,  wire cutting involves preparing new states for the circuit fragments created through circuit cutting.  
As a result,  the subcircuits’ input states are either retained from the original circuit or are prepared as part of wire cutting.  
In Section \ref{sec:wire-and-adv},  we examine how adding adversarial perturbations to the prepared states in wire cutting,  instead of manipulating the classifier's input states,  enables an adversary to plant adversarial gates within intermediate layers of 
the simulated circuit,  constructed 
by merging the results of the circuit fragments (see Fig.~\ref{fig:adv-cut-intro}).   
Similarly,  as discussed in Section \ref{sec:tele-and-adv},  when quantum state teleportation is used to distribute the circuit,  introducing carefully crafted adversarial perturbations before teleporting an state or after receiving it may be viewed as inserting adversarial gates within intermediate layers of the original circuit. 
The connection between adversarial attacks targeting the wire cutting procedure and those manipulating the state teleporation method with insertion of adversarial gates within a quantum classifier's architecture 
motivates us to study the latter problem.  
In Section \ref{sec:theoretical},  we present theorems bounding the 
potential variation 
implementing multiple adversarial gates within intermediate layers of a quantum classifier could cause in its predictive confidence. 
Section \ref{sec:exp} experimentally investigates the impact of planting one or more global or local adversarial gates at different
depths of variational quantum classifiers.  It also provides an experimental evaluation of our theoretical bounds.
 
\section{Related Work}
\label{sec:related-works}

\subsection{Adversarial Robustness of Quantum Classifiers}
There is a rich body of literature that studies the adversarial robustness of quantum classifiers, both theoretically \cite{liu2020vulnerability,  liao2021robust,weber2021optimal,guan2021robustness,du2021quantum,gong2022universal,anil2024generating,dowling2024adversarial} and experimentally \cite{lu2020quantum,ren2022experimental,west2023benchmarking}.
Although the intersection of circuit cutting and QML has garnered some attention in recent years 
\cite{pira2023invitation,guala2023practical, marshall2023high,marchisio2024cutting,sahu2024nac},
to the best of our knowledge,  
this is the first study to explore the impact of partitioning quantum classifier through circuit cutting on their adversarial robustness.
Adversarial robustness 
in quantum federated learning 
has been extensively studied 
 \cite{li2021quantum,xia2021defending,zhang2022federated,
 kumar2023expressive,chu2023cryptoqfl,li2024privacy,
 chen2024robust,papadopoulos2025numerical,maouaki2025qfal}.
The distribution approach in these federated learning works,  however,  is fundamentally different from the 
circuit distribution techniques based on circuit cutting.
Federated learning \cite{konevcny2016federated,konevcny2016federated-b,mcmahan2017communication} allows multiple participants to collaboratively train a shared model without sharing their local private data. 
Quantum computing can be incorporated into federated learning through quantum data \cite{xia2021quantumfed,chehimi2022quantum},  QML models \cite{huang2022quantum,kumar2023expressive}, or quantum communication \cite{li2021quantum,zhang2022federated,xu2023secure}.
However,  unlike classical computing, which benefits from widely available hardware like IoT devices, smartphones, and edge servers for federated training, quantum computing lacks this flexibility due to its limited and costly hardware.
It is therefore imperative to study the distributed execution of a single quantum model across multiple quantum processors, as well as to investigate the adversarial robustness of such distributed models.

Several prior studies have examined quantum teleportation in adversarial settings in the context of secure communication \cite{ahlswede2013quantum,hamdoun2020information,tserkis2020teleportation,
kumar2021state,
gangopadhyay2022controlled,nang2022counterfactual,
unnikrishnan2020authenticated,
bajpayee2024quantum,huang2024quantum,
holmes2025quantum,rahmawati2025encrypted,neves2025experimentally}.
In contrast, we explore a novel adversarial scenario that focuses on the performance of a quantum classifier.  We assume ideal quantum teleportation, in which the teleportation channel acts as the identity channel on the teleported state with unit fidelity.  
Nevertheless, the state may be manipulated through the introduction of adversarial perturbations either prior to transmission or after it is received.
This scenario, which is similar to those studied in classical split learning (see Section~\ref{sec:related-works-classical}), is connected to the adversarial case we analyze for wire cutting.  In both settings, introducing adversarial perturbations to the identity channel implemented through wire cutting or state teleportation results in the incorporation of adversarial gates within the intermediate layers of the circuit.   
Our attack model,  outlined in Section \ref{subsec:q-classifiers-and-adv}, 
provides a more general framework for understanding the implications of implementing adversarial gates within a quantum classifier’s architecture,  
extending beyond those previous works that focus only on 
adversarial perturbations targeting the input state \cite{liao2021robust,dowling2024adversarial}.

\ifthenelse{\boolean{icml-citation-format}}{
Du et al.  \yrcite{du2021quantum}
}{
Du et al.  \cite{du2021quantum}
}
 show that the depolarization noise present in NISQ-era quantum classifiers can make them inherently quantum differentially private \cite{dwork2006calibrating,zhou2017differential} and naturally resilient to adversaries \cite{cohen2019certified}.  
Numerous methods \cite{watkins2023quantum,huang2023certified,
huang2023enhancing,winderl2024quantum,khatun2025classical} have been proposed to improve the adversarial robustness of quantum classifiers, including adversarial training \cite{lu2020quantum,ren2022experimental,
khatun2025quantum,montalbano2025quantum},  as well as randomized and approximate encoding schemes \cite{gong2022enhancing,west2024drastic,
wollschlager2024discrete,saxena2024certifiably}.
In this work, we focus on noise-free quantum classifiers without additional adversarial robustness enhancements,  leaving the exploration of enhancement strategies to future work.

\subsection{Relevant Adversarial Scenarios in Classical Systems}
\label{sec:related-works-classical}

The adversarial scenario explored in this work bears similarities to those observed in split learning frameworks within classical machine learning \cite{gajbhiye2022data,fan2023OnTR,bai2023villain,yu2023backdoor,he2024advusl,shabbir2025taxonomy}.  In split learning \cite{gupta2018distributed,vepakomma2018split,gao2020end},  the model is segmented into sub-networks executed by separate participants to facilitate collaborative training.  
Raw data remains on the client’s device,  with only intermediate results communicated to the server, thereby reducing the computational demands on the client.
Split learning is vulnerable to several privacy threats \cite{shabbir2025taxonomy},  including inference \cite{pasquini2021unleashing,fu2022label}, reconstruction \cite{gawron2022feature,mao2023secure},  and model inversion \cite{erdougan2022unsplit} attacks, as well as adversarial attacks such as evasion \cite{fan2023OnTR,he2024advusl} and poisoning \cite{bai2023villain,yu2024chronic}.
An untrusted server in split learning may tamper with the inputs or outputs of its assigned layers or collect data to train proxy models \cite{fan2023OnTR}.  One possible adversarial scenario involves generating adversarial examples by manipulating the intermediate outputs of the model \cite{fan2023OnTR,he2024advusl}.  
Although this scenario is relevant to our work, the machine learning models and partitioning approaches in quantum computing differ from those in classical systems,  making it essential to conduct a separate study.

\subsection{Wire Cutting and Quantum Teleportation}

Circuit cutting has attracted growing interest and attention from researchers in recent years \cite{tang2021cutqc, majumdar2022error,liu2022classical,casciola2022understanding,chen2022approximate,chen2023efficient,chen2023online,brandhofer2023optimal,nagai2023quantum,perez2023shallow,bhoumik2023distributed,seitz2024multithreaded,gentinetta2024overhead,li2024efficient}.
In this work,  we specifically focus on wire cutting.  This is because cutting wires,  in contrast to cutting gates,  
involves measuring qubits,  transforming quantum information 
into classical information,  and preparing new quantum states.  Studying the robustness of quantum classifiers to adversarial perturbations targeting these new states can be viewed as a 
natural generalization of studying their robustness to adversarial attacks targeting their input states. 

Multiple
studies have focused on improving the original wire cutting decomposition \cite{peng2020simulating} to reduce sampling overhead and minimize the number of channels needed for cutting multiple wires
\cite{brenner2023optimal, lowe2023fast,pednault2023alternative,harada2023doubly,harrow2024optimal}.
To study the adversarial robustness of quantum classifiers 
that undergo wire cutting,  we primarily focus on
the original wire cutting decomposition \cite{peng2020simulating} and 
the decomposition proposed
\ifthenelse{\boolean{icml-citation-format}}{
by Harada et al. \yrcite{harada2023doubly}, 
}{
by Harada et al. \cite{harada2023doubly}, 
}
which achieves the optimal sampling overhead and number of channels required for 
cutting parallel wires. 
However,  other wire cutting approaches based on measure-and-prepare channels could be vulnerable to similar attacks addressed in this paper.  
A range of works has investigated finding optimal cut locations and intelligent strategies for qubit assignment across processors \cite{tang2021cutqc,tang2022scaleqc,smith2023clifford,brandhofer2023optimal}.
Similarly,  when quantum teleportation is used for distributing circuits, minimizing the e-bit cost as well as optimizing partitioning strategies are widely studied \cite{zomorodi2018optimizing,heunen2019automated,
houshmand2020evolutionary,baker2020time,g2021efficient,
nikahd2021automated,ferrari2021compiler,ferrari2023modular,
cuomo2023optimized}.  
Our study applies to distributed classifiers,  regardless of where their circuits are partitioned.

\section{Preliminaries}
\label{sec:background}

Here,  we 
review key  
notations and distance metrics, provide a brief overview of quantum classification,  wire cutting,  and state teleportation,  and introduce our attack model.  

\subsection{Quantum Channels and Diamond Distance}

A linear map $\mathcal{M}: \mathcal{L}(\mathcal{H}_A) \rightarrow \mathcal{L}(\mathcal{H}_B)$ is called trace-preserving (TP) if $\text{Tr}(\mathcal{M}(X)) = \text{Tr}(X)$ for all $X \in \mathcal{L}(\mathcal{H}_A)$,
where $\mathcal{H}_A$ and $\mathcal{H}_B$ are Hilbert spaces and
$\mathcal{L}(\mathcal{H})$ represents the space of square linear operators acting on the Hilbert space $\mathcal{H}$.  
A linear map $\mathcal{M}: \mathcal{L}(\mathcal{H}_A) \rightarrow \mathcal{L}(\mathcal{H}_B)$ is completely positive (CP) if for a reference system $R$ of arbitrary size,  $(\mathcal{I}_R \otimes \mathcal{M})(X)$ is positive semi-definite for all positive semi-definite  $X \in \mathcal{L}(\mathcal{H}_{R} \otimes \mathcal{H}_{A})$, 
where $\mathcal{I}_R$ represents the identity map acting on $\mathcal{H}_R$.
A quantum channel is a linear map that is both 
completely positive and trace-preserving (CPTP).
For a linear map $\mathcal{M}: \mathcal{L}(\mathcal{H}_A) \rightarrow \mathcal{L}(\mathcal{H}_B)$,  the adjoint of this map $\mathcal{M}^\dagger: \mathcal{L}(\mathcal{H}_B) \rightarrow \mathcal{L}(\mathcal{H}_A)$  is the unique linear map satisfying $\langle A, \mathcal{M}(B) \rangle = \langle \mathcal{M}^\dagger(A), B \rangle$ for all $A \in \mathcal{L} (\mathcal{H}_A)$ and $B \in \mathcal{L}(\mathcal{H}_B)$,  where $\langle A, B\rangle := \text{Tr}[A^\dagger B] $ denotes the Hilbert–Schmidt inner product.   
The diamond distance between two quantum channels $\mathcal{N},  \mathcal{M}: \mathcal{L}(\mathcal{H}_A) \rightarrow \mathcal{L}(\mathcal{H}_B)$ is defined as 
\begin{align*}
\Vert \mathcal{N} - \mathcal{M} \Vert_{\diamond} 
:= \underset{\rho}{\sup}\ 
\Vert (\mathcal{I}_R \otimes  \mathcal{N} )(\rho) - (\mathcal{I}_R \otimes  \mathcal{M} )(\rho)\Vert_1,
\end{align*}
where the supremum is taken over all density operators acting on $\mathcal{H}_R \otimes \mathcal{H}_A$,  
and $\Vert . \Vert_1$ denotes the trace norm (also known as the Schatten 1-norm).  

In this paper,  we use $\Vert . \Vert_2$ and $\Vert . \Vert_{op}$
 to denote the Hilbert-Schmidt norm 
  (or the Schatten 2-norm)  
 and the largest singular value 
 (also referred to as the operator norm or spectral norm),  
 respectively.  
For $\rho \in [1,  \infty)$,  the Schatten $\rho$-norm is defined as
 \begin{align*}
 \Vert C \Vert_\rho := \left(\textbf{Tr} \left[  \left(\sqrt{C^\dagger C} \right)^\rho  \right]  \right)^{1/\rho},
 \end{align*}
 where 
 $C$ is a linear 
 operator 
taking 
 $\mathcal{H}$ to another Hilbert space $\mathcal{H}^\prime$.

\subsection{Quantum K-multiclass Classification}
\label{subsection-k-classifier}

The objective of K-multiclass classification is to
learn a model $y(\sigma) = h(\sigma; \theta)$ that assigns a class label $k \in \{0, \cdots, K-1\}$ to each input 
data sample
$\sigma \in S_\sigma$,  
where $h$  refers to 
a function (or hypothesis) parameterized by $\theta$,  and $S_\sigma$  denotes the set of possible input samples. 
Consider a training dataset
$D_M=\{\sigma_i,  Y(\sigma_i)\}_{i=1}^M$,  
with $Y(\sigma_i)$ representing a one-hot K-dimensional vector that indicates the class label of the input state $\sigma_i$.
The model parameters,  denoted as $\theta$,  are typically optimized during the training process to minimize the empirical risk 
$L_M = (1/M) \sum_{i=1}^M L(h(\sigma_i; \theta),  Y(\sigma_i))$,  where $L$ represents a loss function.
We use $\theta^*$ to represent the optimized parameters.

A
quantum circuit can be used for obtaining $y(\sigma)$.  
Let $\mathcal{H}^{\otimes d}$ and $\mathcal{D}(\mathcal{H}^{\otimes d})$ denote the $d$-fold tensor product of the Hilbert space $\mathcal{H}$ and the set of density operators 
acting on $\mathcal{H}^{\otimes d}$,  respectively.  
For a quantum state $\sigma \in S_\sigma$,  
where $S_\sigma \subseteq \mathcal{D}(\mathcal{H}^{\otimes d})$,
we 
define $y_k(\sigma)$ as the probability of obtaining the 
measurement outcome $k$ in the quantum circuit \cite{du2021quantum}.  \begin{align}
y_k(\sigma) \coloneqq \text{Tr}(\Pi_k \mathcal{E} (\sigma \otimes \ket{a}\bra{a})),
\label{equ-prob-k}
\end{align} 
with $\Pi_k$ representing a positive-operator valued measure (POVM),  
$\mathcal{E}$ a completely positive trace-preserving (CPTP) map dependent on the parameter $\theta^*$,   
and  $\ket{a}\bra{a} \in \mathcal{D}(\mathcal{H}^{\otimes  d_a})$ 
an ancilla state.  
Here,  $d$ and $d_a$ denote the number of input and ancilla qubits,  respectively. 
We use 
$d_{+} := d + d_a$ 
to 
indicate 
the total number of qubits in the circuit.
$y_k(\sigma)$  represents  
the probability of assigning label $k$ to $\sigma$,  with $y(\sigma) = \argmax_{k} y_k(\sigma)$ denoting the class label ultimately assigned to this input sample by the learning algorithm.

\subsection{Quantum Classifiers and Adversarial Attacks}
\label{subsec:q-classifiers-and-adv-background}

An adversarial perturbation refers to a small modification to the input of a classifier, aimed at misleading the model into making incorrect predictions.
Such a 
perturbation could be added to the input state of a quantum classifier through a unitary perturbation operator 
$\hat{U}$\footnote{More generally,  the input state could be perturbed by a CPTP map.  
Another setup for adversarial attacks in quantum classifiers that use classical inputs encoded as quantum states,  involves the adversary perturbing classical inputs prior to their encoding, rather than perturbing the quantum states through unitary operations.  This paper leaves such classical perturbations outside its scope. 
}.
In untargeted attacks,  the 
adversary's goal 
is typically to find  
a perturbation 
that 
maximizes the following loss \cite{lu2020quantum}.
\begin{align}
\hat{U} = 
\underset{\hat{U} \in S_{adv}}{\argmax}\,
L(h(\hat{U} \sigma_i \hat{U}^\dagger; \theta^*),  Y(\sigma_i)),
\label{equ-untargeted}
\end{align} 
where $S_{adv} \subseteq U(2^d)$,  with $U(2^d)$ denoting the set of $2^d \times 2^d$ unitary matrices.
In targeted attacks, where the aim is to misclassify an input into a specific class, the optimization objective can be formulated as follows \cite{lu2020quantum}.
\begin{align}
\hat{U} = 
\underset{\hat{U} \in S_{adv}}{\argmin}\,
L(h(\hat{U} \sigma_i \hat{U}^\dagger; \theta^*),  \hat{Y}_i),
\label{equ-targeted}
\end{align}
where $\hat{Y}_i \neq Y(\sigma_i)$.
The set of perturbation operators $S_{adv}$ often consists of unitaries close to the identity operator. 
Approaches to ensure proximity to the identity operator include limiting 
the perturbation operator to 
products of local unitaries 
close to the identity transformation  \cite{lu2020quantum,gong2022universal}
 or 
adding fidelity constraints to the adversarial attacker's loss 
to control the strength of the perturbation \cite{anil2024generating}.

\subsection{Attack Model}
\label{subsec:q-classifiers-and-adv}

\begin{figure}[tb]
    \centering
    \begin{subfigure}
        \centering
			\[
			\Qcircuit @C=.6em @R=.6em {
			& \lstick{} & \qw & \multigate{4}{\ \ \ U\ \ \ } & \qw &\qw \\
			\lstick{\ket{\psi}\ \ }   			
			& \lstick{} & \qw & 						\ghost{\ \ \ U\ \ \ }         & \qw & \qw \\
			& \vdots       &     &            &\vdots      \\
			& \lstick{} & \qw & \ghost{\ \ \ U\ \ \ } & \qw &\meter \\
			\lstick{\ket{a} \,\,} & \lstick{} & \qw & \ghost{\ \ \ U\ \ \ } & \qw &\meter \\
			}
			\]
			{\small (a)}
			
		  \label{fig:quantum-classifier-a}
    \end{subfigure}
    \hfill
    \begin{subfigure}
		  \centering
		 \begin{center}
		  \includegraphics[width=0.3\textwidth]{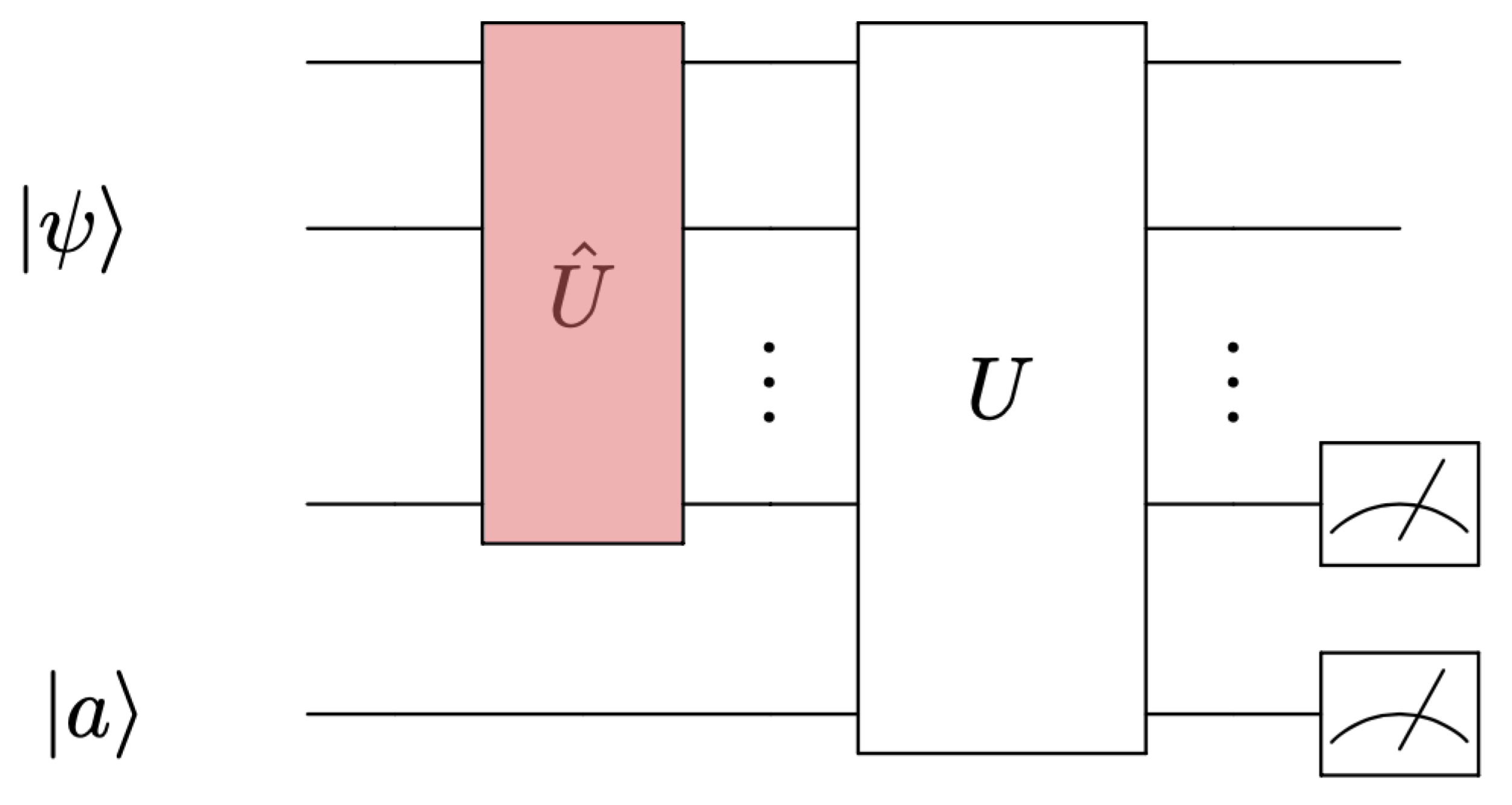}
		  \end{center}
		 
			{\small (b)}
		  \label{fig:quantum-classifier-with-adv}
    \end{subfigure}
    \begin{subfigure}
		  \centering
		   \includegraphics[width=\linewidth]{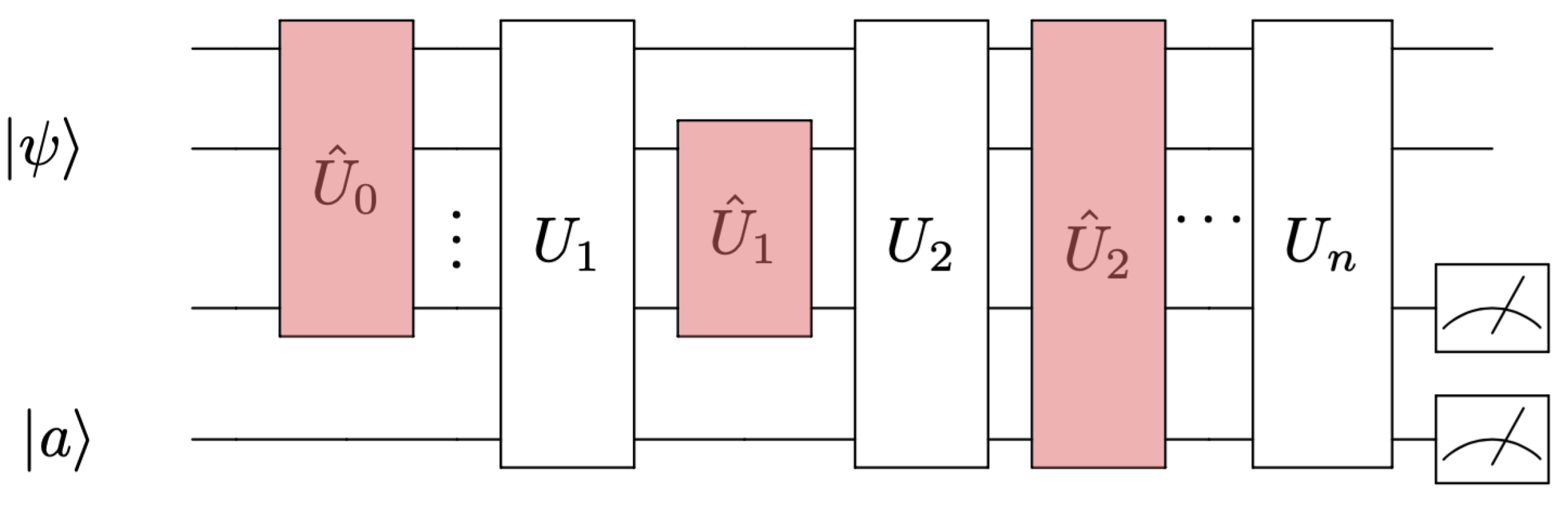}
		   
			{\small (c)}
		  \label{fig:quantum-classifier-multi-adv}
    \end{subfigure}

    \caption{A Quantum classifier (a) without exposure to adversarial 
    perturbations,  
     (b) with an adversarial unitary gate 
     impacting the input state,
     and (c)  under the influence of multiple adversarial gates,  highlighted in red.
    Here,  $\sigma= \ket{\psi}\bra{\psi}$ and $\ket{a}\bra{a}$ denote the input and ancilla states,  respectively.  $U= U_n  \cdots U_2 U_1$ depends on the parameter $\theta^*$ and each $\hat{U}_i \in S_{adv}$ denotes an adversarial perturbation operator.  The 
    adversarial gates may target a few local qubits or all qubits.}  
    \label{fig:main}
\end{figure}

In this paper,  we consider an adversary that can not only perturb the input state but also  insert adversarial gates 
within the intermediate layers of a quantum classifier's circuit (see Fig.  \ref{fig:main}). 
This attack model's connection to wire cutting and state teleportation is explored in Sections \ref{sec:wire-and-adv} and \ref{sec:tele-and-adv},  respectively. 
In the presence of such an adversary,  the probability of assigning 
label $k$ to input $\sigma$ will be modified to
\begin{align}
\hat{y}_k(\sigma) \coloneqq 
\text{Tr}(\Pi_k \hat{\mathcal{E}} (\hat{U}_0\sigma\hat{U}^\dagger_0 \otimes \ket{a}\bra{a})),
\label{equ-prob-k-attacked}
\end{align} 
where
\begin{align}
\label{equ-hat-E-with-unitaries}
\hat{\mathcal{E}}(.) &= \hat{U}_n( \mathcal{E}_{n} \cdots (\hat{U}_1( \mathcal{E}_{1} (.))\hat{U}_1^\dagger ) \cdots )\hat{U}_n^\dagger, \\
\mathcal{E}(.) &= \mathcal{E}_{n} \cdots \mathcal{E}_2(\mathcal{E}_{1} (.)), \nonumber
\end{align}
and each perturbation operator $\hat{U}_i \in S_{adv} \subseteq U(2^d) \cup U(2^{d_+}).$  
Alternatively,  we could represent $\hat{\mathcal{E}}(.)$ as 
\begin{align}
\hat{\mathcal{E}}(.) = \hat{\mathcal{U}}_n \circ \mathcal{E}_{n} \cdots \circ \hat{\mathcal{U}}_1 \circ \mathcal{E}_{1}(.),
\label{equ-hat-E}
\end{align}
 with $\hat{\mathcal{U}}_i(.) = \hat{U}_i(.)\hat{U}_i^\dagger $ 
 denoting the unitary channel corresponding to $ \hat{U}_i$. 
For this type of adversary, 
 the optimization objectives in (\ref{equ-untargeted}) and (\ref{equ-targeted}) can be generalized to
 \begin{align} 
 (\hat{U}_0,  \hat{U}_1,  \cdots,  \hat{U}_n) = 
\underset{\hat{U}_i \in S_{adv}, \forall i}{\argmax}\,
L(\hat{y}(\sigma_i),  Y(\sigma_i)),
\label{equ-untargeted-general}
\end{align} 
and 
\begin{align*}
 (\hat{U}_0,  \hat{U}_1,  \cdots,  \hat{U}_n) = 
\underset{\hat{U}_i \in S_{adv}, \forall i}{\argmin}\,
L(\hat{y}(\sigma_i),  \hat{Y}_i),
\end{align*}
respectively, 
where 
$\hat{y}(\sigma_i) = \argmax_k{\hat{y}(\sigma_i)}$. 

\subsection{Wire Cutting}

In a quantum circuit with the objective of estimating an observable  $O_{out}$,  the
expected value of $O_{out}$ can be expressed as 
$
\langle O_{out} \rangle = \text{Tr}(O_{out} \mathcal{E}(\sigma)), 
$
where 
$\mathcal{E}$
is a channel implemented by this circuit and $\sigma$ denotes a
$d$-qubit
input state.  
The original wire cutting method introduced
\ifthenelse{\boolean{icml-citation-format}}{
by Peng et al. \yrcite{peng2020simulating}, 
}{
by Peng et al. \cite{peng2020simulating}, 
}
is based on replacing 
identity channels with linear combinations of measurement and state preparation operations.  
The decomposition of a single-qubit identity channel
could be expressed as
\begin{align}
\mathcal{I}(\rho) 
= \sum_{i=1}^{8} c_i \rho_i \text{Tr}(O_i \rho).
\label{equ-orig-wire-cut} 
\end{align} 

\begin{figure}[tb]
    \centering
    \begin{subfigure}
        \centering
			\[
			\Qcircuit @C=1em @R=1em {
			 & \qw      & \multigate{1}{U_1} & \qw      & \qw      & \qw &\meter \\
			 & \qw      & \ghost{U_1}        & \qw      & \multigate{1}{U_2} & \qw &\meter \gategroup{2}{4}{2}{4}{1.2em}{--}  \\ 
		  	 & \qw      & \qw                & \qw      & \ghost{U_2}        & \qw  &\meter
			 }
			 \]
			{\small (a) Simple quantum circuit.}
	
	 \label{sub-fig:quantum_circuit_partitioned-a}		
    \end{subfigure}
    \hfill
    \begin{subfigure}
		  \centering
		  \begin{center}
		   \includegraphics[width=0.5\textwidth]{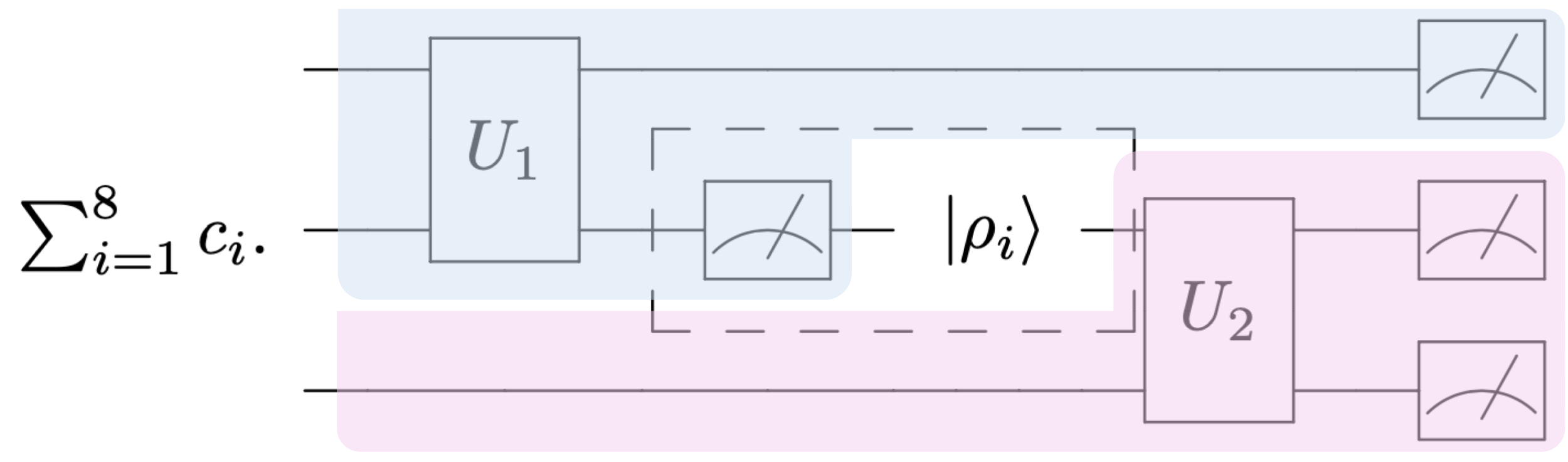}
		   \end{center}
			{\small (b) 
			The circuit after quasiprobabilistically decomposing the identity channel, highlighted by the dashed box in (a),  using Equation  (\ref{equ-orig-wire-cut}).
			}

    \end{subfigure}

 \caption{Quantum circuit partitioned using wire cutting.  The original circuit in (a) could be simulated by running  the subcircuits in (b) and combing the results through classical post-processing.}
 \label{fig:quantum_circuit_partitioned}
\end{figure}

Here,  $\rho$ and $\rho_i$ 
denote 
 density matrices, while $c_i$ and $O_i$ represent a real-valued coefficient and a Hermitian observable corresponding to a measurement, respectively.
\reffig{fig:quantum_circuit_partitioned}
illustrates a simple quantum circuit partitioned using wire cutting. 
After cutting a wire in the circuit, 
the expected value of $O_{out}$ can be expressed as: 
\ifthenelse{\boolean{twocolumnmode}}{
\begin{align*}
\langle O_{out} \rangle = &\text{Tr}(O_{out} \mathcal{E}(\sigma)) \nonumber \\
= &  
\sum_{i=1}^8
c_i \text{Tr}  \big(  
\left( O_i \otimes O_{out} \right)
\big(  \mathcal{E}_{up}(\sigma_{up})  \nonumber \\
&  \otimes \mathcal{E}_{down}(\rho_i \otimes \sigma_{down}) \big)
\big),
\end{align*}
}{
\begin{align*}
\langle O_{out} \rangle = \text{Tr}(O_{out} \mathcal{E}(\sigma))
= 
\sum_{i=1}^8
c_i \text{Tr}  \left(  
\left( O_i \otimes O_{out} \right)
\left(  \mathcal{E}_{up}(\sigma_{up}) \otimes  \mathcal{E}_{down}(\rho_i \otimes \sigma_{down}) \right)
\right),
\end{align*}
}
where  
$\mathcal{E}_{up}$ and  $\mathcal{E}_{down}$ 
represent the channels implemented by the top and bottom subcircuits,  respectively,  
while $\sigma_{up}$ and $\sigma_{down}$ denote marginal states of $\sigma$ corresponding to $\mathcal{E}_{up}$ and  $\mathcal{E}_{down}$: 
$\sigma = \sigma_{up} \otimes \sigma_{down}$.

Equation (\ref{equ-orig-wire-cut}) can be extended to accommodate cutting $m$ parallel wires, which results in the following decomposition \cite{harada2023doubly} for the $m$-qubit identity channel.
\begin{align}
\mathcal{I}^{\otimes m} (\rho) 
= \frac{1}{2^m} \sum_{P \in \{I, X,Y,Z\}^{\otimes m}} \text{Tr}[P \rho]P, 
\label{equ-n-orig-wire-cut}
\end{align}
where $P$ denotes an $m-$qubit Pauli string,  and each term $\text{Tr}[P(.)]P$ can be interpreted as measuring the expectation value of $P$ and subsequently feeding the eigenstates of $P$ into the following subcircuit.  
More efficient decomposition methods for cutting $m-$parallel wires have been proposed  \cite{lowe2023fast,brenner2023optimal,pednault2023alternative,harada2023doubly,harrow2024optimal}.
For instance,  the 
approach proposed 
\ifthenelse{\boolean{icml-citation-format}}{
by Harada et al. \yrcite{harada2023doubly}, 
}{
by Harada et al. \cite{harada2023doubly}, 
}
achieves optimal sampling overhead and minimizes the number of channels required for cutting parallel wires. 
This decomposition is based on removing the redundancy in the number of channels in (\ref{equ-n-orig-wire-cut}) by 
jointly diagonalizing 
the Pauli strings using mutually unbiased bases \cite{wootters1989optimal,lawrence2002mutually,seyfarth2019cyclic,gokhale2020n}. The $4^m -1 $ Pauli strings $\{I, X,Y,Z\}^{\otimes m} \backslash I^{\otimes m}$ in (\ref{equ-n-orig-wire-cut}) could be partitioned into $2^m+1$ disjoint sets $\{ S_i \}_{i=1}^{2^m + 1}$, with each set including $2^m - 1$ mutually commuting Pauli strings \cite{lawrence2002mutually}.  
This is due to the existence of $2^m + 1$ distinct orthonormal bases within an $m$-qubit system that are mutually unbiased \cite{wootters1989optimal}.  Using this partitioning,  we get the following decomposition,  which reduces the cardinality of the sum compared to (\ref{equ-n-orig-wire-cut}).
\ifthenelse{\boolean{twocolumnmode}}{
\begin{align}
\mathcal{I}^{\otimes m}(\rho) &= \frac{1}{2^m} \text{Tr} [I^{\otimes m}\rho] I^{\otimes m} \nonumber \\
&+  \frac{1}{2^m} \sum_{i=1}^{2^m + 1} \sum_{P_{ij} \in S_i} \text{Tr}[P_{ij}\rho]P_{ij}, 
\label{equ-harada-easier-parallel-cut}
\end{align}
}{
\begin{align}
id^{\otimes m}(\rho) = \frac{1}{2^m} \text{Tr} [I^{\otimes m}(\rho)] I^{\otimes m} +  \frac{1}{2^m} \sum_{i=1}^{2^m + 1} \sum_{P_{ij} \in S_i} \text{Tr}[P_{ij}(\rho)]P_{ij}, 
\label{equ-harada-easier-parallel-cut}
\end{align}
}
where $S_i = \{P_{ij} \}_{j=1}^{2^m - 1}$ for each $i$.  
As shown
\ifthenelse{\boolean{icml-citation-format}}{
by Harada et al. \yrcite{harada2023doubly}, 
}{
by Harada et al. \cite{harada2023doubly}, 
}
the following decomposition can be derived from (\ref{equ-harada-easier-parallel-cut}),
where each unitary $U_i$ is implementable 
by a Clifford circuit. 
\ifthenelse{\boolean{twocolumnmode}}{
\begin{align}
&\mathcal{I}^{\otimes m} (\rho) = 
(2^{m+1} - 1) \Bigg(  \nonumber \\
&\sum_{i=1}^{2^m} \frac{1}{2^{m+1} - 1} \sum_{j \in \{0,1 \}^m} \text{Tr} \left[ U_i \ket{j}\bra{j} U_i^\dagger \rho \right] U_i \ket{j}\bra{j} U_i^\dagger \nonumber \\
&- \frac{2^m - 1}{2^{m+1} - 1} \sum_{j \in \{0,1 \}^m} \text{Tr} \left[ \ket{j}\bra{j} \rho \right] \rho_j 
\Bigg),
\label{equ-harada-parallel-cut}
\end{align}
}{
\begin{align}
id^{\otimes m} (\rho) = 
(2^{m+1} - 1) \Bigg(
&\sum_{i=1}^{2^m} \frac{1}{2^{m+1} - 1} \sum_{j \in \{0,1 \}^m} \text{Tr} \left[ U_i \ket{j}\bra{j} U_i^\dagger (\rho) \right] U_i \ket{j}\bra{j} U_i^\dagger \nonumber \\
&- \frac{2^m - 1}{2^{m+1} - 1} \sum_{j \in \{0,1 \}^m} \text{Tr} \left[ \ket{j}\bra{j} (\rho) \right] \rho_j 
\Bigg),
\label{equ-harada-parallel-cut}
\end{align}
}
where
\begin{align*}
\rho_j := \sum_{k \in \{0,1\}^m} \frac{1}{2^m - 1} (1 - \delta_{j,k}) \ket{k}\bra{k},
\end{align*}

the set $\{U_i\}_{i=1}^{2^m} \cup \{I^{\otimes m}\}$ 
consists of operators
 transforming
the computational basis into $2^m + 1$ mutually unbiased bases,  
and $\delta_{j,k}$ denotes the Kronecker delta.

\subsection{Quantum State Teleportation}

Key variants of quantum teleportation include state teleportation \cite{bennett1993teleporting,bouwmeester1997experimental,van2006distributed},  entanglement swapping \cite{zukowski1993event,pan1998experimental}, and gate teleportation \cite{eisert2000optimal,huang2004experimental,meter2006architecture,jiang2007distributed}, which respectively allow one-way transfer of quantum states, bi-directional entanglement distribution, and the remote execution of quantum gates \cite{barral2024review}.
As a fundamental protocol in quantum communication, quantum state teleportation permits the transmission of an unknown quantum state through the use of a single entangled qubit pair (e-bit)  and classical communication channels \cite{bennett1993teleporting,bouwmeester1997experimental,van2006distributed,ren2017ground,cuomo2020towards,cacciapuoti2020entanglement}.
When Alice intends to transmit a quantum state $\rho$ to Bob, the no-cloning theorem \cite{wootters1982single,dieks1982communication,milonni1982photons,barnum1996noncommuting} prohibits her from creating and sending a copy of the state.  The quantum teleportation protocol allows her to conduct this transmission.
A Bell-state measurement (BSM) is carried out by Alice on the qubit representing the state $\rho$ and on one half of the entangled pair shared with Bob.
As a result of the measurement, her two qubits are randomly projected onto one of the four maximally entangled Bell states \cite{nielsen2010quantum}, $\ket{\Phi^{\pm}}$ or $\ket{\Psi^{\pm}}$, with equal probability.  Here,  
 \begin{align*}
\ket{\Phi^{\pm}} = \frac{1}{\sqrt{2}} (\ket{00} \pm \ket{11} ),
\ket{\Psi^{\pm}} = \frac{1}{\sqrt{2}} (\ket{01} \pm \ket{10} ).
 \end{align*}
Bob’s entangled qubit collapses into $B^\dagger \rho B$,  with $B \in \{I,Z,X,ZX\}$ corresponding to the measurement outcome.
Alice subsequently transmits the outcome of her measurement to Bob through two classical bits.  Using the received information,  Bob performs the appropriate operation on his qubit to reconstruct the original state $\rho$.

\begin{figure}[tb]
    \centering
    
	 \begin{center}
	 \includegraphics[width=0.4\textwidth]{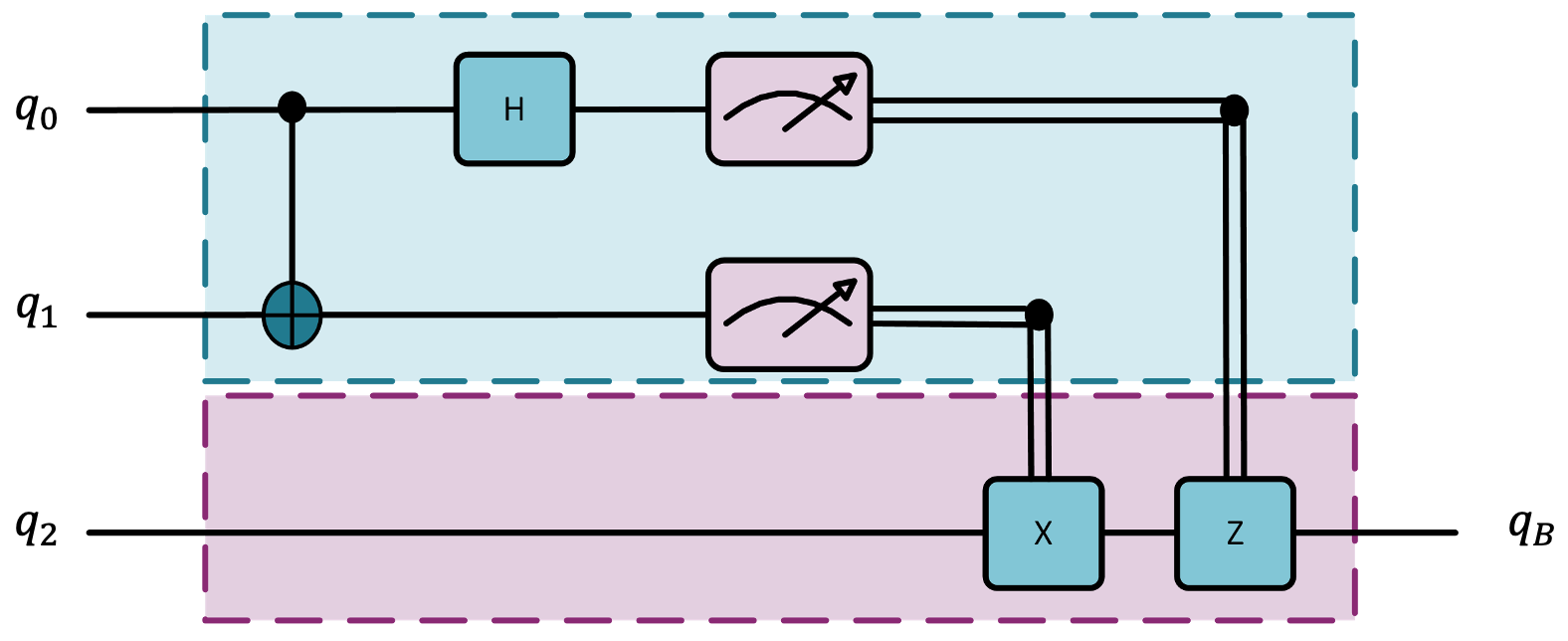}
	 \end{center}
	 \caption{The state teleportation circuit.  $q_0$ is prepared in $\ket{\psi}$ where $\rho = \ket{\psi}\bra{\psi}$ is the state Alice intends to transmit.  $q_1$ and $q_2$ are prepared in the Bell state $\ket{\Phi^{+}}$.  $q_B = \ket{\psi}$ denotes Bob's received qubit. }
	  \label{fig:state-tele}
\end{figure}

\section{Wire Cutting and Adversarial Attacks}
\label{sec:wire-and-adv}

Here, 
we explore how partitioning quantum classifiers by applying wire cutting to their circuits could 
 make them susceptible to adversarial attacks.  
Wire cutting approaches 
may be vulnerable to different adversarial manipulations targeting either
the measurement or state preparation operations.  
Adding adversarial perturbations to the prepared states,  for example,  
could allow an attacker to manipulate a quantum classifier's predictions.
As we uncover in this section,  altering the wire cutting procedure could 
result in the simulated circuit no longer representing a valid quantum circuit.
In this paper,  we assume an adversary’s goal is to influence the output 
without causing this issue.

\begin{figure}[tb]
    \centering
    \begin{subfigure}
		  \centering
		 \begin{center}
		  \includegraphics[width=\linewidth]{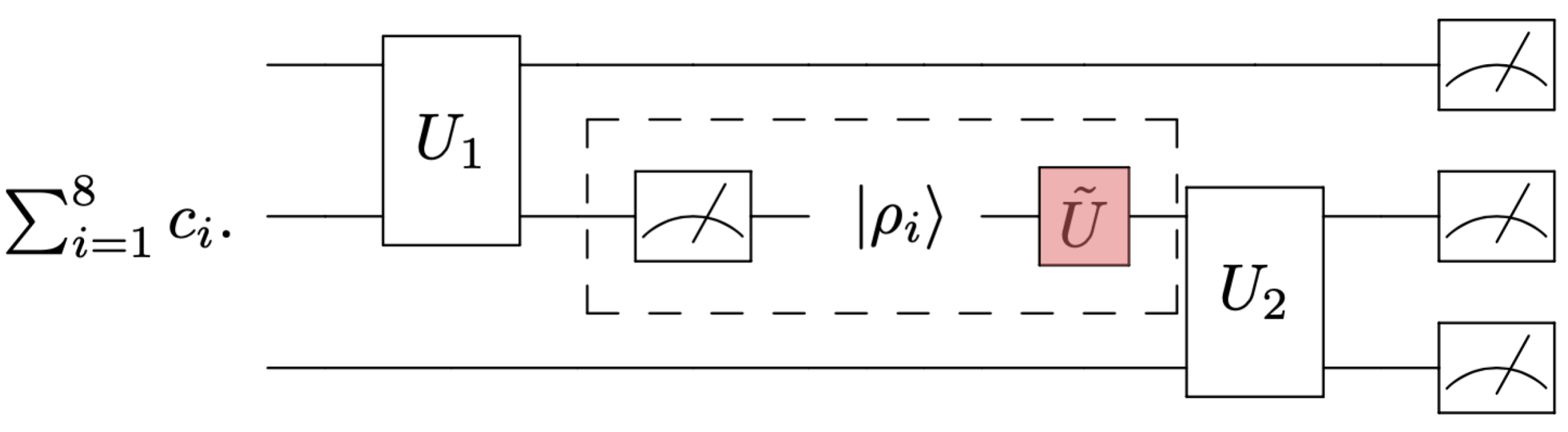}
		  \end{center}
			{\small (a) The quantum circuit in 
			\reffig{fig:quantum_circuit_partitioned}.(a)
			partitioned using the decomposition in (\ref{equ-cut-perturbed-similar-unitaries}). }
    \end{subfigure}
	    \hfill
    \begin{subfigure}
        \centering
         \begin{center}
	    \includegraphics[width=0.6\linewidth]{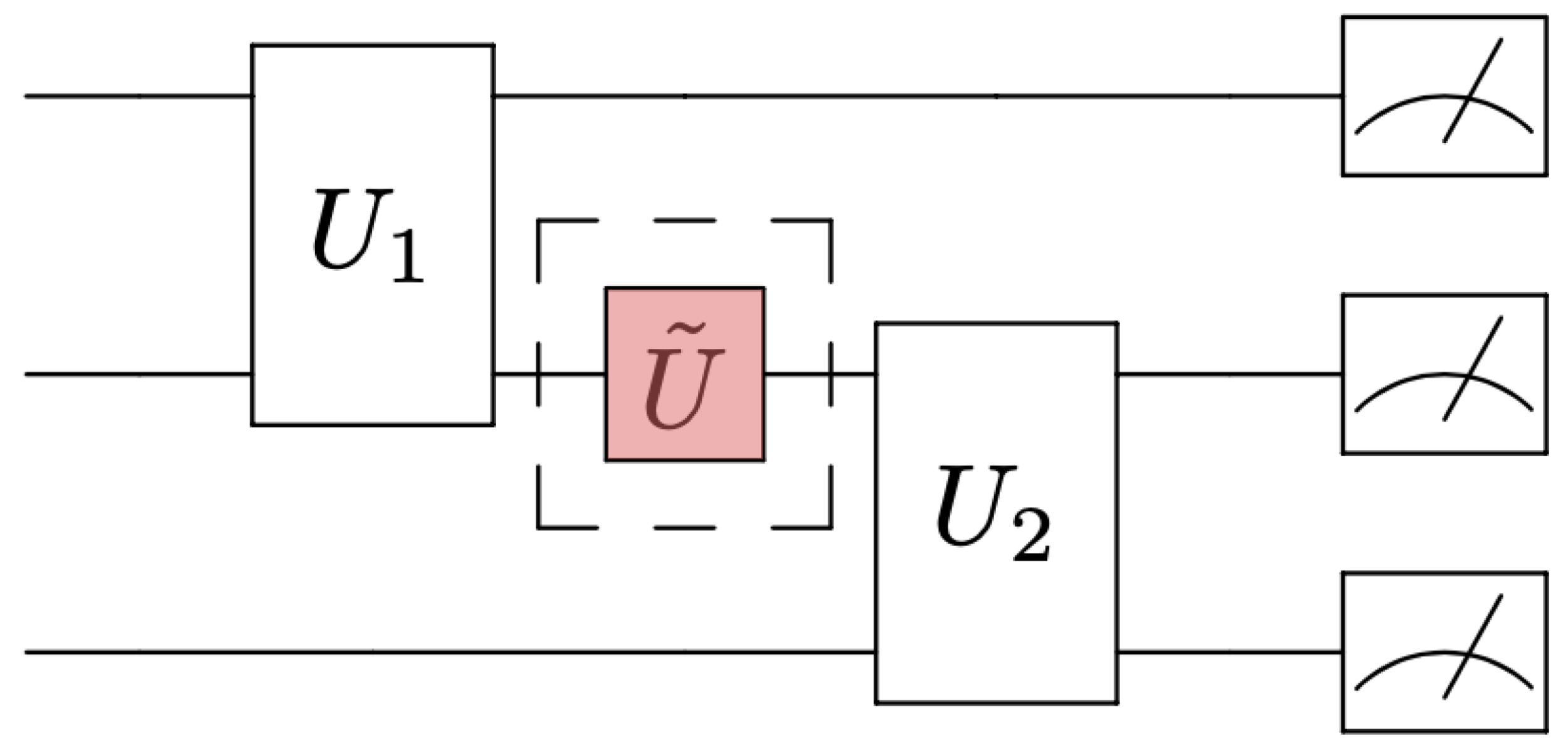}
		\end{center}
			{\small (b) The simulated quantum circuit.}
			
    \end{subfigure}

 \caption{Using the decomposition in (\ref{equ-cut-perturbed-similar-unitaries}) rather than (\ref{equ-orig-wire-cut}) to implement wire cutting would result in a simulated quantum circuit with an additional adversarial gate $\tilde{U}$ compared to the original circuit in 
\reffig{fig:quantum_circuit_partitioned}.(a). }
 \label{fig:quantum_circuit_partitioned_perturbed}
\end{figure}

Consider the original wire cutting decomposition in (\ref{equ-orig-wire-cut}).  
Perturbing one or more of the states within the set $\{ \rho_i \}_{i=1}^8$
would lead to the following decomposition,  with $\{ \tilde{\rho}_i = \tilde{U}_i \rho_i \tilde{U}_i^\dagger \}_{i=1}^8$ and $\{\tilde{U}_i\}_{i=1}^8$ denoting the set of adversarially perturbed states 
and unitary perturbation operators,  respectively. 
\begin{align}
\mathcal{C}_{adv}(\rho) = 
\sum_{i=1}^{8} c_i \tilde{\rho}_i \text{Tr}(O_i \rho) = \sum_{i=1}^{8} c_i \tilde{U}_i \rho_i \tilde{U}_i^\dagger \text{Tr}(O_i \rho), 
\label{equ-cut-perturbed-different-unitaries} 
\end{align} 
where $\mathcal{C}_{adv}: \mathcal{L}(\mathcal{H}) \to \mathcal{L}(\mathcal{H})$,  with  $\mathcal{L}(\mathcal{H})$ denoting the space of 
square 
linear operators acting on $\mathcal{H}$.
Since unitary operators preserve the trace,  it is straightforward to verify 
$\mathcal{C}_{adv}$ 
is trace-preserving.  Nevertheless,  
 it is not necessarily completely positive\footnote{To determine whether $\mathcal{C}_{adv}$ is completely positive (CP), we consider its Choi representation \cite{choi1975completely}.  The corresponding Choi matrix is given by $\mathcal{J}(\mathcal{C}_{adv}) = r.  (\mathcal{I} \otimes \mathcal{C}_{adv}) (\ket{\Phi^+} \bra{\Phi^+}), $  where $\ket{\Phi^+}=\frac{1}{\sqrt{r}} \sum_{i=0}^{r-1} \ket{i} \otimes \ket{i}$ is the maximally entangled state,  and the dimension of the Hilbert space on which $\mathcal{C}_{adv}$ acts is $r=2$.   $\mathcal{C}_{adv}$ is completely positive if and only if $\mathcal{J}(\mathcal{C}_{adv})$ is positive semidefinite.  
Any one-qubit density matrix can be expanded in the normalized Pauli matrices, which form an orthonormal basis for the space of $2 \times 2$ Hermitian matrices: $\rho = \mathcal{I}(\rho) = (1/2) \sum_{P \in \{ I, X,Y,Z \}} \text{Tr}(P\rho) P$. 
Expanding, in this equation, each Pauli matrix outside the trace in its eigenbasis yields an equation of the form (\ref{equ-orig-wire-cut}) \cite{peng2020simulating}.  An example of a map of the form (\ref{equ-cut-perturbed-different-unitaries}) that is not CP can be obtained by expanding each Pauli matrix outside the trace in its eigenbasis in the following equation: $\mathcal{C}_{adv}(\rho) = (1/2) ( \sum_{P \in \{ I, Y,Z \}} \text{Tr}(P\rho) P+  \text{Tr}(X\rho) \tilde{U} X \tilde{U}^\dagger )$,  where $\tilde{U} = Z$.  It is straightforward to check that the Choi representation of this map is not positive semidefinite. 
 }
 and therefore may not constitute a valid quantum channel.
In the special case where a similar perturbation operator is applied to all the $\rho_i$s,  however,  
$\mathcal{C}_{adv}$ 
is a unitary channel:
\ifthenelse{\boolean{twocolumnmode}}{
\begin{align}
\mathcal{C}_{adv}(\rho) 
&= \sum_{i=1}^{8} c_i \tilde{U} \rho_i \tilde{U}^\dagger \text{Tr}(O_i \rho) \nonumber  \\
&=  \tilde{U} \left( \sum_{i=1}^{8} c_i \rho_i \text{Tr}(O_i \rho) \right) \tilde{U}^\dagger \nonumber \\
&= \tilde{U} (\mathcal{I}(\rho)) \tilde{U}^\dagger = \tilde{U}\rho \tilde{U}^\dagger.
\label{equ-cut-perturbed-similar-unitaries} 
\end{align}
}{
\begin{align}
\mathcal{C}_{adv}(\rho) 
= \sum_{i=1}^{8} c_i \tilde{U} \rho_i \tilde{U}^\dagger \text{Tr}(O_i \rho) =  \tilde{U} \left( \sum_{i=1}^{8} c_i \rho_i \text{Tr}(O_i \rho) \right) \tilde{U}^\dagger
= \tilde{U} (id(\rho)) \tilde{U}^\dagger = \tilde{U}\rho \tilde{U}^\dagger.
\label{equ-cut-perturbed-similar-unitaries}
\end{align}
}
In wire cutting, the goal is to simulate a larger quantum circuit by partitioning it into smaller subcircuits. 
The outcomes of these smaller subcircuits are then combined using classical postprocessing.
However,  if the 
wire cutting 
process is adversarially attacked and the subcircuits are fed with perturbed states,  the simulated 
circuit would differ from the intended 
one,
with $\mathcal{C}_{adv}$  replacing an identity channel in the simulated circuit. 
\reffig{fig:quantum_circuit_partitioned_perturbed}
demonstrates a scenario where $\mathcal{C}_{adv}(.) = \tilde{U}(.)\tilde{U}^\dagger$ is a unitary channel,
highlighting
a connection between adversarially
manipulating 
  the state preparation 
  operations
  in wire cutting 
and the adversaries
 in Section \ref{subsec:q-classifiers-and-adv} that are capable of inserting adversarial gates 
 within intermediate layers of a quantum classifier’s circuit.
 
Similar to (\ref{equ-cut-perturbed-different-unitaries}),  
decomposition (\ref{equ-n-orig-wire-cut}) could be adversarially manipulated. 
\begin{align}
\mathcal{C}_{adv}^\prime(\rho) 
= \frac{1}{2^m} \sum_{P \in \{I, X,Y,Z\}^{\otimes m}} \text{Tr}[P(\rho)]\tilde{U}_P P \tilde{U}_P^\dagger,  
\label{equ-n-orig-wire-cut-perturbed}
\end{align}
where
$\mathcal{C}_{adv}^\prime: \mathcal{L}(\mathcal{H}^{\otimes m}) \to \mathcal{L}(\mathcal{H}^{\otimes m})$ and 
$\tilde{U}_P  \in U(2^m)$ 
denote unitary perturbation operators.   
Note that,  in general,  if we decompose each Pauli string $P$ in (\ref{equ-n-orig-wire-cut-perturbed}) using its eigenbasis to feed the eigenstates of $P$ into the subsequent subcircuits,  each eigenstate could be adversarially manipulated using a separate unitary operator.  
Apart from (\ref{equ-n-orig-wire-cut}),  other wire cutting decompositions based on measurement and state preparation operations
may be susceptible to 
similar adversarial attacks.
For instance,  
by adding adversarial perturbations to (\ref{equ-harada-parallel-cut}),  we obtain 
\ifthenelse{\boolean{twocolumnmode}}{
\begin{align}
&C_{adv}^{\prime \prime}( \rho ) = 
(2^{m+1} - 1) \Bigg(  
\sum_{i=1}^{2^m} \frac{1}{2^{m+1} - 1} \nonumber \\
&\sum_{j \in \{0,1 \}^m} \text{Tr} \left[ U_i \ket{j}\bra{j} U_i^\dagger \rho \right] \tilde{U}_{ij}  U_i \ket{j}\bra{j} U_i^\dagger \tilde{U}_{ij}^\dagger \nonumber \\
&- \frac{2^m - 1}{2^{m+1} - 1} \sum_{j \in \{0,1 \}^m} \text{Tr} \left[ \ket{j}\bra{j} \rho \right] \tilde{V}_{j} \rho_j \tilde{V}_{j}^\dagger
\Bigg),
\label{equ-harada-parallel-cut-perturbed}
\end{align}
}{
\begin{align}
C_{adv}^{\prime \prime}( \rho ) = 
(2^{m+1} - 1) \Bigg(
&\sum_{i=1}^{2^m} \frac{1}{2^{m+1} - 1} \sum_{j \in \{0,1 \}^m} \text{Tr} \left[ U_i \ket{j}\bra{j} U_i^\dagger \rho \right]  \tilde{U}_{ij}  U_i \ket{j}\bra{j} U_i^\dagger \tilde{U}_{ij}^\dagger \nonumber \\
&- \frac{2^m - 1}{2^{m+1} - 1} \sum_{j \in \{0,1 \}^m} \text{Tr} \left[ \ket{j}\bra{j} \rho \right] \tilde{V}_{j} \rho_j \tilde{V}_{j}^\dagger 
\Bigg),
\label{equ-harada-parallel-cut-perturbed}
\end{align}
}
where 
$\mathcal{C}_{adv}^{\prime \prime}: \mathcal{L}(\mathcal{H}^{\otimes m}) \to \mathcal{L}(\mathcal{H}^{\otimes m})$ and 
for all $i$ and $j$, 
$\tilde{U}_{ij}$ and $\tilde{V}_j \in U(2^m)$ 
denote adversarial perturbation operators.  
In Equations (\ref{equ-n-orig-wire-cut-perturbed}) and (\ref{equ-harada-parallel-cut-perturbed}),  if a similar perturbation operator is applied to all the prepared states,  meaning 
\begin{align*}
\tilde{U}_P = \tilde{U}^\prime  \,\,\,\,\,\,  &\forall P \text{ in (\ref{equ-n-orig-wire-cut-perturbed})} \\
\tilde{U}_{ij} = \tilde{V}_j = \tilde{U}^{\prime \prime}  \,\,\,\,\,\,  &\forall i,j \text{ in (\ref{equ-harada-parallel-cut-perturbed}),}
\end{align*}
where $\tilde{U}^{\prime}, \tilde{U}^{\prime \prime} \in U(2^m)$,  $C_{adv}^\prime(.) = \tilde{U}^{\prime}(.) \tilde{U}^{\prime \dagger} $ and $C_{adv}^{\prime \prime}(.) = \tilde{U}^{\prime \prime}(.) \tilde{U}^{\prime \prime \dagger}$ 
would represent unitary channels corresponding to $\tilde{U}^{\prime}$ and $\tilde{U}^{\prime \prime}$,  respectively. 
Similar to the single-qubit case,  when the adversarial channels are unitary,  
we could interpret the 
attack 
as equivalent to inserting an $m$-qubit adversarial gate acting within the intermediate layers of the simulated circuit.

We assume the 
adversary's objective is to alter
the wire cutting procedure in a way that the adversarial channels implemented are CPTP maps,  
ensuring the simulated circuit remains a valid quantum circuit, albeit with a manipulated output.  In the wire cutting approaches discussed,  one straightforward method is to use an identical unitary perturbation operator for manipulating all the prepared states.  In the context of attacking quantum classifiers,  this approach enables an adversary to
learn only a single adversarial unitary operator.   
With the connection between adversarial attacks targeting the wire cutting procedure in partitioned quantum classifiers and embedding adversarial gates within intermediate layers of a classifier established,  we now
show a connection between the attack model in  Section \ref{subsec:q-classifiers-and-adv} and quantum classifiers partitioned via state teleporation. 

\section{State Teleportation and Adversarial Attacks}
\label{sec:tele-and-adv}

 \begin{figure*}[tb]
  \centering
  \includegraphics[width=\textwidth]{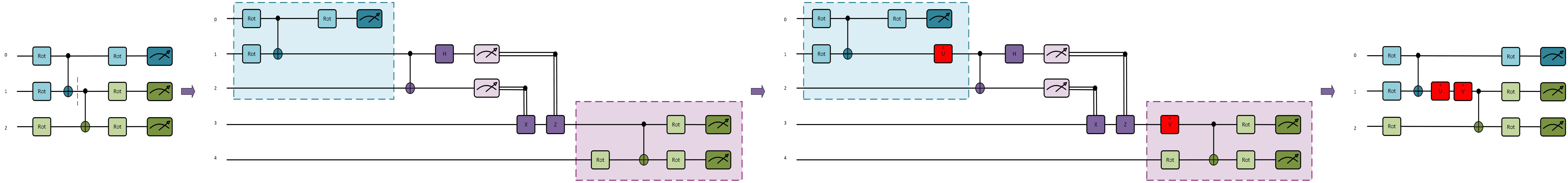}
  \caption{The same circuit from Figure \ref{fig:adv-cut-intro},  partitioned using  state teleportation.  
This figure illustrates a scenario in which the sub-circuits are executed in a distributed manner across different nodes.  If one or both of these nodes are malicious, they could subtly and adversarially influence the circuit's execution.  One way to do this is by executing their assigned sub-circuits correctly,  but introducing small, carefully crafted perturbations to the output or input of their sub-circuits via perturbation gates $\hat{U}$ or $\hat{V}$. 
Such an attack is equivalent to implementing $\hat{U}$, $\hat{V}$,  or both within intermediate layers of the original circuit. }
  \label{fig:adv-tele} 
\end{figure*}

If quantum communication is available,  variational quantum algorithms can be partitioned using teleportation-based methods \cite{diadamo2021distributed,situ2024distributed},  which—unlike circuit cutting—do not introduce exponential sampling overheads and instead consume entangled qubit pairs shared between parties.
As with wire cutting, partitioning quantum classifiers via state teleportation can expose them to various adversarial attacks,including scenarios where the sub-circuits are executed in a distributed manner across multiple nodes, with at least one node being malicious.
As illustrated in Figure \ref{fig:adv-tele}, if a malicious node is responsible for executing a sub-circuit and teleporting its output to another node, it may introduce carefully crafted adversarial perturbations into its output before teleporting it.  Conversely,  a malicious node on the receiving end can adversarially modify the received input state before passing it to the sub-circuit it executes.
These scenarios resemble those explored in classical split learning, where adversarial nodes add perturbations to intermediate outputs of the split model \cite{fan2023OnTR,he2024advusl}.  
As shown in Figure \ref{fig:adv-tele},  such adversarial manipulations to the outputs or inputs of the sub-circuits could be thought of as inserting adversarial gates within intermediate layers of the original circuit.  
Having established the connection between the attack model in Section \ref{subsec:q-classifiers-and-adv} and both wire cutting and state teleporation, we shift our focus in the next section to studying adversarial attacks that implement perturbation gates within intermediate layers of a classifier.

\section{Adversarial Perturbation Operators within Intermediate Layers}
\label{sec:theoretical}

In this section,  
we present theorems bounding the 
predictive 
confidence shift,  $|y_k(\sigma) - \hat{y}_k(\sigma)|$, 
caused by 
implementing multiple adversarial 
gates within  intermediate layers of a quantum classifier.  
A small value of $|y_k(\sigma) - \hat{y}_k(\sigma)|$ indicates robustness and stability against small, malicious changes in the 
classifier's 
architecture,  as it reflects the model's consistent confidence in label \(k\), whereas a large value suggests sensitivity and significant disruption to the prediction.  Especially,  in well-trained classifiers, where decision boundaries are far from data points, bounding this quantity indirectly ensures decision boundary stability.   
Conversely, a large value could indicate  
that the output  
has crossed a decision boundary, leading to misclassification.

Our first theorem establishes an upper bound for the predictive confidence shift 
in 
any quantum classifier in terms of 
the sum of the diamond distances between each unitary perturbation channel in (\ref{equ-hat-E}) and the identity channel.  
This bound can then be further upper-bounded
based on 
  the operator norm.  
This shows that if all the unitary perturbation operators are close to the identity operator,  then the adversarial attack does not trigger a large swing in the classification confidence.
The proof is deferred to Appendix \ref{sec:appendix-theorem-1}.
\begin{theorem}
\label{theorem-confidence-bound}
Consider a quantum classifier attacked by inserting adversarial gates within 
the intermediate layers of its circuit,
where the classifier assigns label $k$ to the input state $\sigma$ with probabilities (\ref{equ-prob-k}) and (\ref{equ-prob-k-attacked}) before and after the attack,  respectively.  Then
\ifthenelse{\boolean{twocolumnmode}}{
\begin{align*}
&|y_k(\sigma) - \hat{y}_k(\sigma)| \nonumber \\
\le & \frac{1}{2} \left(\Vert \mathcal{I}^{\otimes d} - \hat{\mathcal{U}}_0 \Vert_{\diamond} +  \sum_{i=1}^n \Vert \mathcal{I}^{\otimes d_{+}} - \hat{\mathcal{U}}_i \Vert_{\diamond} \right) \nonumber \\
\le &\underset{\phi_0 \in U(1)}{\min}\, \ \Vert I^{\otimes d} - \phi_0 \hat{U}_0 \Vert_{op}  \nonumber \\
&+  \sum_{i=1}^n \underset{\phi_i \in U(1)}{\min}\, \ \Vert I^{\otimes d_+} - \phi_i \hat{U}_i \Vert_{op} , \nonumber \\
\end{align*}
}{
\begin{align*}
&|y_k(\sigma) - \hat{y}_k(\sigma)| \nonumber \\
\le & \frac{1}{2} \left(\Vert \mathcal{I}^{\otimes d} - \hat{\mathcal{U}}_0 \Vert_{\diamond} +  \sum_{i=1}^n \Vert \mathcal{I}^{\otimes d_{+}} - \hat{\mathcal{U}}_i \Vert_{\diamond} \right) \nonumber \\
\le &\underset{\phi_0 \in U(1)}{\min}\, \ \Vert I^{\otimes d} - \phi_0 \hat{U}_0 \Vert_{op}  +  \sum_{i=1}^n \underset{\phi_i \in U(1)}{\min}\, \ \Vert I^{\otimes d_+} - \phi_i \hat{U}_i \Vert_{op}, \nonumber \\
\end{align*}
}
where $I^{\otimes d}$ and $I^{\otimes d_+}$ represent the $d-$qubit and $d_{+}-$qubit identity operators,  respectively,  with $\mathcal{I}^{\otimes d}$ and $\mathcal{I}^{\otimes d_{+}}$ denoting  their associated identity channels. 
\end{theorem}

Since the operator norm can be upper-bounded by the Hilbert-Schmidt norm, we 
immediately 
obtain the following corollary.  The Hilbert-Schmidt norm provides a bridge between Theorem \ref{theorem-confidence-bound} and 
Theorem \ref{theorem-confidence-haar},  
presented after the corollary. 

\begin{corollary}
\label{corollary-confidence-bound}
Under the same setup as Theorem \ref{theorem-confidence-bound},  
\ifthenelse{\boolean{twocolumnmode}}{
\begin{align*}
|y_k(\sigma) - \hat{y}_k(\sigma)| 
\le &\underset{\phi_0 \in U(1)}{\min}\, \ \Vert I^{\otimes d} - \phi_0 \hat{U}_0 \Vert_{2}  \nonumber \\
&+  \sum_{i=1}^n \underset{\phi_i \in U(1)}{\min}\, \ \Vert I^{\otimes d_+} - \phi_i \hat{U}_i \Vert_{2}.\nonumber \\
\end{align*}
}{
\begin{align*}
|y_k(\sigma) - \hat{y}_k(\sigma)|
\le \underset{\phi_0 \in U(1)}{\min}\, \ \Vert I^{\otimes d} - \phi_0 \hat{U}_0 \Vert_{2}  +  \sum_{i=1}^n \underset{\phi_i \in U(1)}{\min}\, \ \Vert I^{\otimes d_+} - \phi_i \hat{U}_i \Vert_{2}, \nonumber \\
\end{align*}
}
\end{corollary}

In the special case  where the adversary perturbs 
the input state only (i.e.,  $\hat{U}_i = I^{\otimes d_+}$ for $i \ge 1$),  Theorem \ref{theorem-confidence-bound} and Corollary \ref{corollary-confidence-bound} can be compared to the analysis in Appendix A of \cite{liao2021robust}\footnote{For additional related theorems,  see Theorem 1 in \cite{dowling2024adversarial} and Lemma 2 in \cite{anil2024generating}.},  which bounds the  predictive confidence difference in terms of the distance between the density matrices of the original and the adversarially perturbed input states.  Our bounds based on the perturbation operators,  as opposed to those based on the density operators,  allow us to provide a theorem for our more general attack model,  outlined in Section \ref{subsec:q-classifiers-and-adv}.

The bound in Theorem \ref{theorem-confidence-bound} weakens as the attack's strength increases and the distances between the perturbation operators and the identity operator grows.  The following theorem provides a probabilistic bound on the predictive confidence shift,  even when the perturbation operators are not close to the identity operator.  Theorem \ref{theorem-confidence-haar} is inspired by Theorem 2 in \cite{dowling2024adversarial}.  However, in comparison,  besides 
our expanded attack model,  our analysis incorporates a more general model for the quantum classifier.   
\ifthenelse{\boolean{icml-citation-format}}{
Dowling et al. \yrcite{dowling2024adversarial}
}{
Dowling et al. \cite{dowling2024adversarial} 
} focus on binary classification,  where the classifier's prediction is determined by the sign of $y(x) = \mathrm{Tr}(Z U \ket{\psi(x)}\bra{\psi(x)} U^\dagger)$,  where $x$ is a classical input,  with $\ket{\psi(x)}$ representing its encoded quantum state.  
Here,  
$Z$ and $U$ denote the Pauli-$Z$ operator acting on a subset of the qubits and a trainable variational unitary,  respectively.   Compared to this setting,  as outlined in 
Section 
\ref{subsection-k-classifier},  we consider $K$-multiclass classification,  a general POVM $\Pi_k$,  
and 
 a CPTP map $\mathcal{E}$ 
which encompasses the specific case of
 the unitary evolution described by $U(.)U^\dagger$.
 Due to our more general setting, the out-of-time-ordered correlator (OTOC) \cite{larkin1969quasiclassical,sekino2008fast,shenker2014black,maldacena2016bound}
 does not appear in our 
 final bound 
  in the same way it does in 
\ifthenelse{\boolean{icml-citation-format}}{
Dowling et al.'s analysis \yrcite{dowling2024adversarial}.
}{
Dowling et al.'s analysis \cite{dowling2024adversarial}.
}  
Instead,  our analysis, 
which can be found in Appendix \ref{sec:appendix-theorem-2},  provides a bound based on the Hilbert-Schmidt distance between $\mathcal{E}(\Pi_k)^\dagger$ and $\hat{\mathcal{E}}(\Pi_k)^\dagger$,  where $\mathcal{E}^\dagger$ and $\hat{\mathcal{E}}^\dagger$ denote the adjoins of the CPTP maps $\mathcal{E}$ and $\hat{\mathcal{E}}$ described in (\ref{equ-prob-k}) and (\ref{equ-hat-E}),  respectively.  
This can be further bounded in terms of the Hilbert-Schmidt distances between the unitary perturbation operators and the identity operator. 

Unlike Theorem \ref{theorem-confidence-bound},  we simplify our analysis for the following theorem by excluding the presence of ancilla bits from the classifier’s input and assuming the circuit's induced CPTP map $\mathcal{E}$ is unital.  
Furthermore,  similar to
\ifthenelse{\boolean{icml-citation-format}}{
Dowling et al.  \yrcite{dowling2024adversarial},
}{
Dowling et al.  \cite{dowling2024adversarial},
}  
we assume the classifier's input $\sigma$ is selected Haar-randomly.  
Note that,  
this
is more of a simplifying assumption 
rather than a realistic one \cite{liao2021robust},  and may lead to 
pessimistic results concerning 
the adversarial vulnerability of quantum classifiers 
(e.g.,  see \cite{liu2020vulnerability}). 
Nevertheless,  it 
provides 
useful tools for theoretical analysis,  which could pave the way for future work that relaxes this assumption. 

Theorem \ref{theorem-confidence-haar} establishes a probabilistic bound for $|y_k(\sigma) - \hat{y}_k(\sigma)|$ by employing Chebyshev’s inequality.  Appendix \ref{sec:appendix-theorem-2} presents the proof of this theorem.
\begin{theorem}
\label{theorem-confidence-haar}
If $\sigma = W \ket{0}\bra{0} W^\dagger \in \mathcal{D}(\mathcal{H}^{\otimes d})$,  with $W$ denoting a unitary operator sampled from the Haar ensemble\footnote{It suffices for our proof that $W$ is drawn from a unitary 2-design \cite{dankert2009exact}.}, 
then for any $\delta > 0$, 
$\vert y_k(\sigma) - \hat{y}_k(\sigma) \vert <  \delta$ holds 
with probability at least 
\begin{align*}
1 - \frac{4 \Vert \Pi_k  \Vert_2^2 \left( \sum_{i=0}^{n} \Vert I^{\otimes d} - \hat{U}_i \Vert_2 \right)^2}{D(D+1) \delta^2},
\end{align*}
where
$D := 2^d$,  
$
y_k(\sigma) = \mathrm{Tr}(\Pi_k \mathcal{E} (\sigma)),
\hat{y}_k(\sigma) = \mathrm{Tr}(\Pi_k \hat{\mathcal{E}} (\hat{U}_0\sigma\hat{U}^\dagger_0)),
$
 and each $\hat{U}_i$ denotes a unitary perturbation operator in (\ref{equ-hat-E-with-unitaries}).

\end{theorem}

\section{Experimental Results}
\label{sec:exp}

This section focuses on experimentally examining the impact of 
planting perturbation operators within layers
of a simulated quantum classifier.
We evaluate 
how inserting one or multiple global or local adversarial gates at 
different circuit depths influences the performance of the model.  
We also experimentally evaluate the bound in Theorem \ref{theorem-confidence-haar} and assess its tightness.

Using Pennylane \cite{bergholm2018pennylane} and Keras \cite{chollet2015keras}, we simulate quantum classifiers in a noiseless setting for 
binary and four-class classification on the MNIST \cite{lecun2010mnist}
   and FMNIST \cite{xiao2017fashion} datasets,  both downsampled to $16 \times 16$ pixels. 
   We further include CIFAR-10 \cite{krizhevsky2009learning},  evaluated only for binary classification.  
We employ quantum classifiers based on hardware-efficient ansätze \cite{kandala2017hardware}, consisting of $\ell$ layers, where each layer contains a rotation unit followed by an entangling unit. The adversarial unitaries we implement within their intermediate layers follow the same structure: each perturbation operator consists of multiple layers of gates, and each layer comprises one rotation block and one entangling block. The full architecture of our classifiers is provided in Appendix \ref{sec:appendix-model-arch}, and their pre-attack performance is reported in Appendix \ref{sec:appendix-classifiers-performance}.

We study the effects of implementing one or multiple adversarial gates within the intermediate layers of our classifiers and investigate two settings for these adversarial gates: global gates that can affect all the qubits, and local gates that can only impact a subset of qubits.  
We assess the effectiveness of each adversarial scenario by measuring the misclassification rate it induces at a given attack strength,  where the attack strength is defined as the sum of the distances between the adversarial perturbation operators and the identity operator.  
For each scenario, we record the misclassification rate and the corresponding attack strength at every training epoch, and then plot misclassification rate versus attack strength to analyze how their relationship evolves over time.  Appendix \ref{sec:appendix-scenarios} details the implemented attack scenarios, the training procedure for the adversarial layers, their hyperparameters, and our evaluation methodology.

Our experiments indicate that partitioned classifiers could be more vulnerable to adversarial attacks than unpartitioned ones in two ways.  First, 
unlike unpartitioned classifiers which can only be attacked by manipulating thier input states,  partitioned ones can by manipulated by adding adversarial gates within their layers,  and we observe that
in some cases implementing adversarial gates within intermediate layers of quantum classifiers can produce more potent adversarial effects than manipulating their input states.   
Second,  partitioned classifiers expose multiple attack surfaces, as adversaries can deploy several adversarial gates within their layers.  Appendix \ref{sec:appendix-interpreting-results} analyzes our experimental results in greater detail, while detailed figures corresponding to our experiments are presented in Appendices \ref{sec:appendix-global-adv} and \ref{sec:appendix-local-adv}.

The bound provided by Theorem  \ref{theorem-confidence-haar} is particularly useful when the pre-attack probabilities assigned by the classifier to each label are known. Using this bound, especially in the binary classification setting, we can predict whether an adversarial attack is likely to flip the classifier’s predicted label.  We experimentally investigate the tightness of the bound provided by this theorem and its practical usefulness. Our experiments, detailed in Appendix \ref{sec:appendix-evaluating-theoretical}, show that the theoretical bound on 
$\vert y_k(\sigma) - \hat{y}_k(\sigma) \vert$ often provides useful  bounds on the values of $\hat{y}_k(\sigma)$ during training the adversarial gates in practice, particularly when the attack strength is small. Since increasing the attack strength generally results in less stealthy attacks, these results demonstrate the practical utility of our bound in scenarios where adversaries implement stealthy attacks.  Recall that $\hat{y}_k(\sigma)$ and  $y_k(\sigma)$ denote the probabilities of assigning label $k$ to  an input sample $\sigma$ before and after an adversarial attack, respectively.

\section{Conclusions}
In this work,  we 
shed light on the adversarial robustness of 
quantum classifiers when partitioned using wire cutting or state teleportation
and 
demonstrated a connection between attacks targeting partitioned classifiers and implementing adversarial gates within intermediate layers of quantum classifiers.  We bound the shift in quantum classifiers' confidence resulting from inserting multiple adversarial gates within their architecture and empirically studied the effects of planting these gates at different circuit depths.  
Our 
findings contribute to a deeper understanding of quantum classifiers' adversarial robustness,  paving the way for further exploration into 
their 
resilience to   
attacks targeting various quantum circuit distribution methods. 

\section*{Acknowledgements}
This research has been funded by the research project "Quantum Software Consortium: Exploring Distributed Quantum Solutions for Canada" (QSC).  QSC is financed under the National Sciences and Engineering Research Council of Canada (NSERC) Alliance Consortia Quantum Grants \#ALLRP587590-23.

\bibliography{ref}

\newpage
\appendix 
\onecolumn  

\section{Proof of Theorem \ref{theorem-confidence-bound}}
\label{sec:appendix-theorem-1}

We use the following lemmata in our analysis.

\begin{lemma}[Subadditivity of diamond distance; Proposition 3.48 in \cite{watrous2018theory}]
For CPTP maps $\mathcal{A}$ and $\mathcal{C}$ from $d$-qubit to $d^\prime$-qubit systems and CPTP maps $ \mathcal{B}$ and $\mathcal{D}$ from $d^\prime$-qubit to $d^{\prime\prime}$-qubit systems, 
\begin{align*}
\Vert \mathcal{A}\mathcal{B} -  \mathcal{C}\mathcal{D} \Vert_{\diamond} 
\le \Vert \mathcal{A} -  \mathcal{C}  \Vert_{\diamond} +  \Vert \mathcal{B} -  \mathcal{D}  \Vert_{\diamond}.
\end{align*}
\label{diamond-subadditivity-lemma}
\end{lemma}

\begin{lemma}[Diamond and operator distance of unitaries; Proposition I.6 in \cite{haah2023query}]
For unitary channels $\mathcal{U}$ and $\mathcal{V}$ 
associated 
to 
unitary matrices 
$U, V \in U(d)$, 
\begin{align}
\frac{1}{2} \Vert \mathcal{U} - \mathcal{V} \Vert_{\diamond} 
\le \min_{\phi \in U(1)} \Vert \phi U - V  \Vert_{op}
\le \Vert \mathcal{U} - \mathcal{V} \Vert_{\diamond},
\label{equ-haah-lemma}
\end{align}
where $\mathcal{U}(.) = U(.)U^\dagger$,  $\mathcal{V}(.) = V(.)V^\dagger$,  and $\Vert . \Vert_{\diamond}$ and $\Vert . \Vert_{op}$ denote  the diamond norm and operator norm,  respectively.  The intermediate term in (\ref{equ-haah-lemma}) represents the distance between the unitary matrices up to a global phase. 
\label{haah-lemma}
\end{lemma}

With these in place,  we are now prepared to proceed with the proof of Theorem \ref{theorem-confidence-bound}.

\begin{proof}
Before delving into the proof,  we define $\tilde{\mathcal{E}} := \hat{\mathcal{E}} \circ (\hat{\mathcal{U}}_0 \otimes \mathcal{I}^{\otimes d_a})$,  where $\hat{\mathcal{U}}_0$ and $ \mathcal{I}^{\otimes d_a}$ denote the unitary channel associated with $\hat{U}_0$ and the the identity channel acting on a $d_a$-qubit system,  receptively.  Using this definition,  we have $\hat{\mathcal{E}} (\hat{U}_0\sigma\hat{U}^\dagger_0 \otimes \ket{a}\bra{a})) = \tilde{\mathcal{E}} (\sigma \otimes \ket{a}\bra{a}))$.  To establish a bound on $|y_k(\sigma) - \hat{y}_k(\sigma)|$,  we begin by deriving the following inequality.
\begin{align} 
\Vert  \mathcal{E} (\sigma \otimes \ket{a}\bra{a}) - \hat{\mathcal{E}} (\hat{U}_0\sigma\hat{U}^\dagger_0 \otimes \ket{a}\bra{a})) \Vert_1 
&= \Vert  \mathcal{E} (\sigma \otimes \ket{a}\bra{a}) - \tilde{\mathcal{E}}(\sigma \otimes \ket{a}\bra{a})) \Vert_1 \nonumber \\
&\le 
\underset{\rho}{\sup}\ \Vert (\mathcal{I}_R \otimes  \mathcal{\mathcal{E}} )(\rho) - (\mathcal{I}_R \otimes  \tilde{\mathcal{E}} )(\rho)\Vert_1
\nonumber \\
&= \Vert \mathcal{E} -   \tilde{\mathcal{E}} \Vert_{\diamond} \nonumber \\
&= \Vert \mathcal{E}_{n} \circ \cdots \circ \mathcal{E}_2 \circ \mathcal{E}_{1} -  \hat{\mathcal{U}}_n \circ \mathcal{E}_{n} \cdots \circ \hat{\mathcal{U}}_1 \circ \mathcal{E}_{1} \circ (\hat{\mathcal{U}}_0 \otimes \mathcal{I}^{\otimes d}) \Vert_{\diamond} \nonumber \\
&\le \Vert \mathcal{I}^{\otimes d_+} - \hat{\mathcal{U}}_0 \otimes \mathcal{I}^{\otimes d} \Vert_{\diamond} +  \sum_{i=1}^n \Vert \mathcal{I}^{\otimes d_{+}} - \hat{\mathcal{U}}_i \Vert_{\diamond} \nonumber \\
&\le \Vert \mathcal{I}^{\otimes d} - \hat{\mathcal{U}}_0 \Vert_{\diamond} +  \sum_{i=1}^n \Vert \mathcal{I}^{\otimes d_{+}} - \hat{\mathcal{U}}_i \Vert_{\diamond} \nonumber \\
&\le 2 \left( \underset{\phi_0 \in U(1)}{\min}\, \ \Vert I^{\otimes d} - \phi_0 \hat{U}_0 \Vert_{op}  +  \sum_{i=1}^n \underset{\phi_i \in U(1)}{\min}\, \ \Vert I^{\otimes d_+} - \phi_i \hat{U}_i \Vert_{op} \right), \nonumber \\
\label{equ-ounding-trace-norm-diamond}
\end{align}
where $I^{\otimes d_+}$ represents the $d_{+}-$qubit identity operator and $\mathcal{I}^{\otimes d_{+}}$ denotes  the associated identity channel. 
The second and third inequalities are due to Lemma \ref{diamond-subadditivity-lemma},  while the forth inequality follows from Lemma \ref{haah-lemma}.
To bound $|y_k(\sigma) - \hat{y}_k(\sigma)|$ using the above inequality,  
we use the following property \cite{nielsen2010quantum} of the trace distance,  which can be derived using Hölder duality for Schatten norms. This property provides a physical interpretation of the trace distance, indicating that it represents the maximum possible difference in measurement outcome probabilities between two states, optimized over all measurement setups.
\begin{align*}
D(\rho_1,  \rho_2)  = \max_{P} \, \text{Tr}[P(\rho_1 - \rho_2)],
\end{align*}
where $D(\rho_1,  \rho_2) = 1/2 \Vert \rho_1 - \rho_2 \Vert_1$ denotes the trace distance between two quantum states $\rho_1$ and $\rho_2$ and the maximization could be taken either over all projectors $P$ or all positive operators such that ${P \le I}$.   
Using this property,  we have
\begin{align}
|y_k(\sigma) - \hat{y}_k(\sigma)| &= \left\vert \text{Tr}(\Pi_k \mathcal{E} (\sigma \otimes \ket{a}\bra{a})) - \text{Tr}(\Pi_k \hat{\mathcal{E}} (\hat{U}_0\sigma\hat{U}^\dagger_0 \otimes \ket{a}\bra{a})) \right\vert \nonumber \\
&= \left\vert \text{Tr}\left(\Pi_k \left[ \mathcal{E} (\sigma \otimes \ket{a}\bra{a}) - \hat{\mathcal{E}} (\hat{U}_0\sigma\hat{U}^\dagger_0 \otimes \ket{a}\bra{a})\right] \right) \right\vert \nonumber \\
&\le 
 \max_{0 \le P \le I} 
\, \text{Tr}\left(P \left[ \mathcal{E} (\sigma \otimes \ket{a}\bra{a}) - \hat{\mathcal{E}} (\hat{U}_0\sigma\hat{U}^\dagger_0 \otimes \ket{a}\bra{a})\right] \right) \nonumber \\
&= \frac{1}{2} \Vert \mathcal{E} (\sigma \otimes \ket{a}\bra{a}) - \hat{\mathcal{E}} (\hat{U}_0\sigma\hat{U}^\dagger_0 \otimes \ket{a}\bra{a}) \Vert_1
\label{equ-using-phyis-interpret-of-trace-dist-new}
\end{align}
where 
$P$ belongs to the set of positive operators and 
the inequality in line 3 follows from the property that $|x| \le y$ if $x \le y$ and $-x \le y$ for two real numbers $x$ and $y$.  
Combining (\ref{equ-ounding-trace-norm-diamond}) and (\ref{equ-using-phyis-interpret-of-trace-dist-new}) completes the proof.
\end{proof}

\section{Proof of Theorem \ref{theorem-confidence-haar}}
\label{sec:appendix-theorem-2}

\begin{proof}
Using Chebyshev’s inequality,  for the random variable $y_k(\sigma) - \hat{y}_k(\sigma)$ and a real number $\delta > 0$,  we have
\begin{align*}
\text{Pr}\{\vert y_k(\sigma) - \hat{y}_k(\sigma) - \mathbb{E}_{W \sim \mu \mathbb{H}}[y_k(\sigma) - \hat{y}_k(\sigma)] \vert \ge  \delta \sqrt{\text{Var}(y_k(\sigma) - \hat{y}_k(\sigma))} \} \le \frac{1}{\delta^2},
\end{align*}
where $\text{Var}(y_k(\sigma) - \hat{y}_k(\sigma))$ denotes the variance.
By replacing $\delta^\prime = \delta  \sqrt{\text{Var}(y_k(\sigma) - \hat{y}_k(\sigma))}$,  
we get: 
\begin{align}
\text{Pr}\{\vert y_k(\sigma) - \hat{y}_k(\sigma) - \mathbb{E}_{W \sim \mu \mathbb{H}}[y_k(\sigma) - \hat{y}_k(\sigma)] \vert \ge  \delta^\prime \} \le \frac{ \text{Var}(y_k(\sigma) - \hat{y}_k(\sigma)) }{(\delta^\prime)^2}.
\label{equ-chebyshev-modified}
\end{align}
We can calculate the expected value as follows. 
\begin{align}
\mathbb{E}_{W \sim \mu \mathbb{H}}[y_k(\sigma) - \hat{y}_k(\sigma)] &= 
\mathbb{E}[
\text{Tr}(\Pi_k \mathcal{E} (\sigma)) 
- \text{Tr}(\Pi_k \hat{\mathcal{E}} (\hat{U}_0\sigma\hat{U}^\dagger_0))] \nonumber \\
& = 
\mathbb{E}[
\text{Tr}(\mathcal{E}^\dagger( \Pi_k) \sigma) 
- \text{Tr}(\hat{\mathcal{E}}^\dagger( \Pi_k)  \hat{U}_0\sigma\hat{U}^\dagger_0)]  \nonumber \\
& = 
\text{Tr}(\mathcal{E}^\dagger( \Pi_k) \mathbb{E}[\sigma]) 
- \text{Tr}(\hat{\mathcal{E}}^\dagger( \Pi_k)  \mathbb{E}[\hat{U}_0\sigma\hat{U}^\dagger_0])  \nonumber \\
& = 
\text{Tr}(\mathcal{E}^\dagger( \Pi_k) \mathbb{E}[\sigma] ) 
- \text{Tr}(\hat{\mathcal{E}}^\dagger( \Pi_k) \mathbb{E}[\sigma]).
\label{equ-exp-value}
\end{align}
Here,  $\mathcal{E}^\dagger$ and  $\hat{\mathcal{E}}^\dagger$ denote the adjoints of the channels $\mathcal{E}$ and $\hat{\mathcal{E}}$,  respectively.  
Line 4 is a result of the invariance of  Haar measure under left and right multiplication by unitary matrices.  
By replacing $\mathbb{E}_{W \sim \mu \mathbb{H}}[\sigma] = \mathbb{E}_{W \sim \mu \mathbb{H}}[W \ket{0}\bra{0} W^\dagger] = (1/D)I$ in (\ref{equ-exp-value}),  
where $D= 2^d$ denotes the dimension of the $D \times D$ unitary $W$ and $I$ is used,  for simplicity,  instead of $I^{\otimes d}$, 
we get
\begin{align}
\mathbb{E}_{W \sim \mu \mathbb{H}}[y_k(\sigma) - \hat{y}_k(\sigma)] = 
\frac{1}{D}\text{Tr}(\mathcal{E}^\dagger( \Pi_k))
- \frac{1}{D}\text{Tr}(\hat{\mathcal{E}}^\dagger( \Pi_k)).
\label{equ-exp-value-2}
\end{align}
It is straightforward to show $\hat{\mathcal{E}}^\dagger(.) = \mathcal{E}_{1}^\dagger \circ \hat{\mathcal{U}}_1^\dagger \cdots \circ \mathcal{E}_n^\dagger \circ \hat{\mathcal{U}}_n^\dagger(.)$.  To see why,  for two operators 
$A,  B \in \mathcal{L}(\mathcal{H}^{\otimes d})$, 
we have: 
\begin{align*}
\text{Tr}(\hat{\mathcal{E}}^\dagger(A) B) = 
&\text{Tr}(A \hat{\mathcal{E}}(B)) \\
=&\text{Tr}(A (\hat{\mathcal{U}}_n \circ \mathcal{E}_{n} \cdots \circ \hat{\mathcal{U}}_1 \circ \mathcal{E}_{1}(B))) \\
=& \text{Tr}((\hat{\mathcal{U}}_n^\dagger (A)) ( \mathcal{E}_{n} \circ \hat{\mathcal{U}}_{n-1} \circ \mathcal{E}_{n-1}  \cdots \circ \hat{\mathcal{U}}_1 \circ \mathcal{E}_{1}(B))) \\
=& \text{Tr}((\mathcal{E}_{n}^\dagger \circ \hat{\mathcal{U}}_n^\dagger (A)) (\hat{\mathcal{U}}_{n-1} \circ \mathcal{E}_{n-1}   \cdots \circ \hat{\mathcal{U}}_1 \circ \mathcal{E}_{1}(B))) \\
&\vdots \\
=& \text{Tr}((\mathcal{E}_{1}^\dagger \circ \hat{\mathcal{U}}_1^\dagger \cdots \circ \mathcal{E}_n^\dagger \circ \hat{\mathcal{U}}_n^\dagger  (A)) B).
\end{align*}  
Given the invariance of the trace under unitary operations,  CPTP maps,  and adjoints of unital maps,
we have $\text{Tr}(\mathcal{E}^\dagger( \Pi_k)) = \text{Tr}(\Pi_k)$ and $\text{Tr}(\hat{\mathcal{E}}^\dagger( \Pi_k)) = \text{Tr}(\mathcal{E}_{1}^\dagger \circ \hat{\mathcal{U}}_1^\dagger \cdots \circ \mathcal{E}_n^\dagger \circ \hat{\mathcal{U}}_n^\dagger( \Pi_k))  = \text{Tr}(\Pi_k)$.  Combining this with (\ref{equ-exp-value-2}),  we obtain
\begin{align}
\mathbb{E}_{W \sim \mu \mathbb{H}}[y_k(\sigma) - \hat{y}_k(\sigma)] = 0.
\label{equ-exp-value-3}
\end{align}
To determine the variance, we can expand 
the following expression  
and  address each term separately.
\begin{align}
\text{Var}(y_k(\sigma) - \hat{y}_k(\sigma))
&= \mathbb{E}_{W \sim \mu \mathbb{H}}[(y_k(\sigma) - \hat{y}_k(\sigma))^2] - \mathbb{E}_{W \sim \mu \mathbb{H}}[y_k(\sigma) - \hat{y}_k(\sigma)]^2 \nonumber \\
&= \mathbb{E}_{W \sim \mu \mathbb{H}}[(y_k(\sigma) - \hat{y}_k(\sigma))^2] \nonumber \\
&= \mathbb{E}_{W \sim \mu \mathbb{H}}[y_k(\sigma)^2 + \hat{y}_k(\sigma)^2 - 2y_k(\sigma)\hat{y}_k(\sigma)] \nonumber \\
&= \mathbb{E}_{W \sim \mu \mathbb{H}}[(\text{Tr}(\Pi_k \mathcal{E} (\sigma)))^2
+ (\text{Tr}(\Pi_k \hat{\mathcal{E}} (\hat{U}_0\sigma\hat{U}^\dagger_0)))^2 - 2\text{Tr}(\Pi_k \mathcal{E} (\sigma))\text{Tr}(\Pi_k \hat{\mathcal{E}} (\hat{U}_0\sigma\hat{U}^\dagger_0)) ]. 
\label{equ-variance}
\end{align}
Since $\text{Tr}(A)\text{Tr}(B) = \text{Tr}(A \otimes B)$ and 
$(AB) \otimes (A^\prime B^\prime) = (A \otimes A^\prime)(B \otimes B^\prime)$ 
for 
operators
$A,B,A^\prime$ and $B^\prime$, 
 we have  
\begin{align*}
(\text{Tr}(\Pi_k \mathcal{E} (\sigma)))^2 = (\text{Tr}( \mathcal{E}^\dagger(\Pi_k) \sigma))^2 = \text{Tr}( \mathcal{E}^\dagger(\Pi_k) \sigma \otimes  \mathcal{E}^\dagger(\Pi_k) \sigma) = \text{Tr}((\mathcal{E}^\dagger(\Pi_k))^{\otimes 2} \sigma^{\otimes 2} ).
\end{align*}
Therefore,
\begin{align}
\mathbb{E}_{W \sim \mu \mathbb{H}}[(\text{Tr}(\Pi_k \mathcal{E} (\sigma)))^2] = 
\mathbb{E}_{W \sim \mu \mathbb{H}}[\text{Tr}((\mathcal{E}^\dagger(\Pi_k))^{\otimes 2} \sigma^{\otimes 2} )] =
\text{Tr}((\mathcal{E}^\dagger(\Pi_k))^{\otimes 2} \mathbb{E}[ \sigma^{\otimes 2}] ).
\label{equ-var-first-term-init}
\end{align}
For an operator $O$ acting on $\mathcal{H} \otimes \mathcal{H}$,    
the following holds  \cite{roberts2017chaos}
\begin{align}
\mathbb{E}_{U \sim \mu \mathbb{H}}[U^\dagger \otimes U^\dagger O U \otimes U] &= 
\int_{U \sim \mu \mathbb{H}} U^\dagger \otimes U^\dagger O U \otimes U \,\mathrm{d}U \nonumber \nonumber \\
&= \frac{1}{D^2 - 1}\left( I\,  \text{Tr}[O] + S\,  \text{Tr}[SO] - \frac{1}{D}S\,  \text{Tr}[O] -\frac{1}{D}I\,  \text{Tr}[SO] \right),
\label{equ-twirl}
\end{align}   
where $S$ denotes the SWAP operator.  Replacing $O$ with $(\ket{0}\bra{0})^{\otimes 2}$ in (\ref{equ-twirl}),  we get  
\begin{align}
\mathbb{E}[\sigma^{\otimes 2}] 
&= \frac{1}{D^2 - 1}\left( I \text{ Tr}[(\ket{0}\bra{0})^{\otimes 2}] +  S \text{ Tr}[(\ket{0}\bra{0})^{2}] - \frac{1}{D}S\text{ Tr}[(\ket{0}\bra{0})^{\otimes 2}] - \frac{1}{D}I \text{ Tr}[(\ket{0}\bra{0})^{2}] \right) \nonumber \\
&= \frac{1}{D^2 - 1}(1 - \frac{1}{D})(I + S) = \frac{I + S}{D(D+1)},
\label{equ-first-use-of-twirl}
\end{align}
where we used the property that $\text{Tr}(S \rho \otimes \rho) = \text{Tr}(\rho^2)$ for a density operator $\rho$.  
Combining (\ref{equ-var-first-term-init}) and (\ref{equ-first-use-of-twirl}),  we get:
\begin{align}
\mathbb{E}_{W \sim \mu \mathbb{H}}[(\text{Tr}(\Pi_k \mathcal{E} (\sigma)))^2]
= \frac{1}{D(D+1)} \left( \text{Tr}((\mathcal{E}^\dagger(\Pi_k))^{\otimes 2}) + \text{Tr}((\mathcal{E}^\dagger(\Pi_k))^{2})  \right).
\label{equ-var-first-term}
\end{align}
We define $\tilde{\mathcal{E}} := \hat{\mathcal{E}} \circ \hat{\mathcal{U}}_0$. 
Similar to (\ref{equ-var-first-term}),  we can show
\begin{align}
\mathbb{E}_{W \sim \mu \mathbb{H}}[(\text{Tr}(\Pi_k \hat{\mathcal{E}} (\hat{U}_0\sigma\hat{U}^\dagger_0)))^2] 
= \mathbb{E}_{W \sim \mu \mathbb{H}}[(\text{Tr}(\Pi_k \tilde{\mathcal{E}} (\sigma)))^2] 
= \frac{1}{D(D+1)} \left( \text{Tr}((\tilde{\mathcal{E}}^\dagger(\Pi_k))^{\otimes 2}) + \text{Tr}((\tilde{\mathcal{E}}^\dagger(\Pi_k))^{2})  \right).
\label{equ-var-2nd-term}
\end{align}
For the third term in (\ref{equ-variance}),  we have:
\begin{align*}
-2 \text{Tr}(\Pi_k \mathcal{E} (\sigma))\text{Tr}(\Pi_k \hat{\mathcal{E}} (\hat{U}_0\sigma\hat{U}^\dagger_0))
&= -2 \text{Tr}(\Pi_k \mathcal{E} (\sigma))\text{Tr}(\Pi_k \tilde{\mathcal{E}} (\sigma)) \\
&= -2 \text{Tr}( \mathcal{E}^\dagger(\Pi_k) \sigma)\text{Tr}(\tilde{\mathcal{E}}^\dagger( \Pi_k)  \sigma) \\
&= -2 \text{Tr}( \mathcal{E}^\dagger(\Pi_k) \sigma \otimes \tilde{\mathcal{E}}^\dagger( \Pi_k)  \sigma) \\
&= -2 \text{Tr}( (\mathcal{E}^\dagger(\Pi_k) \otimes \tilde{\mathcal{E}}^\dagger( \Pi_k)) \sigma^{\otimes 2}),
\end{align*}
Taking the expectation of both sides gives us:
\begin{align}
\mathbb{E}_{W \sim \mu \mathbb{H}}[-2 \text{Tr}(\Pi_k \mathcal{E} (\sigma))\text{Tr}(\Pi_k \hat{\mathcal{E}} (\hat{U}_0\sigma\hat{U}^\dagger_0))]
&= \mathbb{E}_{W \sim \mu \mathbb{H}}[-2 \text{Tr}( (\mathcal{E}^\dagger(\Pi_k) \otimes \tilde{\mathcal{E}}^\dagger( \Pi_k)) \sigma^{\otimes 2})] \nonumber \\
&= -2 \text{Tr}( (\mathcal{E}^\dagger(\Pi_k) \otimes \tilde{\mathcal{E}}^\dagger( \Pi_k)) \mathbb{E} [\sigma^{\otimes 2}]).
\label{equ-var-3rd-term-init}
\end{align}
Utilizing (\ref{equ-first-use-of-twirl}) and (\ref{equ-var-3rd-term-init}),  we obtain
\begin{align}
\mathbb{E}_{W \sim \mu \mathbb{H}}[-2 \text{Tr}(\Pi_k \mathcal{E} (\sigma))\text{Tr}(\Pi_k \hat{\mathcal{E}} (\hat{U}_0\sigma\hat{U}^\dagger_0))] = 
\frac{-2}{D(D+1)} \left( \text{Tr}( \mathcal{E}^\dagger(\Pi_k) \otimes \tilde{\mathcal{E}}^\dagger( \Pi_k)) + \text{Tr}( \mathcal{E}^\dagger(\Pi_k) \tilde{\mathcal{E}}^\dagger( \Pi_k))  \right).
\label{equ-var-3rd-term}
\end{align}
By combining (\ref{equ-variance}) with (\ref{equ-var-first-term}),  (\ref{equ-var-2nd-term}),  and (\ref{equ-var-3rd-term}),  we get
\begin{align*}
\text{Var}(y_k(\sigma) - \hat{y}_k(\sigma)) = \frac{1}{D(D+1)} \Big[ &\text{Tr}((\mathcal{E}^\dagger(\Pi_k))^{\otimes 2}) + \text{Tr}((\mathcal{E}^\dagger(\Pi_k))^{2}) + \text{Tr}((\tilde{\mathcal{E}}^\dagger(\Pi_k))^{\otimes 2}) + \text{Tr}((\tilde{\mathcal{E}}^\dagger(\Pi_k))^{2}) \\
&-2 \left( \text{Tr}( \mathcal{E}^\dagger(\Pi_k) \otimes \tilde{\mathcal{E}}^\dagger( \Pi_k)) + \text{Tr}( \mathcal{E}^\dagger(\Pi_k) \tilde{\mathcal{E}}^\dagger( \Pi_k)) \right)
\Big].
\end{align*}
By rearranging the terms,  we have
\begin{align*}
\text{Var}(y_k(\sigma) - \hat{y}_k(\sigma)) = \frac{1}{D(D+1)} \Big[
&\text{Tr}((\mathcal{E}^\dagger(\Pi_k))^{\otimes 2}) + \text{Tr}((\tilde{\mathcal{E}}^\dagger(\Pi_k))^{\otimes 2}) - 2\text{Tr}( \mathcal{E}^\dagger(\Pi_k) \otimes \tilde{\mathcal{E}}^\dagger( \Pi_k)) \\
&+ \text{Tr}((\mathcal{E}^\dagger(\Pi_k))^{2})  + \text{Tr}((\tilde{\mathcal{E}}^\dagger(\Pi_k))^{2}) -2   \text{Tr}( \mathcal{E}^\dagger(\Pi_k) \tilde{\mathcal{E}}^\dagger( \Pi_k))
\Big].
\end{align*}
By replacing $\text{Tr}((\mathcal{E}^\dagger(\Pi_k))^{\otimes 2})$ with $(\text{Tr}(\mathcal{E}^\dagger(\Pi_k)))^2$,  $\text{Tr}((\tilde{\mathcal{E}}^\dagger(\Pi_k))^{\otimes 2})$ with $(\text{Tr}(\tilde{\mathcal{E}}^\dagger(\Pi_k)))^2$,  and $ - 2\text{Tr}( \mathcal{E}^\dagger(\Pi_k) \otimes \tilde{\mathcal{E}}^\dagger( \Pi_k))$ with $ - 2\text{Tr}( \mathcal{E}^\dagger(\Pi_k))\text{Tr}(\tilde{\mathcal{E}}^\dagger( \Pi_k))$,  we obtain
\begin{align*}
\text{Var}(y_k(\sigma) - \hat{y}_k(\sigma)) = \frac{1}{D(D+1)} \Big[
&(\text{Tr}(\mathcal{E}^\dagger(\Pi_k)) -  \text{Tr}(\tilde{\mathcal{E}}^\dagger(\Pi_k)))^2 \\
&+ \text{Tr}((\mathcal{E}^\dagger(\Pi_k))^{2})  + \text{Tr}((\tilde{\mathcal{E}}^\dagger(\Pi_k))^{2}) -2   \text{Tr}( \mathcal{E}^\dagger(\Pi_k) \tilde{\mathcal{E}}^\dagger( \Pi_k))
\Big].
\end{align*}
As we previously argued,  $\text{Tr}(\mathcal{E}^\dagger(\Pi_k)) = \text{Tr}(\tilde{\mathcal{E}}^\dagger(\Pi_k)) = \text{Tr}(\Pi_k)$.  Therefore,  $(\text{Tr}(\mathcal{E}^\dagger(\Pi_k)) -  \text{Tr}(\tilde{\mathcal{E}}^\dagger(\Pi_k)))^2 = 0$,  and 
\begin{align*}
\text{Var}(y_k(\sigma) - \hat{y}_k(\sigma)) &= \frac{1}{D(D+1)} \Big[
\text{Tr}((\mathcal{E}^\dagger(\Pi_k))^{2})  + \text{Tr}((\tilde{\mathcal{E}}^\dagger(\Pi_k))^{2}) -2   \text{Tr}( \mathcal{E}^\dagger(\Pi_k) \tilde{\mathcal{E}}^\dagger( \Pi_k))
\Big] \\
& = \frac{1}{D(D+1)} \text{Tr}\left( (\mathcal{E}^\dagger(\Pi_k))^{2} + (\tilde{\mathcal{E}}^\dagger(\Pi_k))^{2} - 2\mathcal{E}^\dagger(\Pi_k) \tilde{\mathcal{E}}^\dagger( \Pi_k) \right) \\
& = \frac{1}{D(D+1)} \text{Tr}\left( (\mathcal{E}^\dagger(\Pi_k) - \tilde{\mathcal{E}}^\dagger(\Pi_k))^2  \right).
\end{align*}
Since $\Pi_k$ is Hermitian and adjoints of CPTP maps preserve Hermiticity, 
$(\mathcal{E}^\dagger(\Pi_k) - \tilde{\mathcal{E}}^\dagger(\Pi_k))^2 = (\mathcal{E}^\dagger(\Pi_k) - \tilde{\mathcal{E}}^\dagger(\Pi_k))^\dagger (\mathcal{E}^\dagger(\Pi_k) - \tilde{\mathcal{E}}^\dagger(\Pi_k))$,  and 
\begin{align}
\text{Var}(y_k(\sigma) - \hat{y}_k(\sigma)) &=  \frac{1}{D(D+1)} \text{Tr}\left( (\mathcal{E}^\dagger(\Pi_k) - \tilde{\mathcal{E}}^\dagger(\Pi_k))^\dagger (\mathcal{E}^\dagger(\Pi_k) - \tilde{\mathcal{E}}^\dagger(\Pi_k))  \right) \nonumber \\
&= \frac{1}{D(D+1)} D_{HS}(\mathcal{E}^\dagger(\Pi_k),  \tilde{\mathcal{E}}^\dagger(\Pi_k) )^2,
\label{equ-var-bound-1}
\end{align}
where $D_{HS}$ denotes the Hilbert-Schmidt distance.  We can bound this distance as follows.
\begin{align}
D_{HS}(\mathcal{E}^\dagger(\Pi_k),  \tilde{\mathcal{E}}^\dagger(\Pi_k) )
= &D_{HS}(\mathcal{E}_{1}^\dagger \circ \mathcal{E}_{2}^\dagger  \cdots \circ \mathcal{E}_n^\dagger( \Pi_k),  \hat{\mathcal{U}}_0^\dagger \circ \mathcal{E}_{1}^\dagger \circ \hat{\mathcal{U}}_1^\dagger \cdots \circ \mathcal{E}_n^\dagger \circ \hat{\mathcal{U}}_n^\dagger( \Pi_k) ) \nonumber \\
= &\Vert \mathcal{E}_{1}^\dagger \circ \mathcal{E}_{2}^\dagger  \cdots \circ \mathcal{E}_n^\dagger( \Pi_k) - \hat{\mathcal{U}}_0^\dagger \circ \mathcal{E}_{1}^\dagger \circ \hat{\mathcal{U}}_1^\dagger \cdots \circ \mathcal{E}_n^\dagger \circ \hat{\mathcal{U}}_n^\dagger( \Pi_k) \Vert_2  \nonumber \\
= &\Vert \mathcal{E}_{1}^\dagger \circ  \cdots \circ \mathcal{E}_n^\dagger( \Pi_k) -  \hat{U}_0^\dagger ( \mathcal{E}_1^\dagger \circ \cdots \circ \mathcal{E}_n^\dagger \circ \hat{\mathcal{U}}_n^\dagger( \Pi_k) ) \hat{U}_0 \Vert_2  \nonumber \\
= &\Vert \mathcal{E}_{1}^\dagger \circ  \cdots \circ \mathcal{E}_n^\dagger( \Pi_k) 
- \hat{U}_0^\dagger ( \mathcal{E}_{1}^\dagger \circ  \cdots \circ \mathcal{E}_n^\dagger( \Pi_k) ) \nonumber \\
&+ \hat{U}_0^\dagger ( \mathcal{E}_{1}^\dagger \circ  \cdots \circ \mathcal{E}_n^\dagger( \Pi_k) )
-   \hat{U}_0^\dagger ( \mathcal{E}_1^\dagger \circ \cdots \circ \mathcal{E}_n^\dagger \circ \hat{\mathcal{U}}_n^\dagger( \Pi_k) ) \hat{U}_0  \Vert_2  \nonumber \\
\le &\Vert \mathcal{E}_{1}^\dagger \circ  \cdots \circ \mathcal{E}_n^\dagger( \Pi_k) 
- \hat{U}_0^\dagger ( \mathcal{E}_{1}^\dagger \circ  \cdots \circ \mathcal{E}_n^\dagger( \Pi_k) )  \Vert_2  \nonumber \\
&+  \Vert \hat{U}_0^\dagger ( \mathcal{E}_{1}^\dagger \circ  \cdots \circ \mathcal{E}_n^\dagger( \Pi_k) )
-   \hat{U}_0^\dagger ( \mathcal{E}_1^\dagger \circ \cdots \circ \mathcal{E}_n^\dagger \circ \hat{\mathcal{U}}_n^\dagger( \Pi_k) ) \hat{U}_0  \Vert_2  \nonumber \\
\le &\Vert I - \hat{U}_0^\dagger \Vert_2 \Vert \mathcal{E}_{1}^\dagger \circ  \cdots \circ \mathcal{E}_n^\dagger( \Pi_k)  \Vert_2  \nonumber \\
&+  \Vert \mathcal{E}_{1}^\dagger \circ  \cdots \circ \mathcal{E}_n^\dagger( \Pi_k) 
-  ( \mathcal{E}_1^\dagger \circ \cdots \circ \mathcal{E}_n^\dagger \circ \hat{\mathcal{U}}_n^\dagger( \Pi_k) ) \hat{U}_0  \Vert_2  \nonumber \\
\le &\Vert I - \hat{U}_0^\dagger \Vert_2 \Vert \Pi_k  \Vert_2  \nonumber \\
&+  \Vert \mathcal{E}_{1}^\dagger \circ  \cdots \circ \mathcal{E}_n^\dagger( \Pi_k) 
-  ( \mathcal{E}_1^\dagger \circ \cdots \circ \mathcal{E}_n^\dagger \circ \hat{\mathcal{U}}_n^\dagger( \Pi_k) ) \hat{U}_0  \Vert_2,  \nonumber \\
\label{equ-bounding-D_HS-1}
\end{align}
where $\Vert .  \Vert_2$ denotes Schatten $2$-norm (also called the Hilbert-Schmidt norm).  
The first inequality is a result of the triangle inequality.  
The second inequality follows from the sub-multiplicative property of the Schatten $2$-norm and its invariance under unitary transformations.  
The third inequality above holds because the Schatten $2$-norm is contractive under the adjoint of unital CPTP maps.
We can bound the second term in (\ref{equ-bounding-D_HS-1}) using a similar approach:
\begin{align}
&\Vert \mathcal{E}_{1}^\dagger \circ  \cdots \circ \mathcal{E}_n^\dagger( \Pi_k) 
-  ( \mathcal{E}_1^\dagger \circ \cdots \circ \mathcal{E}_n^\dagger \circ \hat{\mathcal{U}}_n^\dagger( \Pi_k) ) \hat{U}_0  \Vert_2 \nonumber \\
= & \Vert \mathcal{E}_{1}^\dagger \circ  \cdots \circ \mathcal{E}_n^\dagger( \Pi_k) 
- (\mathcal{E}_{1}^\dagger \circ  \cdots \circ \mathcal{E}_n^\dagger( \Pi_k)) \hat{U}_0 \nonumber \\
&+  (\mathcal{E}_{1}^\dagger \circ  \cdots \circ \mathcal{E}_n^\dagger( \Pi_k)) \hat{U}_0
-  ( \mathcal{E}_1^\dagger \circ \cdots \circ \mathcal{E}_n^\dagger \circ \hat{\mathcal{U}}_n^\dagger( \Pi_k) ) \hat{U}_0  \Vert_2 \nonumber \\
\le & \Vert (\mathcal{E}_{1}^\dagger \circ  \cdots \circ \mathcal{E}_n^\dagger( \Pi_k)) (I - \hat{U}_0) \Vert_2 \nonumber \\
&+ \Vert (\mathcal{E}_{1}^\dagger \circ  \cdots \circ \mathcal{E}_n^\dagger( \Pi_k)
-  \mathcal{E}_1^\dagger \circ \cdots \circ \mathcal{E}_n^\dagger \circ \hat{\mathcal{U}}_n^\dagger( \Pi_k) ) \hat{U}_0  \Vert_2 \nonumber \\
\le & \Vert (\mathcal{E}_{1}^\dagger \circ  \cdots \circ \mathcal{E}_n^\dagger( \Pi_k))  \Vert_2 \Vert I - \hat{U}_0 \Vert_2 \nonumber \\
&+ \Vert \mathcal{E}_{1}^\dagger \circ  \cdots \circ \mathcal{E}_n^\dagger( \Pi_k)
-  \mathcal{E}_1^\dagger \circ \cdots \circ \mathcal{E}_n^\dagger \circ \hat{\mathcal{U}}_n^\dagger( \Pi_k) \Vert_2 \nonumber \\
\le & \Vert \Pi_k  \Vert_2 \Vert I - \hat{U}_0 \Vert_2 \nonumber \\
&+ D_{HS}( \mathcal{E}_{1}^\dagger \circ  \cdots \circ \mathcal{E}_n^\dagger( \Pi_k)
,  \mathcal{E}_1^\dagger \circ \cdots \circ \mathcal{E}_n^\dagger \circ \hat{\mathcal{U}}_n^\dagger( \Pi_k) )
\label{equ-bounding-D_HS-2}
\end{align}
Note that,  $\Vert I - \hat{U}_0 \Vert_2 = \Vert I - \hat{U}_0^\dagger \Vert_2 $ since the Schatten $2$-norm of an operator is the same as the Schatten $2$-norm of its adjoint.
Combining (\ref{equ-bounding-D_HS-1}) and (\ref{equ-bounding-D_HS-2}),  
we have
\begin{align*}
D_{HS}(\mathcal{E}^\dagger(\Pi_k),  \tilde{\mathcal{E}}^\dagger(\Pi_k) ) 
&\le 2 \Vert I - \hat{U}_0 \Vert_2 \Vert \Pi_k  \Vert_2 + 
D_{HS}( \mathcal{E}_{1}^\dagger \circ  \cdots \circ \mathcal{E}_n^\dagger( \Pi_k)
,  \mathcal{E}_1^\dagger \circ \hat{\mathcal{U}}_1^\dagger \circ \cdots \circ \mathcal{E}_n^\dagger \circ \hat{\mathcal{U}}_n^\dagger( \Pi_k) ) \nonumber \\
&\le 2 \Vert I - \hat{U}_0 \Vert_2 \Vert \Pi_k  \Vert_2 + 
D_{HS}( \mathcal{E}_{2}^\dagger \circ  \cdots \circ \mathcal{E}_n^\dagger( \Pi_k)
,  \hat{\mathcal{U}}_1^\dagger  \circ \mathcal{E}_2^\dagger \circ \cdots \circ \mathcal{E}_n^\dagger \circ \hat{\mathcal{U}}_n^\dagger( \Pi_k) ),
\end{align*}
where the second inequality holds since the Schatten $2$-norm is contractive under the adjoint of unital CPTP maps.
It is straightforward to inductively show
\begin{align}
D_{HS}(\mathcal{E}^\dagger(\Pi_k),  \tilde{\mathcal{E}}^\dagger(\Pi_k) ) 
\le 
2 \Vert \Pi_k  \Vert_2 \left( \sum_{i=0}^{n} \Vert I - \hat{U}_i \Vert_2 \right).
\label{equ-var-bound-2}
\end{align}
By combining (\ref{equ-chebyshev-modified}) and (\ref{equ-exp-value-3}) with (\ref{equ-var-bound-1}) and (\ref{equ-var-bound-2}),  we get
\begin{align*}
\text{Pr}\{\vert y_k(\sigma) - \hat{y}_k(\sigma) \vert \ge  \delta^\prime \} 
&\le \frac{D_{HS}(\mathcal{E}^\dagger(\Pi_k),  \tilde{\mathcal{E}}^\dagger(\Pi_k) )^2 }{D(D+1) (\delta^\prime)^2} \nonumber \\
&\le \frac{4 \Vert \Pi_k  \Vert_2^2 \left( \sum_{i=0}^{n} \Vert I - \hat{U}_i \Vert_2 \right)^2}{D(D+1) (\delta^\prime)^2}.
\end{align*}
This completes the proof.  
\end{proof}

\section{Experimental Results}
\label{sec:appendix-experiments}

\subsection{Model Architecture}
\label{sec:appendix-model-arch}

\begin{figure}[h] 
  \centering
\includegraphics[width=\linewidth]{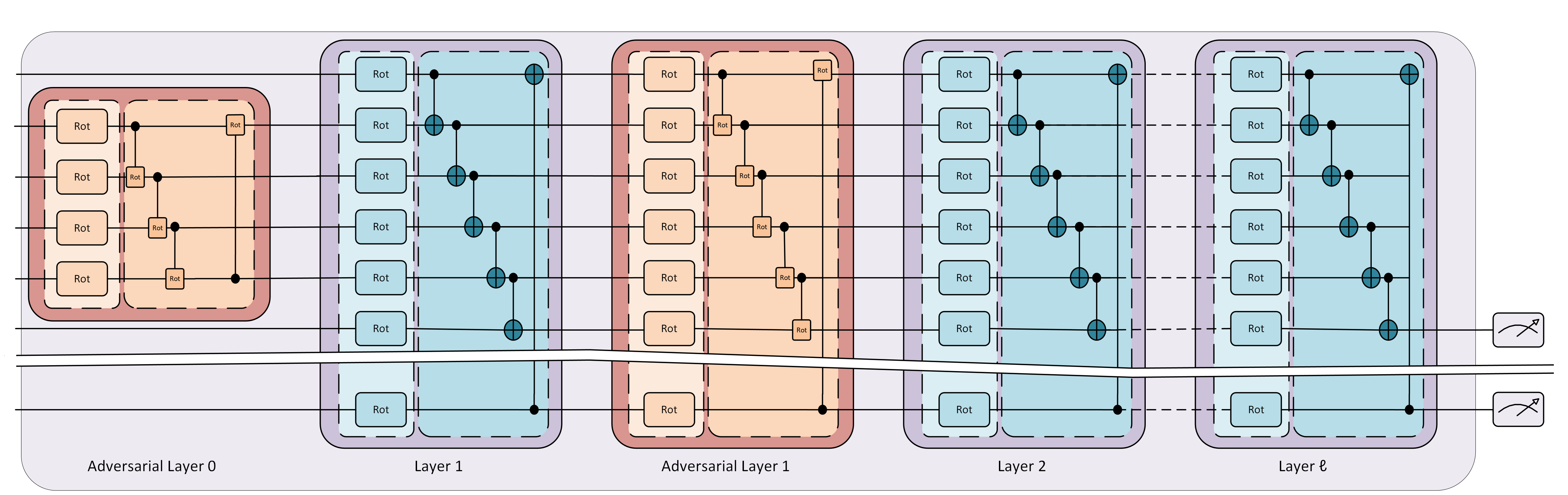}
    \caption{Architecture of the quantum classifier employed in our experiments.  
The top $d$-qubits  correspond to the input state, while the  $d_a$-qubits at the bottom represent the ancilla bits.   The measurements at the output
are performed on the bottom $ \lceil \log_2{K} \rceil$ qubits.   
    Depending on the experimental setup, a number of adversarial layers are 
     added either within the classifier's architecture or before its first layer to perturb the input state. 
    These adversarial layers 
    can target 
    all qubits or act locally on a subset of qubits.
Throughout the rest of the paper, 
    the qubit at the topmost wire will be referred to as qubit $1$,  the qubit on the next wire below as qubit $2$,  and so on, with the qubit at the bottommost wire labeled as qubit $d_+$.
    }
    
  \label{fig:experimental-settup}
\end{figure}

We employ parametrized quantum circuit (PQC)-based classifiers using hardware-efficient ansätze \cite{kandala2017hardware} composed of $\ell$ layers, with each layer including a rotation unit and an entangling unit.  Each rotation unit consists of single-qubit rotation gates with three trainable parameters: $Rot(\omega_1, \omega_2,\omega_3)=RZ(\omega_1)\cdot RY(\omega_2)\cdot RZ(\omega_3)$.  
Each entangling layer is structured such that the qubits are sequentially interconnected in a cyclic manner. Specifically, qubit \(i\) is entangled with qubit \(i+1\) for \(i = 1, 2, \dots, (d_+)-1\) using CNOT gates, and the final qubit, \(d_+\), is entangled with the first qubit, \(1\), thereby forming a closed loop of entanglements.  Recall that $d_+ = d + d_a$ denotes the total number of qubits in the classifier.  
We employ amplitude encoding to map classical data into a quantum state.  Consequently, \( d = \lceil \log_2{c} \rceil \) qubits are required to represent \( c \)-dimensional classical data.
For $K$-class classification, each classifier includes $d_a =  \lceil \log_2{K} \rceil$ ancilla bits initialized to $\ket{0}$\footnote{The number of ancilla bits is inspired by the settings used
\ifthenelse{\boolean{icml-citation-format}}{
by Anil et al. \yrcite{anil2024generating}.
}{
by Anil et al. \cite{anil2024generating}.
}}.  
Fig.  \ref{fig:experimental-settup} shows a classifier with adversarial layers implemented within its intermediate layers.  
We begin by training the classifier without any adversarial layers.  Once the classifier is trained, we freeze its weights, add the desired adversarial layers based on our different experimental settings, and then train only the new layers.
The adversarial layers have a similar architecture to the classifier's layer,  with CNOT gates replaced 
by controlled phase-shift gates $CRZ(\phi)$  
that apply a phase shift of angle \(\phi\) to the target qubit around the $Z$-axis when the control qubit is in the \(\ket{1}\) state.  
When \(\phi\) is set to zero, this controlled gate behaves as an identity gate, irrespective of the control qubit's state.
This allows us to initially set the adversarial layers to function as identity gates, and observe their impact on the model's performance as we train them.

To calculate gradients for our simulated quantum classifier, we employ backpropagation due to its computational efficiency within simulation environments that support automatic differentiation. When working with real quantum hardware, the parameter-shift rule \cite{guerreschi2017practical, mitarai2018quantum,bergholm2018pennylane,schuld2019evaluating}
is required because backpropagation relies on the explicit knowledge of the system's internal operations, which 
could be inaccessible.
The parameter-shift rule,
which enables gradient estimation by systematically varying parameters and measuring the resulting changes in the output, requires at least two forward passes for each parameter to estimate its gradient, significantly increasing computational overhead compared to backpropagation as the number of parameters grows. 

As outlined in 
Section \ref{sec:wire-and-adv},  
a key motivation for studying the robustness of quantum classifiers against adversarial unitaries within their intermediate layers is its connection to 
partitioned quantum classifiers' robustness 
to attacks targeting the wire cutting procedure. 
However,  applying wire cutting to quantum classifiers with strongly entangled ansätze 
 comes with its own difficulties.   
Since all the qubits are interconnected in these ansätze,  at least $d_+$ wires 
must be cut
 to obtain
two separate subcircuits.   
To simulate the original circuit, 
the number of subcircuits that need to be executed grows exponentially with the number of cuts,  
resulting in a very large number of circuit evaluations required to accurately reconstruct the original circuit.
When running the 
circuit on quantum hardware,  one can resort to approximation methods
to reduce the computational overhead while sacrificing some accuracy  \cite{marshall2023high}.
Applying circuit cutting to a simulated quantum circuit,  however,   does not increase the asymptotic simulation cost.  
Nevertheless,  to cut a 
strongly entangled circuit, 
it is necessary to generate and run the subcircuits in parallel,  
 as doing so sequentially would make the implementation time impractical.
Since the tools for parallelizing the implementation of circuit cutting are  not as readily available as those for  
simulating quantum circuits,  and parallelizing the circuit cutting procedure is beyond the scope of our paper, we focus 
our experiments on 
evaluating the robustness of quantum classifiers to adversarial layers implemented within their architecture without 
 incorporating those layers into the circuit via 
circuit cutting. 
As discussed in Section \ref{sec:wire-and-adv},  for partitioned classifiers,  
applying adversarial perturbations to the state preparation channels in wire cutting is equivalent to implementing adversarial gates within intermediate layers of the original circuit before partitioning it (see Fig.~\ref{fig:adv-cut-intro}).  Consequently,  it suffices to study the impact of adding adversarial gates within the quantum classifier’s architecture  experimentally, allowing us to avoid the substantial overhead associated with implementing circuit cutting.  
Note that if the circuit cutting process is carried out under ideal conditions,  free from noise, errors, and approximations,  
planting the adversarial layers through circuit cutting 
produces the exact same effects as directly 
 implementing these layers 
 in the simulation.
Due to the connection between adversarial perturbations targeting state teleporation and implementing adversarial gates within intermediate layers of a classifier,  as explored in Section \ref{sec:tele-and-adv}, studying the latter problem enables us to analyze adversarial perturbations targeting both wire cutting and state teleporation without incurring the additional overhead of implementing state teleporation as well.

As noted 
\ifthenelse{\boolean{icml-citation-format}}{
by Guala et al. \yrcite{guala2023practical}, 
}{
by Guala et al. \cite{guala2023practical}, 
}
there are ansätze,  
such as those based on tree tensor networks \cite{shi2006classical,tagliacozzo2009simulation},
that are better-suited than strongly entangled ansätze 
for integration with circuit cutting techniques.
These circuits can be cut in a way that each tensor block corresponds to a circuit fragment,  resulting in a number of circuits to evaluate that increases polynomially with the number of tensor blocks.   
However,  recent arguments suggest that circuit architectures such as tree tensor networks and quantum convolutional neural networks (QCNNs) \cite{cong2019quantum} that are free of barren plateaus \cite{pesah2021absence,miao2024isometric} can be classically simulable in polynomial time by a classical algorithm augmented by classical shadows \cite{cerezo2023does,bermejo2024quantum}.
Therefore,  
one might consider these ansätze suboptimal choices for a quantum classifier.
We chose strongly entangled classifiers for our experiments as they are widely used and allow us to investigate the robustness to adversarial unitaries inserted at different depths in classifiers that maintain a consistent number of qubits across all layers, preserving dimensionality throughout their depth.

\subsection{Classifiers' Performance}
\label{sec:appendix-classifiers-performance}

Table \ref{tab:classification-performance} summarizes
the classifiers' performance for MNIST and FMNIST before they are 
subjected to adversarial attacks.  
The classifiers are trained with depths of 10 and 20 for binary classification, and with depths of 20 and 40 for four-class classification. 
For binary classification, classes 0 and 1 from both the MNIST and FMNIST datasets are used, while for four-class classification, classes 0, 1, 2, and 3 from both datasets are utilized.
In the FMNIST dataset, these classes correspond to images of T-shirts/tops, trousers, pullovers, and dresses.
The
models are trained with the Adam optimizer \cite{Kingma2017AdamAM} and a batch size of $64$ for 5 epochs with a learning rate of 
$0.001$, followed by 5 epochs with a learning rate of 
$0.0001$ for binary classification.
 For four-class classification,  they 
 are trained for 30 epochs, with learning rates 
$0.001$,  
$0.0001$,  and 
$0.00001$ used during the first, middle, and final 10 epochs, respectively.
Some of the settings here,  such as the choice of classes from MNIST and FMNIST for binary and four-class training, the optimizer, batch size, number of epochs, and learning rates, are inspired by the settings employed
\ifthenelse{\boolean{icml-citation-format}}{
by Anil et al. \yrcite{anil2024generating}.
}{
by Anil et al. \cite{anil2024generating}.}

\begin{table}[h]
\centering
\caption{
Test accuracy comparison across model depths for binary and multi-class classification on MNIST and FMNIST datasets.}
\vskip 0.15in
\begin{tabular}{c|cc|cc}
\hline
& \multicolumn{2}{c|}{MNIST} & \multicolumn{2}{c}{FMNIST} \\
\cline{2-5}
Number of layers & Binary & Four-class & Binary & Four-class \\
\hline
10 & 99.67\% & -  & 94.80\% & - \\
20 & 99.86\% & 90.40\% & 94.75\% & 79.80\% \\
40 & - & 92.18\% & - & 84.18\% \\
\hline
\end{tabular}
\label{tab:classification-performance}
\end{table}

For CIFAR-10,  inspired by the settings used 
\ifthenelse{\boolean{icml-citation-format}}{
by Winderl et al. \yrcite{winderl2024quantum},   
}{
by Winderl et al. \cite{winderl2024quantum},  
}
the classifier is trained only for binary classification between classes zero and nine.  We refer to this two-class subset as CIFAR-2.
The images are resized from $32 \times 32 \times 3$ to $64 \times 64$, after which histogram of oriented gradients (HOG) features are extracted using $8 \times 8$ pixel cells and an inner block size of $1 \times 1$ cells,  resulting in a 576-dimensional feature vector. 
We employ a classifier with depth 80,  trained using Adam optimizer for 40 epochs with a learning rate of 0.0005 and a batch size of 64.
Our model achieves a test accuracy of $83.05\%$.

\subsection{Adversarial Attacks}

Here,  we examine the robustness of the classifiers with different depths  to adversarial perturbations targeting their intermediate layers.  

\subsubsection{Attack Scenarios and Parameters}
\label{sec:appendix-scenarios}

\paragraph{Global and Local Adversarial Layers:} 
We explore three different scenarios for the number of qubits on which the adversarial layers can act.
In the first settings, we assume the adversarial layers can impact all the qubits in the circuit, similar to adversarial layer 1 in Fig.  \ref{fig:experimental-settup},  
and 
study the effects of adding one or more blocks of 
such 
layers to the architecture.  
In the other two settings, the adversarial layers are restricted to acting only on local qubits,  similar to adversarial layer 0 in Fig.  \ref{fig:experimental-settup}. 
In the two settings where we study local adversarial gates, we consider gates acting on either three or four qubits. For each setting, we select a random block of consecutive qubits on which the adversarial gates act: three neighboring qubits for the three-qubit gates, and four neighboring qubits for the four-qubit gates. To ensure consistency across all experiments, we fix these selections: all three-qubit adversarial gates act on qubits $3, 4,  5$, and all four-qubit gates act on qubits $5,6,7,8$.

\paragraph{Single versus Multiple Adversarial Layers:} 
For all three scenarios described above,  
our goal is to compare the effects of adding a single block of adversarial layers to different depths, as well as incorporating multiple blocks into the model's architecture.
For the case where a single 
block of adversarial layers
is inserted within the layers of the classifier's circuit,  
we consider four scenarios.  In the first, an adversarial block is inserted between the input and the first layer of the circuit,  perturbing the classifier's input state.  In the second, an adversarial block is added between the $\lceil \ell / 4 \rceil-$th and $(\lceil \ell / 4 \rceil + 1)-$th layers,  where $\ell$ denotes the number of classifier's layers. 
The third scenario involves inserting an adversarial block after the $\lceil \ell / 2 \rceil-$th layer, before the $(\lceil \ell / 2 \rceil+1)-$th.  Finally, in the fourth scenario, an adversarial block is introduced after the $\lceil 3( \ell / 4) \rceil-$th layer, just before the  $(\lceil 3(\ell / 4) \rceil + 1)-$th.  
We compare the effects of these adversarial blocks with the simultaneous insertion of three blocks in place of those described in the second, third, and fourth scenarios.
Regarding the scenario in which the adversarial layers can act on all qubits and are inserted between the input and the first layer of the circuit, 
note that in Equation (\ref{equ-prob-k-attacked}) and in earlier sections of the paper, we assume an adversarial unitary perturbing the input state does not affect the ancilla bits.  This assumption is made to align with prior literature in this area,  which often assumes that the ancillary bits remain unaffected. 
However, in our experiments, we assume that the adversarial block of layers between the input and the first layer acts on all qubits in a manner similar to the other adversarial blocks examined,  in order to maintain consistency.  Note that the analysis in Section \ref{sec:theoretical} could be easily extended to a scenario where $U_0$ also affects the ancilla bits. 

\paragraph{Training the Adversarial Layers:}
Consider a quantum classifier 
with multiple adversarial gates within its intermediate layers, 
similar to the one shown in Fig.  \ref{fig:main}.(c), where the classifier is trained,  but the adversarial gates 
have not
been trained yet and act as identity gates.  
Let $\hat{\theta}$ denote the parameters the adversarial unitaries $\{ \hat{U}_i \}_{i=0}^n$ depend on.  
Our experiments focus on untargeted attacks (see Section \ref{subsec:q-classifiers-and-adv-background}). Therefore, our objective is to train $\hat{\theta}$ using an optimization criterion similar to the one presented in (\ref{equ-untargeted-general}).
We employ an iterative process that progressively refines the adversarial unitaries to increase their effectiveness. 
To control the strength of the attack,  i.e.,  the sum of the distances between the perturbation operators and the identity operator,  we employ a constraint-based loss function.
Following our analysis in Section \ref{sec:theoretical}, 
one possible approach 
 involves using the Hilbert-Schmidt norm:
\begin{align*} 
(\hat{U}_0,  \hat{U}_1,  \cdots,  \hat{U}_n) = 
\underset{\hat{\theta}}{\argmax}\,
\left( L(\hat{y}(\sigma_i),  Y(\sigma_i)) + \gamma \sum_{i=1} \Vert I - \hat{U}_i \Vert_2 \right),
\end{align*}
where $\gamma$ is a hyperparameter controlling the trade-off between the attack's strength and its effectiveness.  
However,  when $\hat{\theta}$ represents a weights matrix,  it is possible to 
opt 
for a simpler yet practically effective method,  which relies on adding $\Vert \hat{\theta} \Vert_{\ell_2}$ to the optimization objective,  
with $||.||_{\ell_2}$ denoting the $\ell_2$-norm: 
\begin{align} 
(\hat{U}_0,  \cdots,  \hat{U}_n) =  
\underset{\hat{\theta}}{\argmax}\left(L(\hat{y}(\sigma_i),  Y(\sigma_i))
+ \gamma \Vert \hat{\theta} \Vert_{\ell_2}\right),
\label{equ-adv-loss-l2}
\end{align} 
where for the loss function $L$ in our experiments, we employ  
the cross-entropy loss function, which is the same loss function used to train our 
classifiers before subjecting them to adversarial attacks. 
Using this optimization objective eliminates the need to calculate the Hilbert-Schmidt distances between the perturbation operators and the identity operator during training,  
resulting in a more computationally efficient training process.

We consider a white-box setting,   where the adversary has complete knowledge of the target classifier.   
To train the adversarial layers,  we freeze the classifier's weights,  and employ a loss function similar to (\ref{equ-adv-loss-l2}).  
Since each single-qubit rotation gate requires three trainable parameters and each controlled phase-shift gate needs one trainable parameter,
for a circuit with $\ell_{adv}$ adversarial layers,
the trainable parameters can be represented by a $d_{adv} \times \ell_{adv} \times  4$ matrix,  where $d_{adv}$ is the number of qubits the adversarial layers act on.  We employ the $\ell_2$ norm of this matrix in place of $\Vert \hat{\theta} \Vert_2$ in (\ref{equ-adv-loss-l2}).
The trainable parameters are initialized to zero before training begins, causing the adversarial layers to behave as identity operators initially.

\paragraph{Evaluating the Adversarial Attacks:}
To evaluate the effectiveness of the adversarial layers in causing incorrect predictions for each classifier,   we compare the misclassification rates across different scenarios.
Specifically, we record the misclassification rate and the corresponding attack strength at each epoch during the training of each adversarial scenario, plotting the misclassification rate versus attack strength to analyze their relationship over time.
This 
visualization helps identify the strength levels needed to maximize misclassification
for each attack,  facilitating 
comparisons of attack efficiency and convergence dynamics across different scenarios.
Here, by misclassification rate, we refer to the percentage of incorrect predictions made by the attacked classifier after each epoch, when evaluated on the entire test set.  
Furthermore, to evaluate the attack strength, we use the sum of the normalized Hilbert Schmidt distances between the adversarial unitaries inserted into the architecture and the identity operator: 
\( \left(1/ \sqrt{2^{d_{adv}}}\right)  
\sum_{i=0}^n || \hat{U}_i - I^{\otimes d_{adv}} ||_2
\),  where $n$ denotes the total number of adversarial unitaries,  and each $\hat{U}_i  \in U(2^{d_{adv}})$.
Normalizing the distance 
prevents 
larger operators 
from artificially dominating the metric due to their size.

\paragraph{Hyperparameters for Training the Adversarial Layers:}
For MNIST and FMNIST,  when inserting a single adversarial unitary operator into the architecture,  we use a block consisting of 10 adversarial layers in all our experiments.  As explained earlier,  we compare the effect of adding one adversarial unitary operator versus simultaneously inserting three adversarial unitaries into the circuits.  When three unitaries are inserted,  we consider two scenarios:  In the first,  the three adversarial unitaries consist of 3, 3,  and 4 adversarial layers,  respectively,  resulting in a total of 10 layers.  In the second scenario,  each adversarial unitary consists of 10 layers.  
For CIFAR-2, when inserting a single unitary operator, we conduct experiments with two configurations: one in which the unitary consists of 10 layers and another in which it consists of 40 layers.  This choice is motivated by the fact that the classifier used for CIFAR-2 is deeper than those used for MNIST and FMNIST. Therefore, we aim to examine whether an increased number of adversarial layers can have a greater impact on the model.
The parameter $\gamma$ in (\ref{equ-adv-loss-l2}) controls whether the attack strength (i.e.,  the sum of the normalized Hilbert Schmidt distances) is allowed to increase significantly during training.  In our experiments,  we set $\gamma=0$.  As explained in Section $\ref{sec:appendix-hyperparameters}$,  constraining the attack strength yields learning curves that closely match the early epochs of the curves obtained when the attack strength is allowed to grow freely.  Consequently, it suffices to run the experiments with $\gamma=0$.
The adversarial layers are trained using stochastic gradient descent,  
a batch size of 64 for binary classification and a batch size of 256 for four-class classification.  The larger batch size for four-class classification helps accelerate training,  as the dataset is larger for this task.
The learning rate for training the adversarial layers is set to 0.001 for MNIST,  matching the rate used for training the classifier. For FMNIST, as detailed in Section $\ref{sec:appendix-hyperparameters}$,  we experimented with learning rates of 0.001 and 0.005, ultimately selecting 0.005 due to its superior performance.  Similarly, for CIFAR-2,  we adopted a higher learning rate of 0.005 for the adversarial layers, compared to the classifier’s learning rate of 0.0005.  Section $\ref{sec:appendix-hyperparameters}$ provides the rationale for these hyperparameter choices and reports the comparative results that motivated the selected learning rates.

\subsubsection{Interpreting the Results}
\label{sec:appendix-interpreting-results}
Here,  we analyze the experimental results; the associated plots are presented in the following Sections \ref{sec:appendix-global-adv} and \ref{sec:appendix-local-adv} .
\paragraph{Vulnerability of Partitioned vs Unpartitioned Classifiers:} 
In the figures presented in Sections \ref{sec:appendix-global-adv} and \ref{sec:appendix-local-adv},  plots that exhibit higher misclassification rates for a given attack strength correspond to more successful adversarial attacks.   Across these figures, we observe that introducing a single adversarial block can, in some instances,  yield more successful attacks when the perturbations are implemented within intermediate layers of the circuit than when they target the input layer (see Figs.~\ref{fig-exp-1}.(a),  \ref{fig-exp-3}.(b),  \ref{fig-exp-4}.(b),  \ref{fig-experiments-cifar-1},  \ref{fig-exp-5}.(a),  \ref{fig-exp-8},  \ref{fig-exp-9},  \ref{fig-exp-10}.(b),  or \ref{fig-exp-11}.(b)).
This suggests that partitioning and distributing quantum classifiers can increase their vulnerability to adversarial attacks.  In a distributed setting,  adversarial perturbations can be introduced within intermediate layers of the partitioned classifiers by manipulating the wire cutting or state teleportation procedures,  whereas unpartitioned classifiers can only be attacked by perturbing their input states.  A second way in which partitioned quantum classifiers are more vulnerable than centralized ones is that an adversary can inject multiple perturbations across their intermediate layers by manipulating multiple wire cutting or state teleportation processes simultaneously.

Comparing the effects of inserting three adversarial blocks versus a single one,  we observe that when the attack strength is measured using the (normalized) sum of Hilbert Schmidt distances between the adversarial unitaries implemented within the architecture and the identity operator,  single adversarial blocks are often able to achieve a higher misclassification rate at a given attack strength.  
This,  however,  depends on how we define the attack strength and would change if we defined it in terms of the (normalized) average of Hilbert Schmidt distances. 
In such cases,  the plots associated with single adversarial blocks would remain unchanged, whereas those corresponding to multiple blocks would often show improved performance, surpassing that of single blocks.  This makes sense intuitively: when the attack strength is defined based on the average of Hilbert Schmidt distances,  using multiple adversarial blocks generally leads to better outcomes.
For clarity, all plots are based on the sum of Hilbert Schmidt distances.  However, one can infer how they might differ if the average were used instead.

\paragraph{Global vs Local Adversarial Layers:}
By comparing the figures in Section \ref{sec:appendix-global-adv} with those in Section \ref{sec:appendix-local-adv},  we observe that when the adversarial layers are allowed to act on all qubits,  they often perform better than local adversarial gates.  However,  this is not always the case: in some instances,  restricting the adversarial attacks to local qubits can yield more successful attacks.  For example,  compare Figs.~\ref{fig-exp-5} and \ref{fig-exp-9},  which correspond to local adversarial attacks,  with Fig.  \ref{fig-exp-1},  which corresponds to global attacks.  

\paragraph{Effects of the Number of Layers:}
Figures in Sections \ref{sec:appendix-global-adv} and \ref{sec:appendix-local-adv} show that adversarial blocks consisting of more layers tend to produce more successful attacks (compare subfigures (a) and (b) in Figs. ~\ref{fig-experiments-cifar-1},  \ref{fig-experiments-cifar-2},  and \ref{fig-experiments-cifar-3}).  
Conversely, although deeper classifiers (i.e., classifiers with more layers) are generally expected to be more resilient to adversarial attacks, this is not always observed in practice.  For instance, Figs. ~\ref{fig-exp-3} and \ref{fig-exp-4} illustrate a case where the deeper classifier (Fig. \ref{fig-exp-4}) is not more robust.

\newcommand{\myscale}{0.7}
\newcommand{\scaleA}{1}
\newcommand{\scaleB}{0.8}
\subsubsection{Global Adversarial layers}
\label{sec:appendix-global-adv}

Here,  we present the result for the case where the adversarial layers act on all qubits.  In the following plots,  each line represents the average of three runs,  with the shaded regions surrounding the lines indicating the variance across these runs. The individual points correspond to the actual data collected from the experiments,  and the dotted lines connect these points to illustrate the trends observed.  
The fluctuations in the plots result from the inherent randomness of stochastic gradient descent and the shuffling of batches at each epoch, which alters the data order used for gradient calculation, causing slight variations in the optimization path and final model parameters.

\begin{figure}[H]
    \centering
    \begin{minipage}{0.47\textwidth}  
        \centering
        \scalebox{\scaleA}{
        \includegraphics[width=\textwidth]{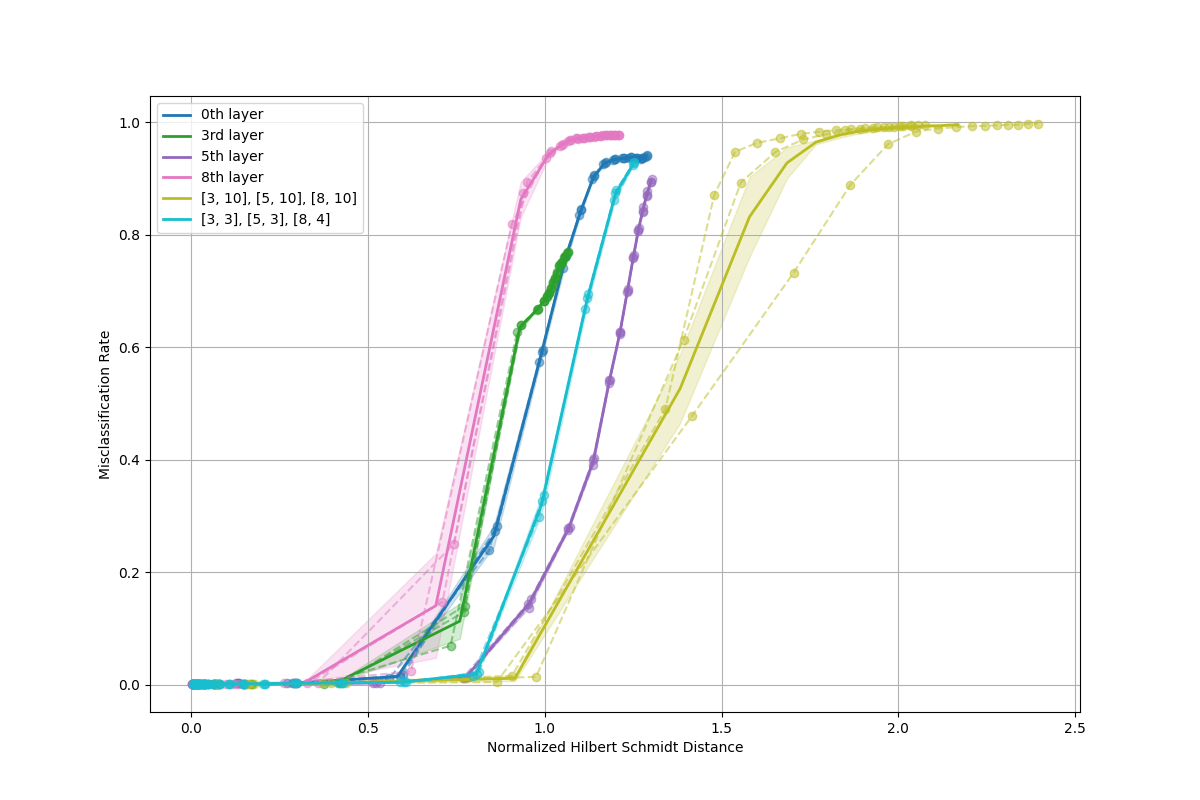}}  
        \caption*{(a)}
    \end{minipage} \hfill
    \begin{minipage}{0.47\textwidth}  
        \centering
        \scalebox{\scaleA}{
        \includegraphics[width=\textwidth]{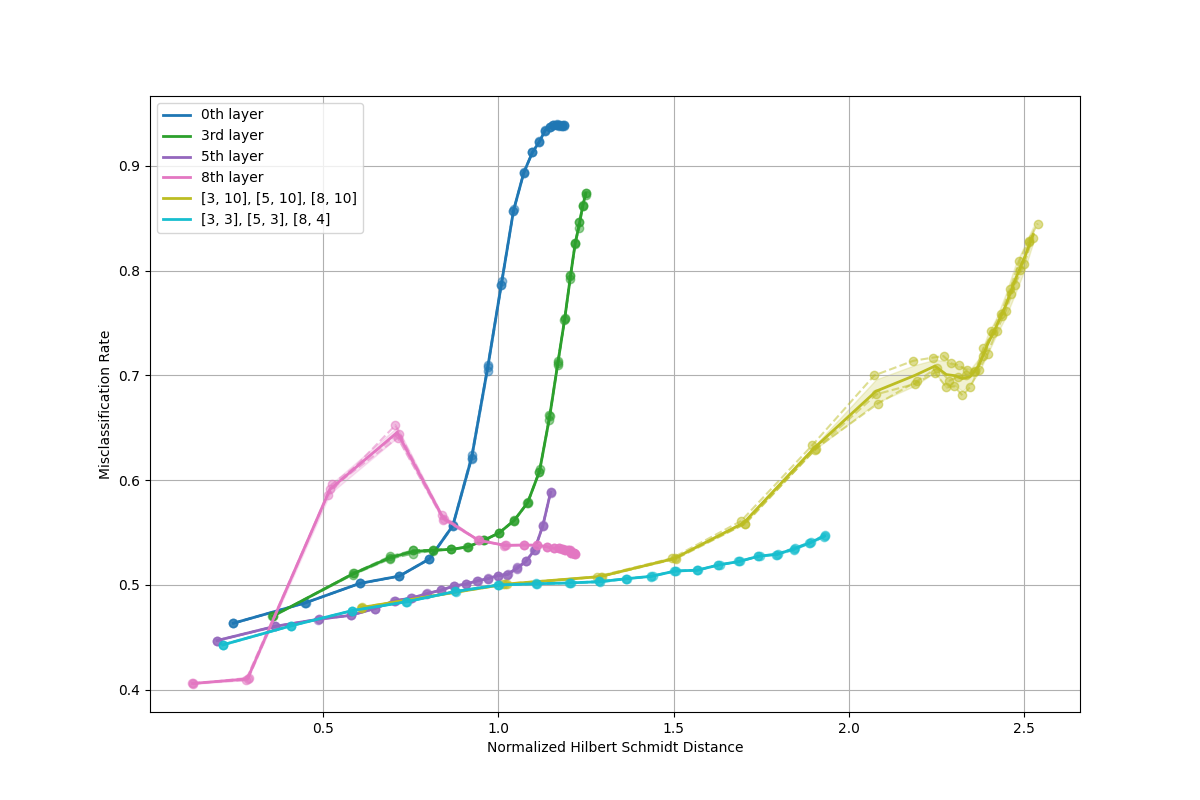}}  
         \caption*{(b)}
    \end{minipage}
    \caption{
    These plots depict misclassification rate (y-axis) against attack strength (x-axis) for a binary classifier,  where  an adversarial block consisting of 10  layer is incorporated into a model with 10 existing layers.
The plots on the left show the results for the MNIST dataset, with the plots on the right displaying the results for the FMNIST dataset.
    The performance of these adversarial blocks is compared with two cases where multiple adversarial blocks are  inserted at different depths within the architecture.  
    In the first case, the total number of adversarial layers is 10,  whereas in the second case,  there are 30 adversarial layer,  organized into three blocks with 10 layers each.
    The attack strength is determined by the sum of Hilbert Schmidt distances between the unitary operators the adversarial blocks induce and the identity operator.
In the legend,  each plot labeled '$q-$th layer' corresponds to an adversarial block located between the $q-$th and $(q+1)-$th layers of the classifier.  In contrast,  plots labeled '$[q_1,  r_1],  [q_2,  r_2],  [q_3,  r_3]$' represent three adversarial blocks inserted between the $q_1-$th and $(q_1+1)-$th layers,  the $q_2-$th and $(q_2+1)-$th layers, and the $q_3-$th and $(q_3+1)-$th layers,  where the first,  second,  and third block consist of  
$r_1,  r_2$,  and $r_3$ adversarial layers,  respectively.  
Note that the maximum Hilbert Schmidt distance between two unitary operators is $\sqrt{2}$. Consequently, the sum of the distances between three unitary perturbation operators and the identity operator is at most $3\sqrt{2}$.
}
\label{fig-exp-1}
\end{figure}

\begin{figure}[H]
    \centering
    \begin{minipage}{0.47\textwidth}  
        \centering
         \scalebox{\scaleB}{
        \includegraphics[width=\textwidth]{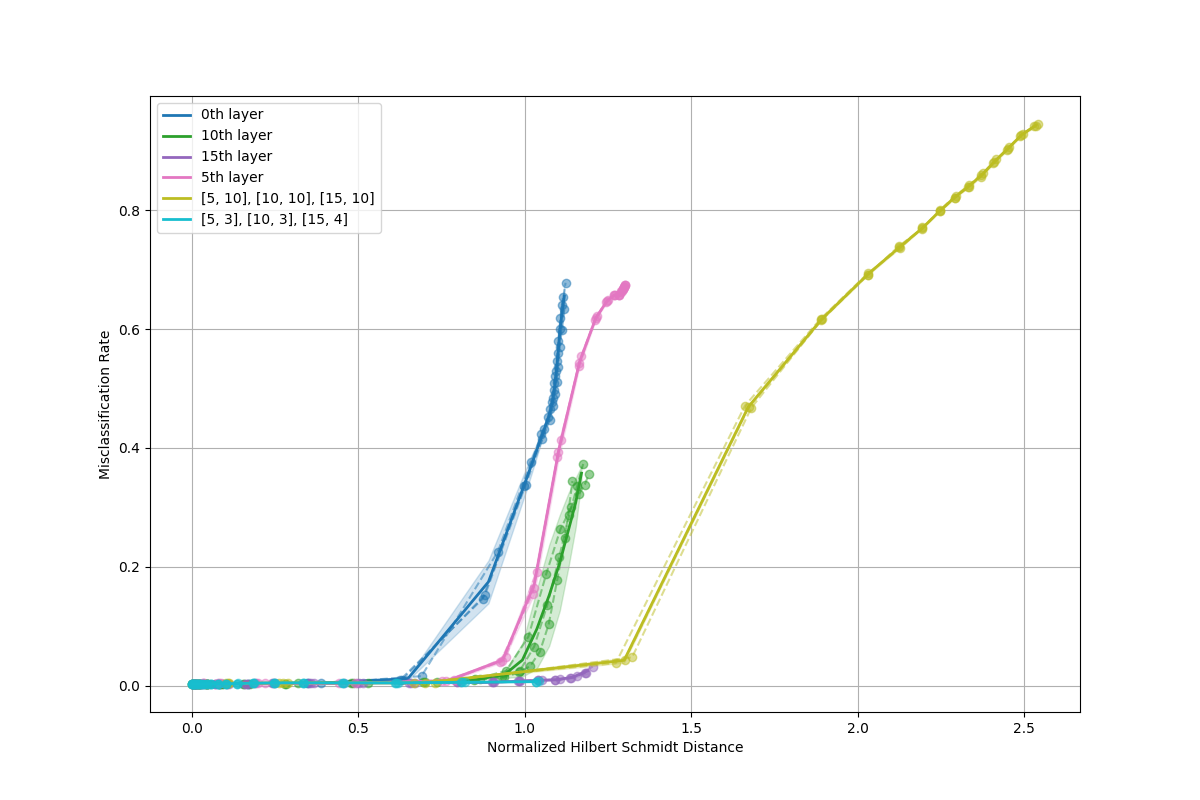}}  
         \caption*{(a)}
    \end{minipage} \hfill
    \begin{minipage}{0.47\textwidth}  
        \centering
         \scalebox{\scaleB}{
        \includegraphics[width=\textwidth]{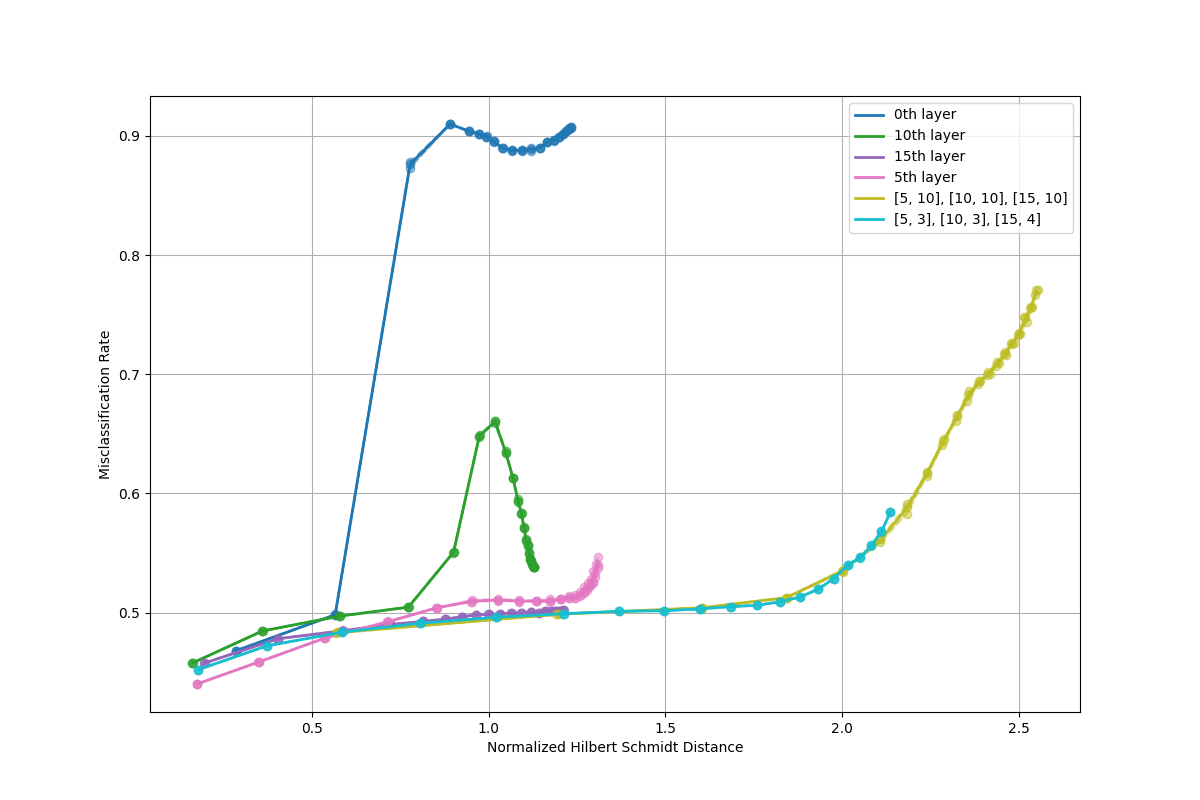}}  
         \caption*{(b)}
    \end{minipage}
    \caption{
Comparing
the effects of inserting an adversarial block consisting of 10 layers into a binary classifier with 20 existing layers versus incorporating three adversarial blocks.  
In the multiple-block settings,  the number of layers for each block is detailed in the legends,
where the
legends are organized similarly to that of 
    Fig.  \ref{fig-exp-1}.
    }
    \label{fig-exp-2}
\end{figure}

\begin{figure}[H]
    \centering
    \begin{minipage}{0.47\textwidth}  
        \centering
         \scalebox{\scaleB}{
        \includegraphics[width=\textwidth]{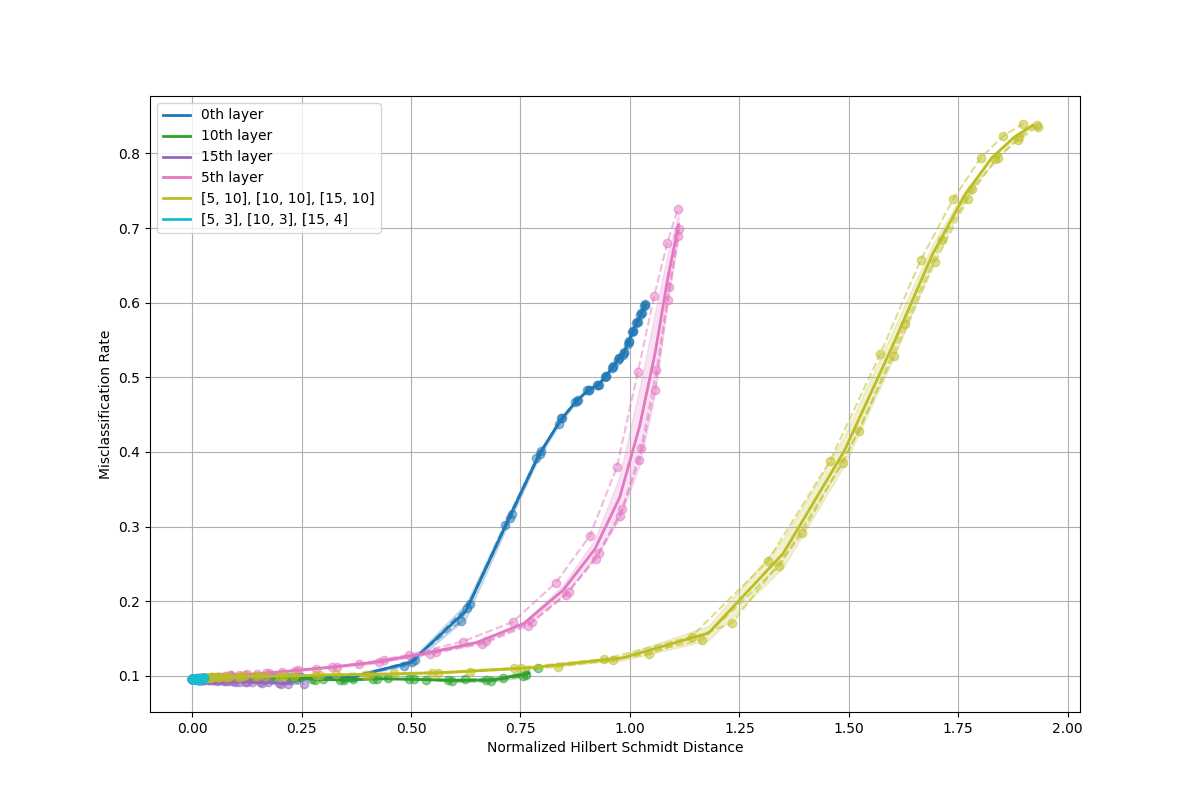}}  
         \caption*{(a)}
    \end{minipage} \hfill
    \begin{minipage}{0.47\textwidth}  
        \centering
         \scalebox{\scaleB}{
        \includegraphics[width=\textwidth]{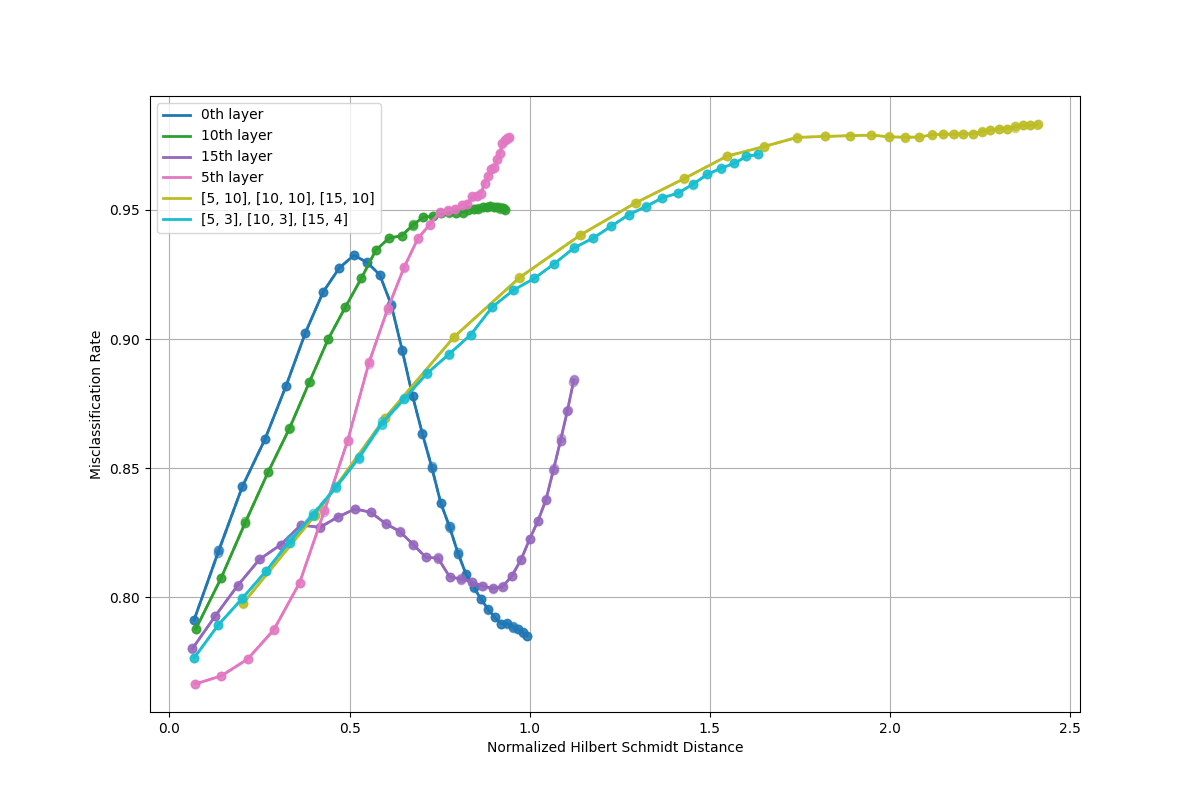}}  
         \caption*{(b)}
    \end{minipage}
    \caption{
Comparing
the effects of inserting an adversarial block consisting of 10 layers into a four-class classifier with 20 existing layers versus incorporating three adversarial blocks.  
}
\label{fig-exp-3}
\end{figure}

\begin{figure}[H]
    \centering
    \begin{minipage}{0.47\textwidth}  
        \centering
        \scalebox{\scaleB}{
        \includegraphics[width=\textwidth]{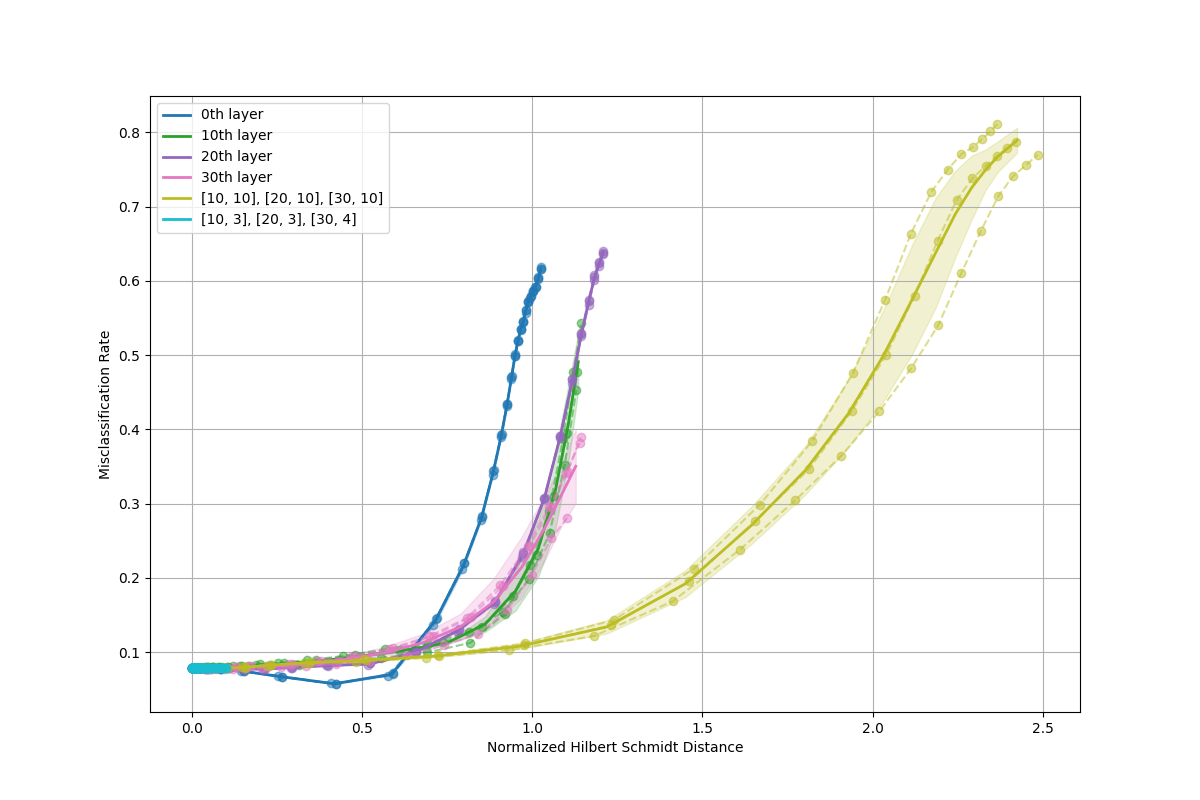}}  
         \caption*{(a)}
    \end{minipage} \hfill
    \begin{minipage}{0.47\textwidth}  
        \centering
        \scalebox{\scaleB}{
        \includegraphics[width=\textwidth]{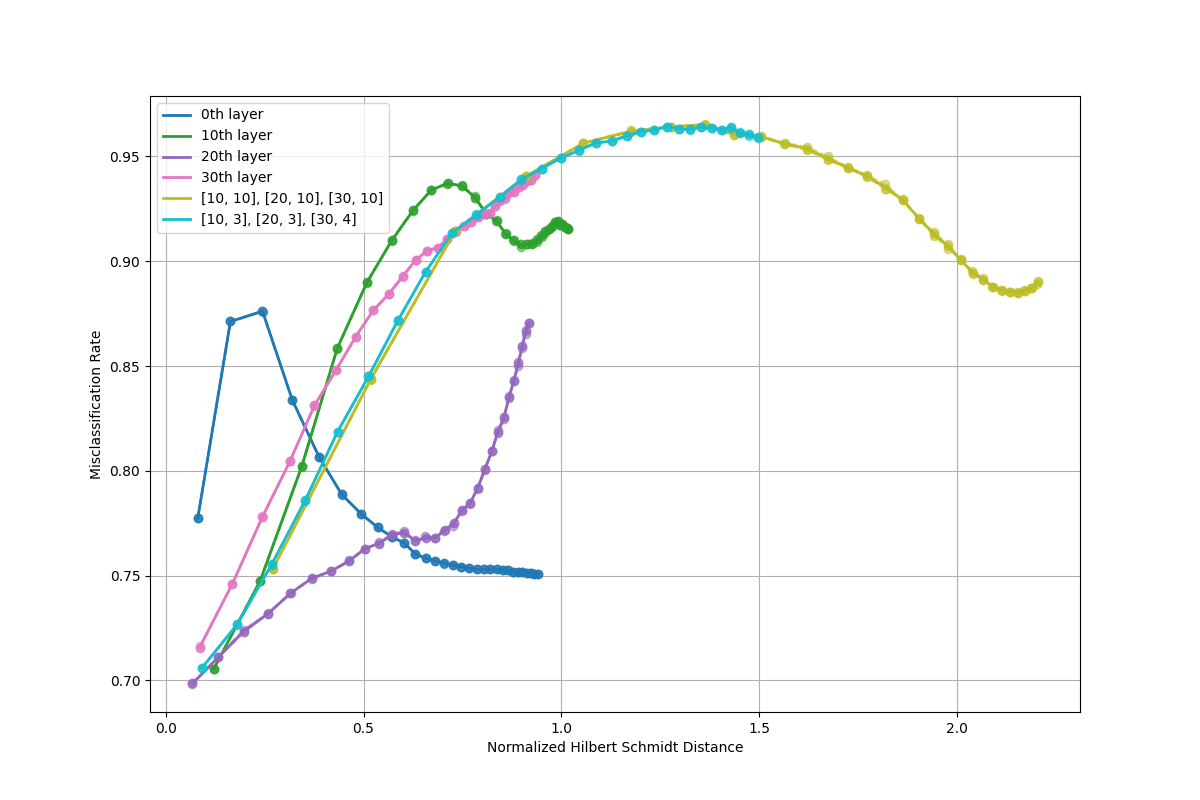}}  
         \caption*{(b)}
    \end{minipage}
    \caption{
Comparing
the effects of inserting an adversarial block consisting of 10 layers into a four-class classifier with 40 existing layers versus incorporating three adversarial blocks.  
    }
    \label{fig-exp-4}
\end{figure}

\paragraph{MNIST and FMNIST:}
In Figures~\ref{fig-exp-1} to \ref{fig-exp-4},  the plots on the left show the results for the MNIST dataset, while those on the right show the results for the FMNIST dataset.
In all these figures,  
the number of adversarial layers is indicated in the caption for plots corresponding to a single adversarial block.
For plots showing the results for multiple blocks of adversarial layers,  the legends displays both the position of the adversarial blocks and the number of layers in each block.

\begin{figure}[H]
    \centering
    \begin{minipage}{0.47\textwidth}  
        \centering
        \scalebox{\myscale}{
        \includegraphics[width=\textwidth]{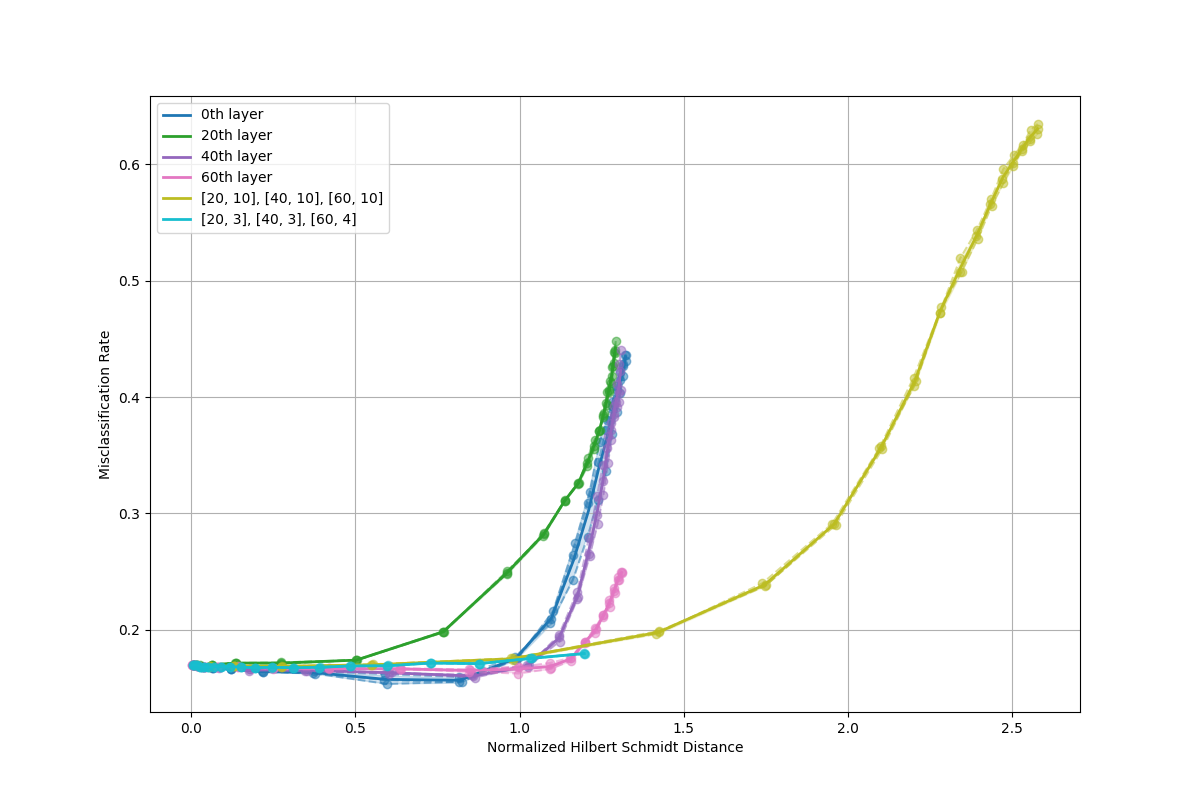}}  
         \caption*{(a)}
    \end{minipage} \hfill
    \begin{minipage}{0.47\textwidth}  
        \centering
        \scalebox{\myscale}{
        \includegraphics[width=\textwidth]{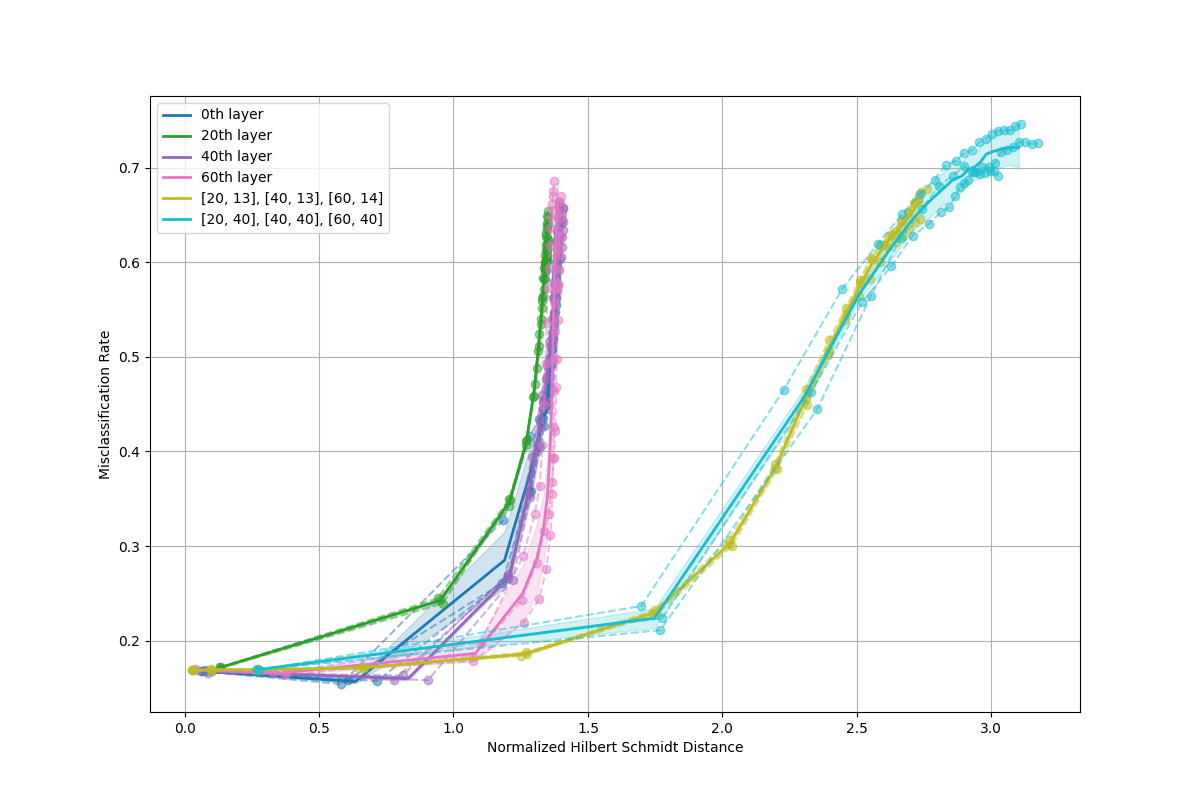}}  
         \caption*{(b)}
    \end{minipage}
    \caption{Impact of inserting adversarial blocks into an 80-layer classifier trained on CIFAR-2.  The left plot shows the effect of adding a 10-layer block, while the right plot shows the effect of adding a 40-layer block.  In contrast to Figure~\ref{fig-exp-1},  where the left and right plots correspond to classifiers trained on MNIST and FMNIST,  both plots here present results for a classifier trained on CIFAR-2.}
    \label{fig-experiments-cifar-1}
\end{figure}

\paragraph{CIFAR-2:} 
The left plot in Figure~\ref{fig-experiments-cifar-1} illustrates the effects of modifying an 80-layer quantum classifier by inserting a 10-layer adversarial block within its layers,  while the right plot shows the results when a 40-layer adversarial block is inserted.  For the cases where multiple blocks of adversarial layers are inserted, the legends specifies both the insertion depth and the number of layers in each added block.

\subsubsection{Local Adversarial layers}
\label{sec:appendix-local-adv}

\begin{figure}[H]
    \centering
    \begin{minipage}{0.47\textwidth}  
        \centering
        \scalebox{\myscale}{
        \includegraphics[width=\textwidth]{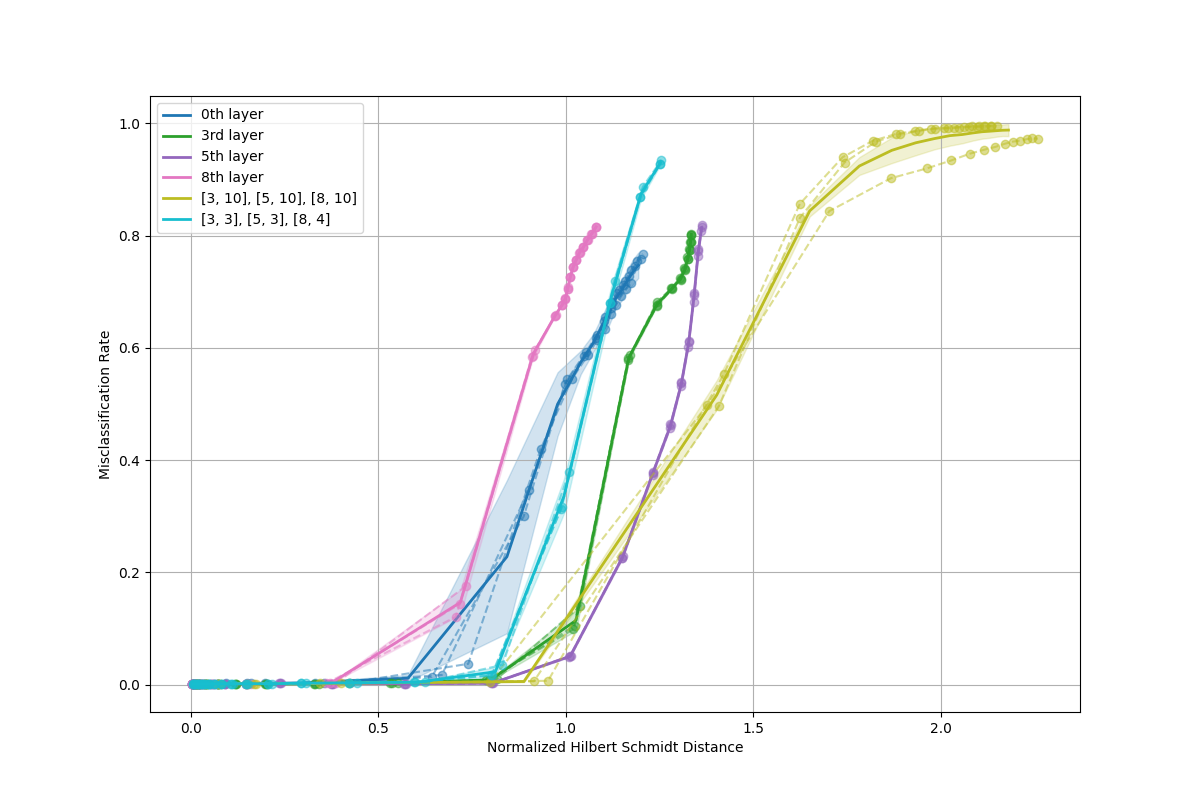}}  
         \caption*{(a)}
    \end{minipage} \hfill
    \begin{minipage}{0.47\textwidth}  
        \centering
        \scalebox{\myscale}{
        \includegraphics[width=\textwidth]{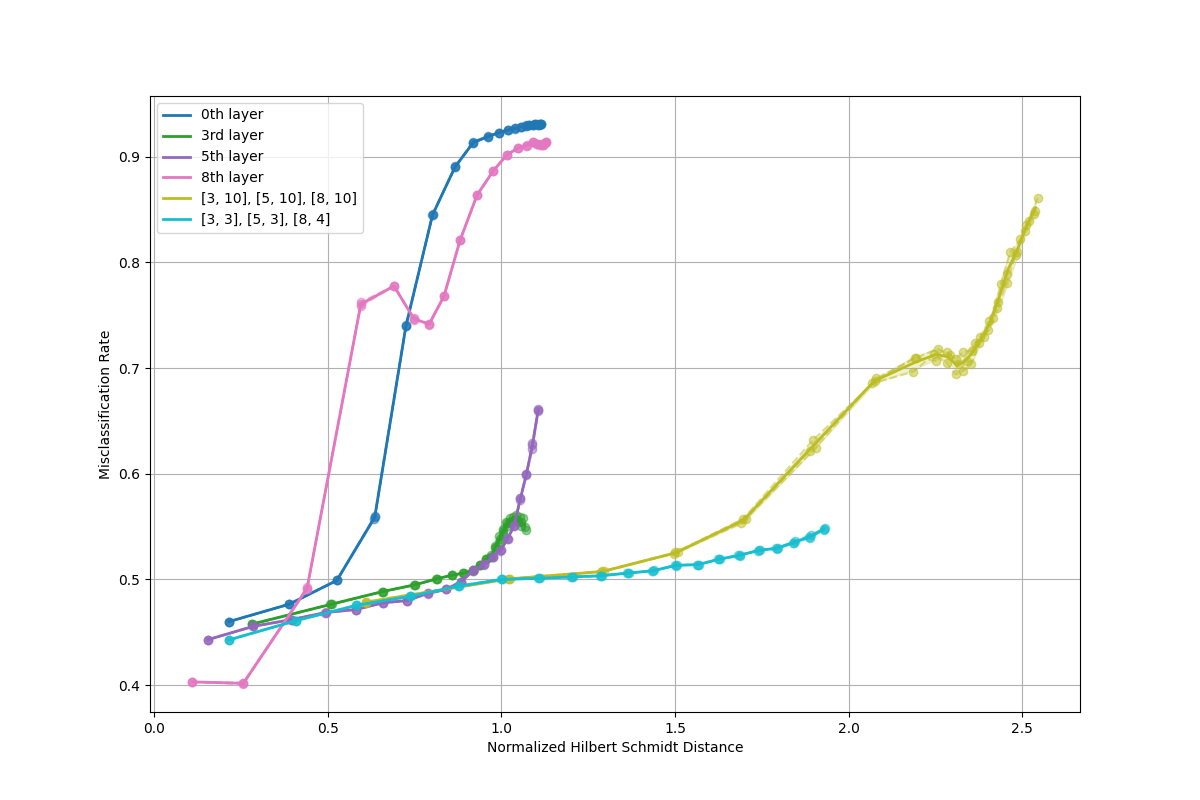}}  
         \caption*{(b)}
    \end{minipage}
    \caption{
Comparing
the effects of inserting an adversarial block consisting of 10 layers into a binary classifier with 10 existing layers versus incorporating three adversarial blocks.  
    In contrast to Fig.  \ref{fig-exp-1}, where the adversarial layers act on all qubits,   here
the adversarial layers 
    act only on qubits number $3 ,4$ and $5$.  
    }
    \label{fig-exp-5}
\end{figure}

\begin{figure}[H]
    \centering
    \begin{minipage}{0.47\textwidth}  
        \centering
        \scalebox{\myscale}{
        \includegraphics[width=\textwidth]{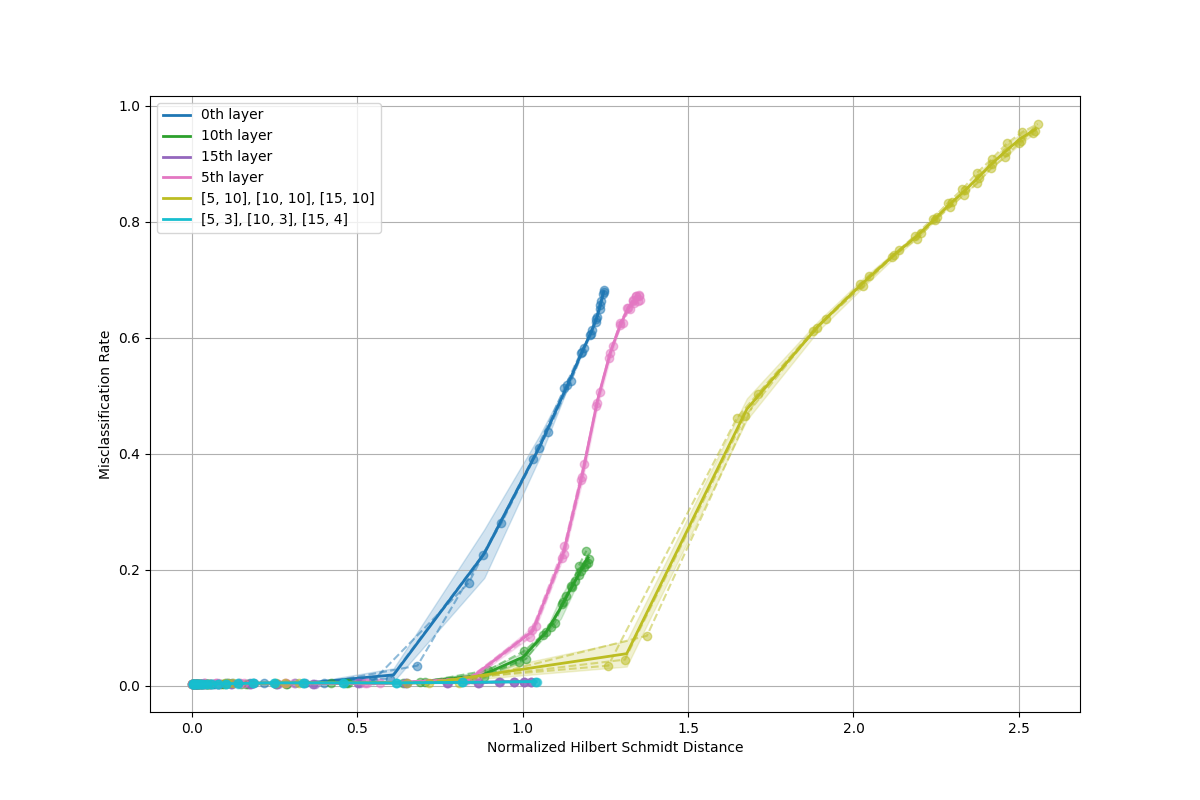}}  
         \caption*{(a)}
    \end{minipage} \hfill
    \begin{minipage}{0.47\textwidth}  
        \centering
        \scalebox{\myscale}{
        \includegraphics[width=\textwidth]{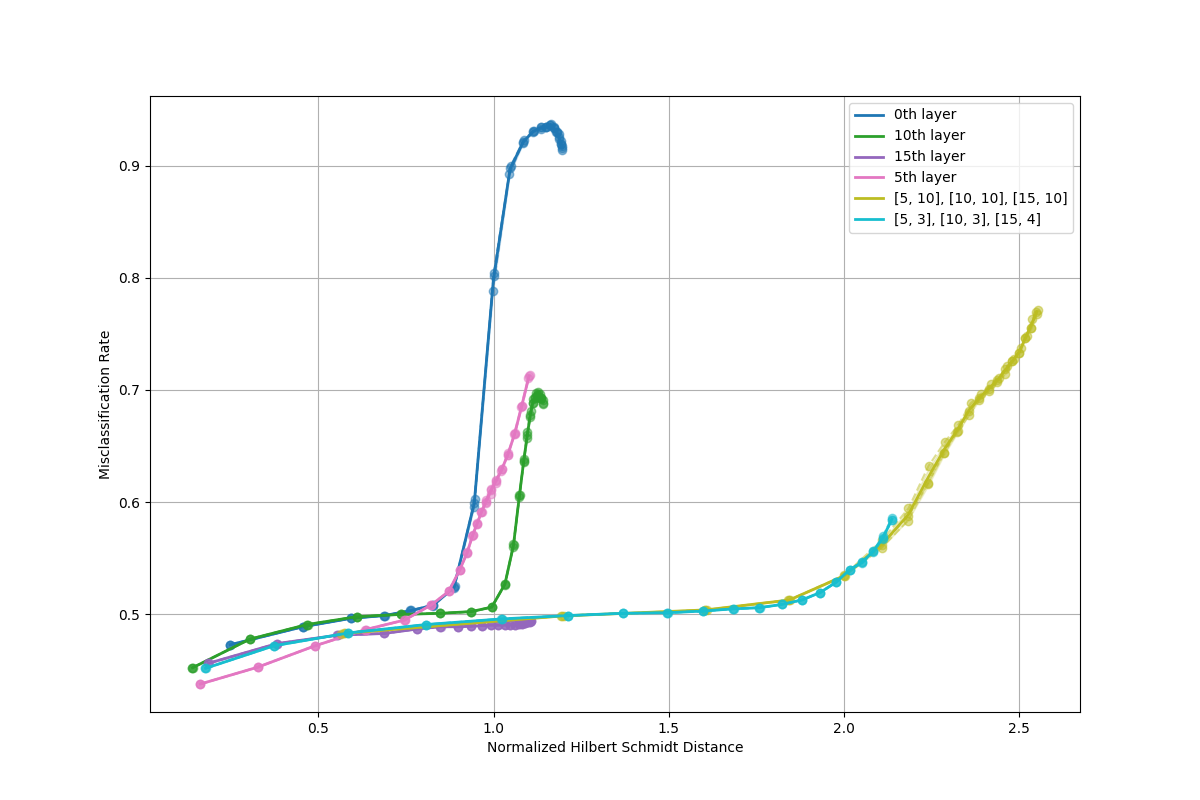}}  
         \caption*{(b)}
    \end{minipage}
    \caption{
Comparing
the effects of inserting an adversarial block consisting of 10 layers into a binary classifier with 20 existing layers versus incorporating three adversarial blocks.  
    Unlike 
    Fig.  \ref{fig-exp-2}, where the adversarial layers act on all qubits,   here
the adversarial layers 
    act only on qubits number $3 ,4$ and $5$.  
    }
    \label{fig-exp-6}
\end{figure}

\begin{figure}[H]
    \centering
    \begin{minipage}{0.47\textwidth}  
        \centering
        \scalebox{\myscale}{
        \includegraphics[width=\textwidth]{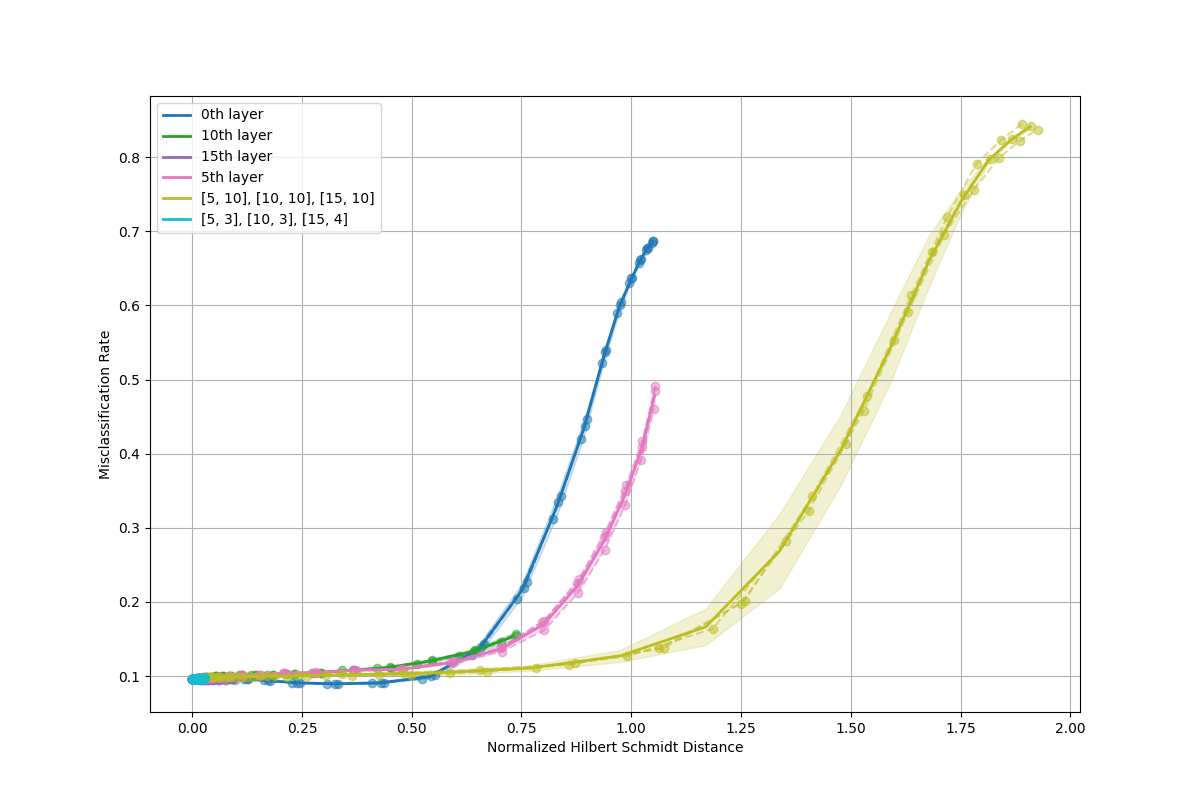}}  
         \caption*{(a)}
    \end{minipage} \hfill
    \begin{minipage}{0.47\textwidth}  
        \centering
        \scalebox{\myscale}{
        \includegraphics[width=\textwidth]{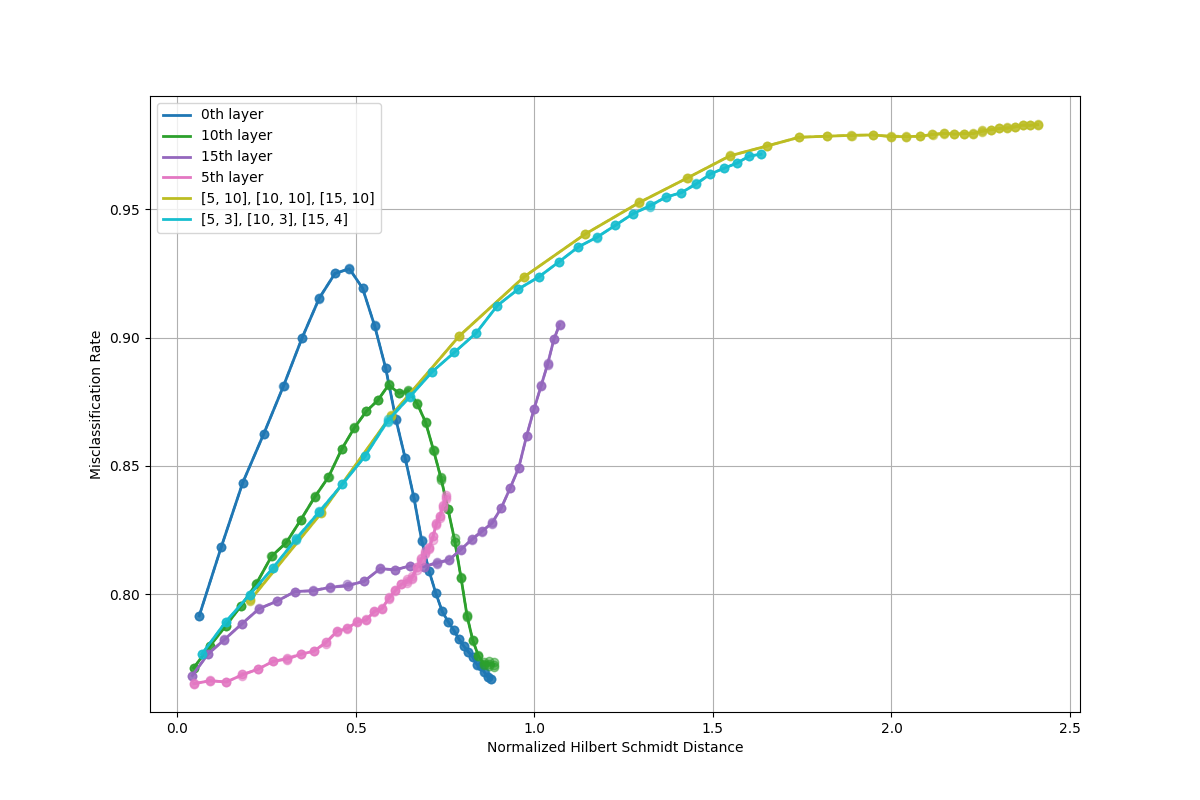}}  
         \caption*{(b)}
    \end{minipage}
    \caption{
Comparing
the effects of inserting an adversarial block consisting of 10 layers into a four-class classifier with 20 existing layers versus incorporating three adversarial blocks.  
    Contrary to 
    Fig.  \ref{fig-exp-3}, where the adversarial layers act on all qubits,   here
the adversarial layers 
    act only on qubits number $3 ,4$ and $5$.  
    }
    \label{fig-exp-7}
\end{figure}

\begin{figure}[H]
    \centering
    \begin{minipage}{0.47\textwidth}  
        \centering
        \scalebox{\myscale}{
        \includegraphics[width=\textwidth]{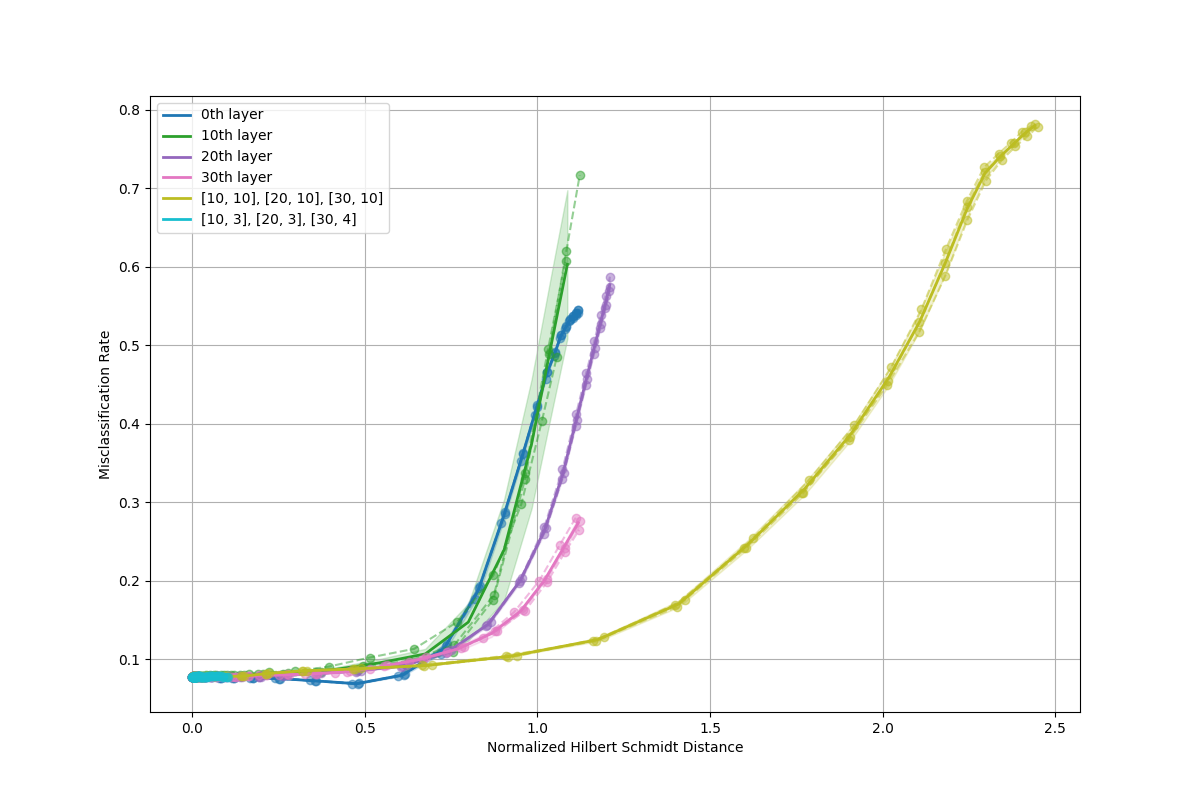}}  
         \caption*{(a)}
    \end{minipage} \hfill
    \begin{minipage}{0.47\textwidth}  
        \centering
        \scalebox{\myscale}{
        \includegraphics[width=\textwidth]{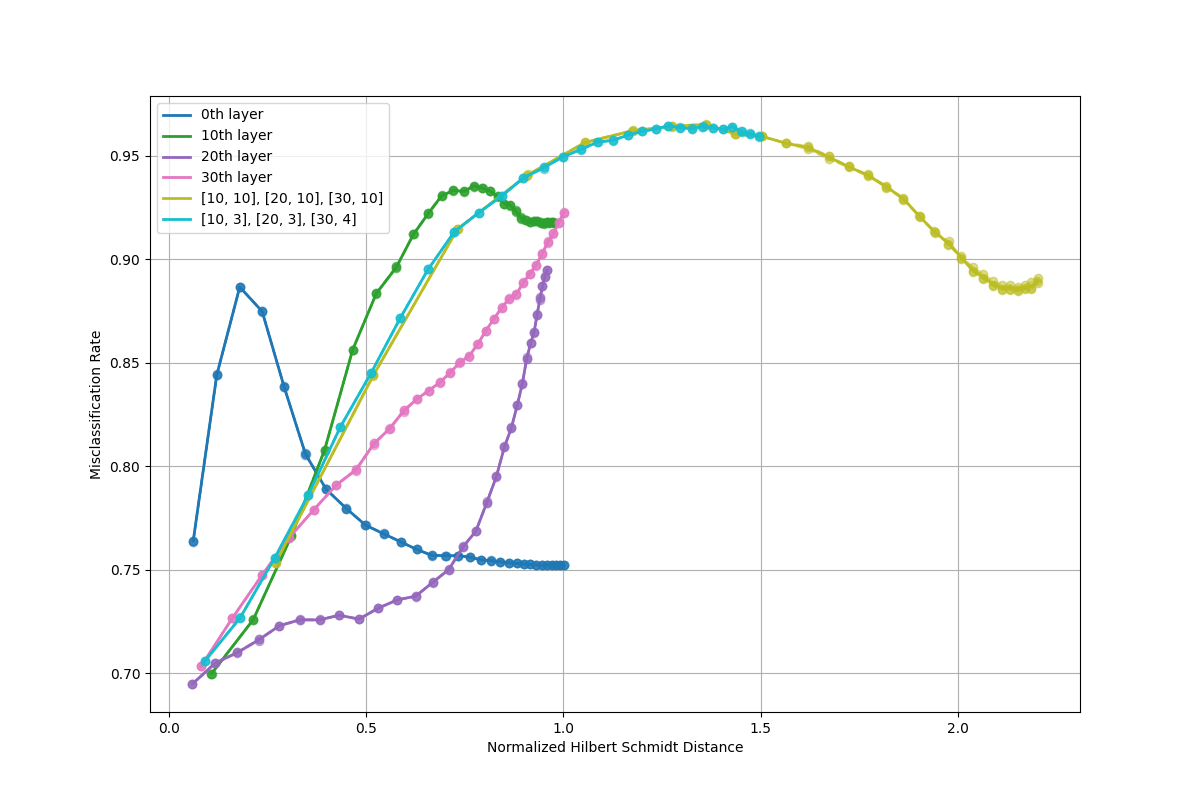}}  
         \caption*{(b)}
    \end{minipage}
    \caption{
Comparing
the effects of inserting an adversarial block consisting of 10 layers into a four-class classifier with 40 existing layers versus incorporating three adversarial blocks.  
    In contrast to Fig.  \ref{fig-exp-4}, where the adversarial layers act on all qubits,   here
the adversarial layers 
    act only on qubits number $3 ,4$ and $5$.  
    }
    \label{fig-exp-8}
\end{figure}

\begin{figure}[H]
    \centering
    \begin{minipage}{0.47\textwidth}  
        \centering
        \scalebox{\myscale}{
        \includegraphics[width=\textwidth]{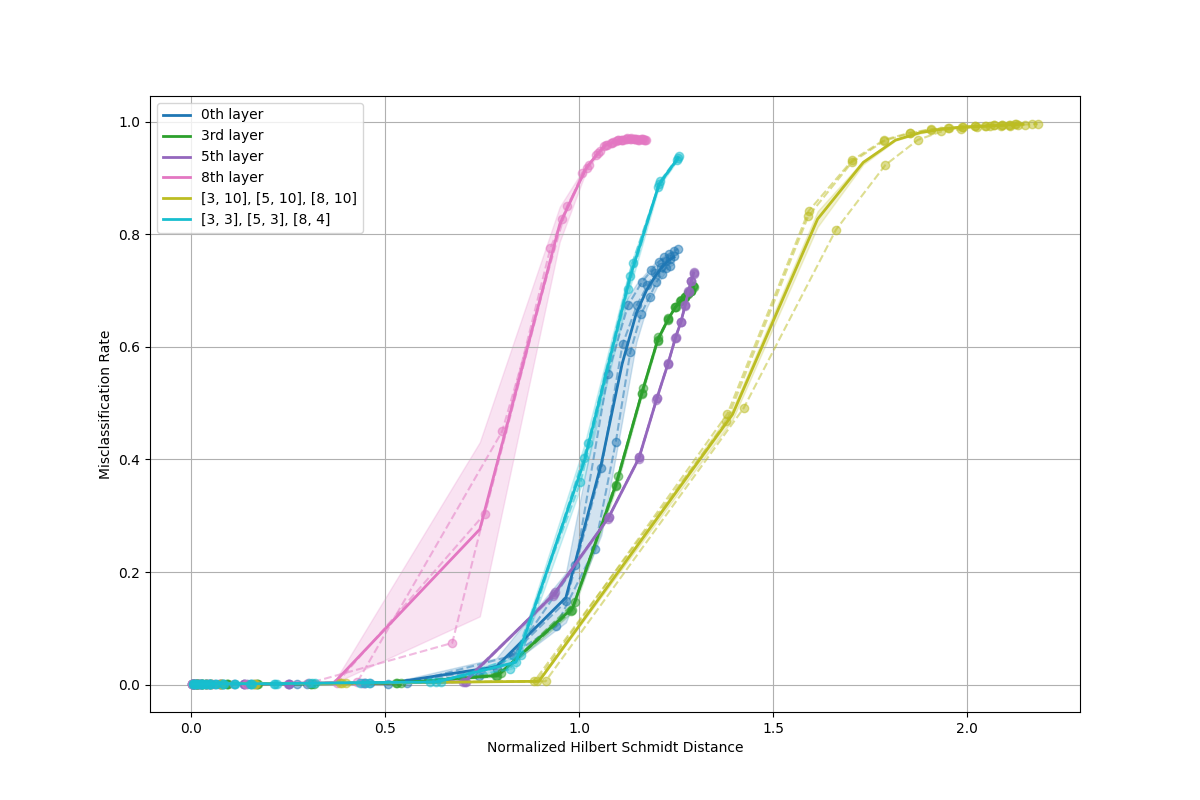}}  
         \caption*{(a)}
    \end{minipage} \hfill
    \begin{minipage}{0.47\textwidth}  
        \centering
        \scalebox{\myscale}{
        \includegraphics[width=\textwidth]{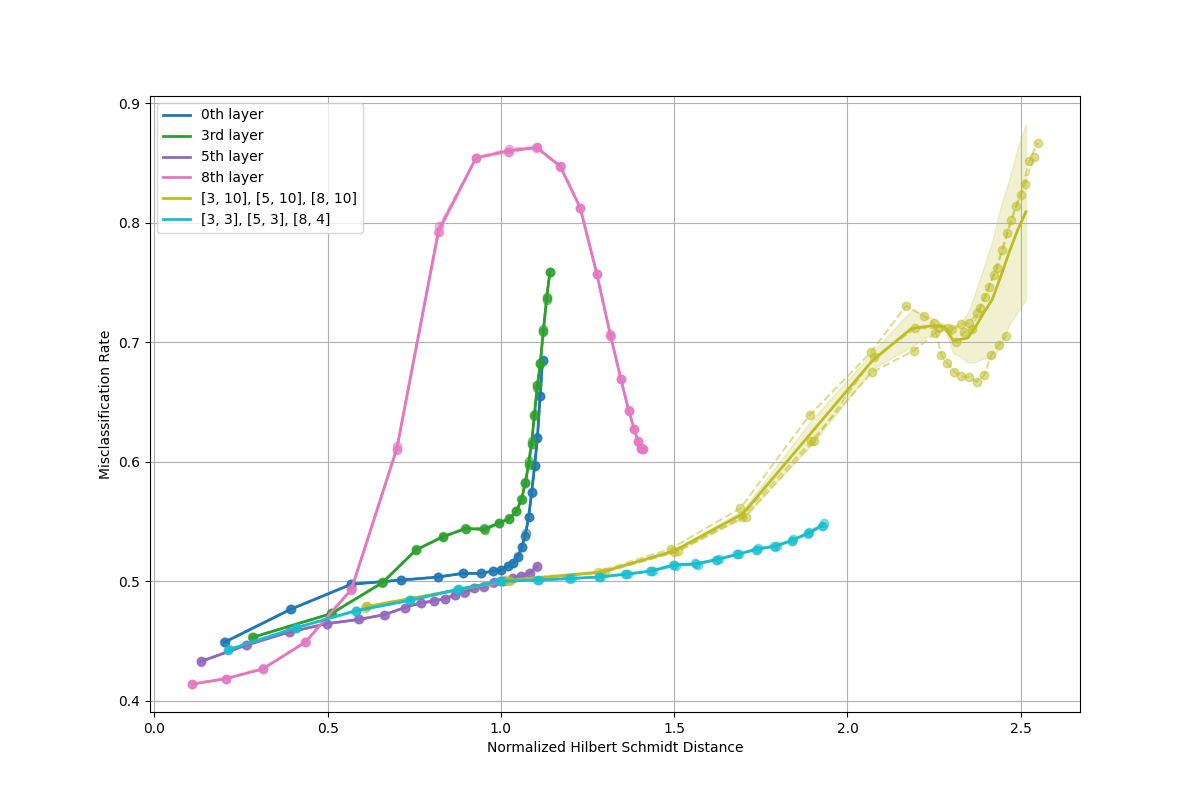}}  
         \caption*{(b)}
    \end{minipage}
    \caption{
Comparing
the effects of inserting an adversarial block consisting of 10 layers into a binary classifier with 10 existing layers versus incorporating three adversarial blocks.  
    Unlike 
    Fig.  \ref{fig-exp-1}, where the adversarial layers act on all qubits,   here
the adversarial layers 
    act only on qubits number $5 ,6, 7$ and $8$. 
    }
    \label{fig-exp-9}
\end{figure}

\begin{figure}[H]
    \centering
    \begin{minipage}{0.47\textwidth}  
        \centering
        \scalebox{\scaleB}{
        \includegraphics[width=\textwidth]{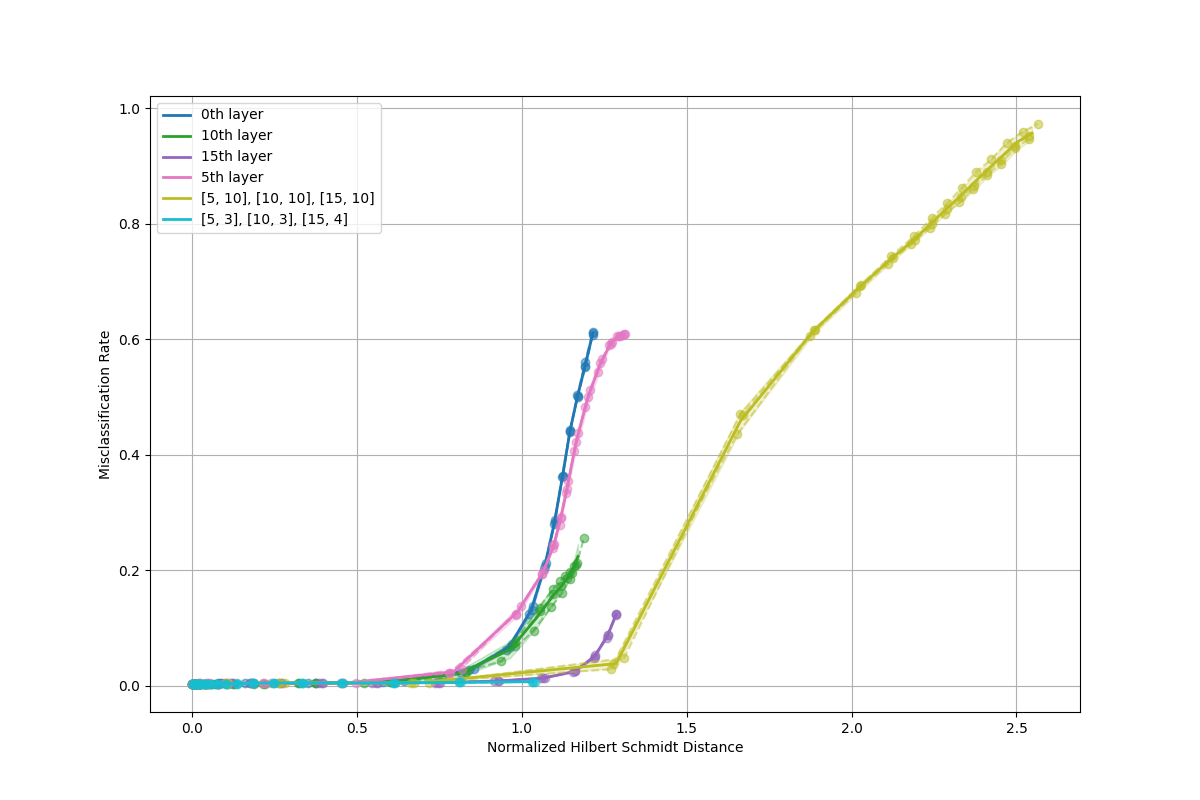}}  
         \caption*{(a)}
    \end{minipage} \hfill
    \begin{minipage}{0.47\textwidth}  
        \centering
        \scalebox{\scaleB}{
        \includegraphics[width=\textwidth]{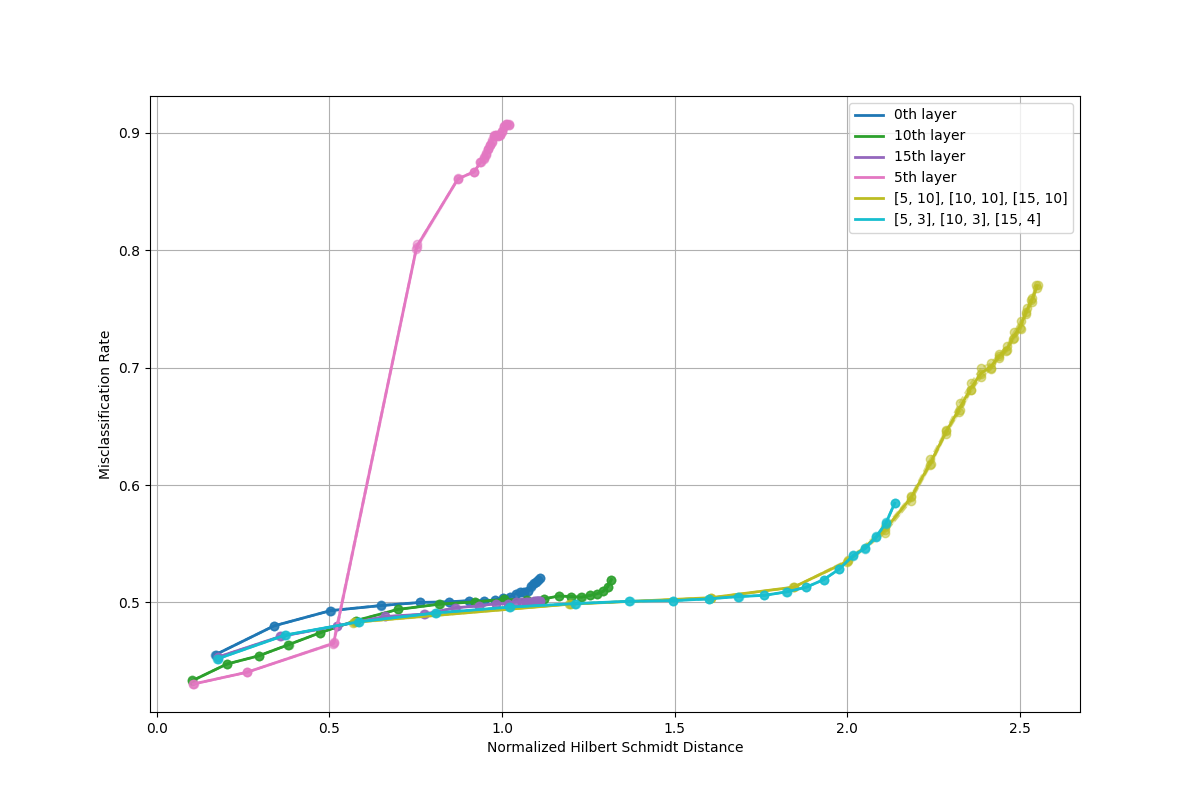}}  
        \caption*{(b)}
    \end{minipage}
    \caption{
Comparing
the effects of inserting an adversarial block consisting of 10 layers into a binary classifier with 20 existing layers versus incorporating three adversarial blocks.  
    In contrast to Fig.  \ref{fig-exp-2}, where the adversarial layers act on all qubits,   here
the adversarial layers 
    act only on qubits number $5 ,6, 7$ and $8$. 
    }
    \label{fig-exp-10}
\end{figure}

\begin{figure}[H]
    \centering
    \begin{minipage}{0.47\textwidth}  
        \centering
        \scalebox{\scaleB}{
        \includegraphics[width=\textwidth]{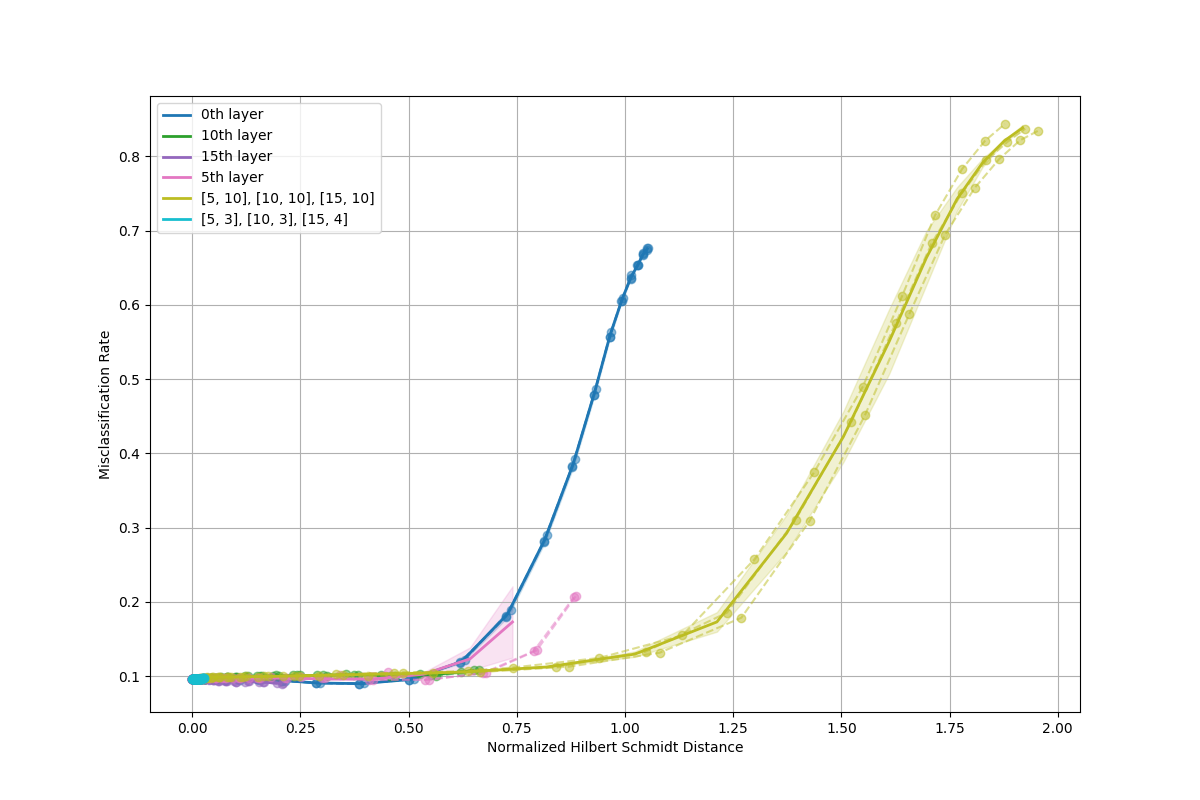}}  
        \caption*{(a)}
    \end{minipage} \hfill
    \begin{minipage}{0.47\textwidth}  
        \centering
        \scalebox{\scaleB}{
        \includegraphics[width=\textwidth]{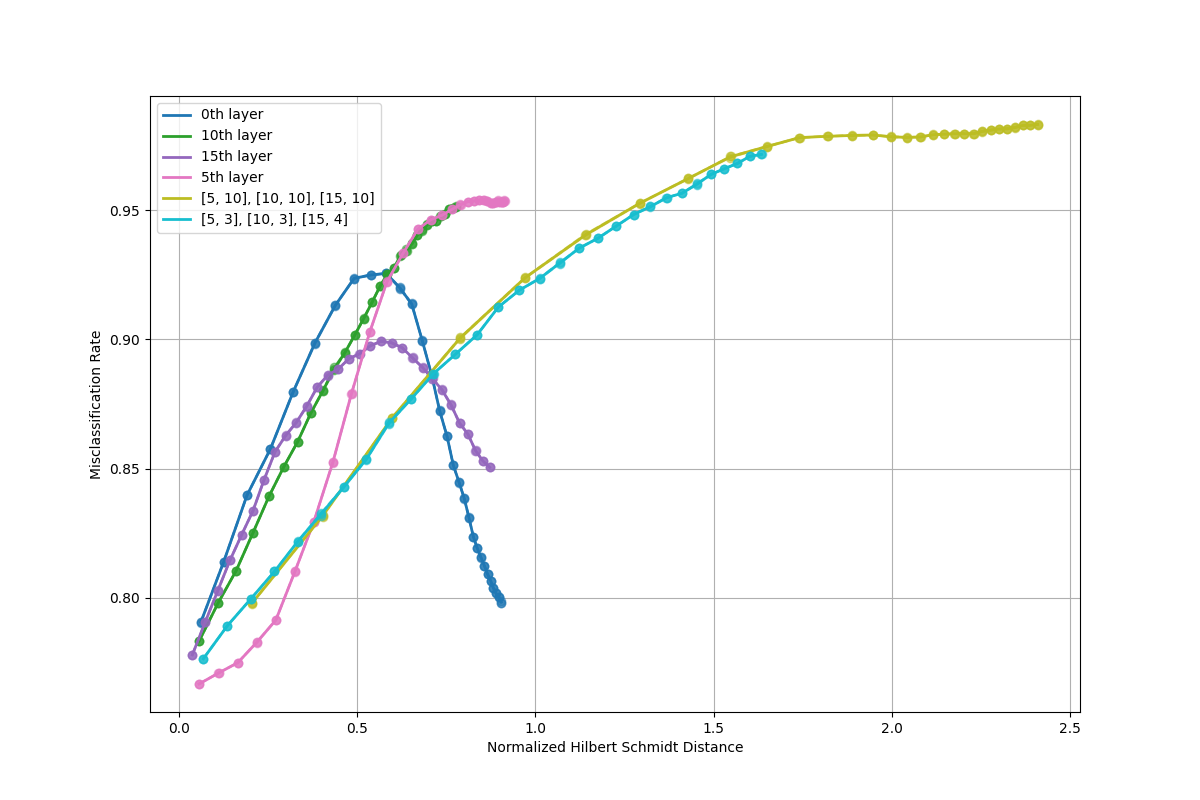}}  
        \caption*{(b)}
    \end{minipage}
    \caption{
Comparing
the effects of inserting an adversarial block consisting of 10 layers into a four-class classifier with 20 existing layers versus incorporating three adversarial blocks.  
    Contrary to 
    Fig.  \ref{fig-exp-3}, where the adversarial layers act on all qubits,   here
the adversarial layers 
    act only on qubits number $5 ,6, 7$ and $8$. 
    }
    \label{fig-exp-11}
\end{figure}

\begin{figure}[H]
    \centering
    \begin{minipage}{0.47\textwidth}  
        \centering
        \scalebox{\scaleB}{
        \includegraphics[width=\textwidth]{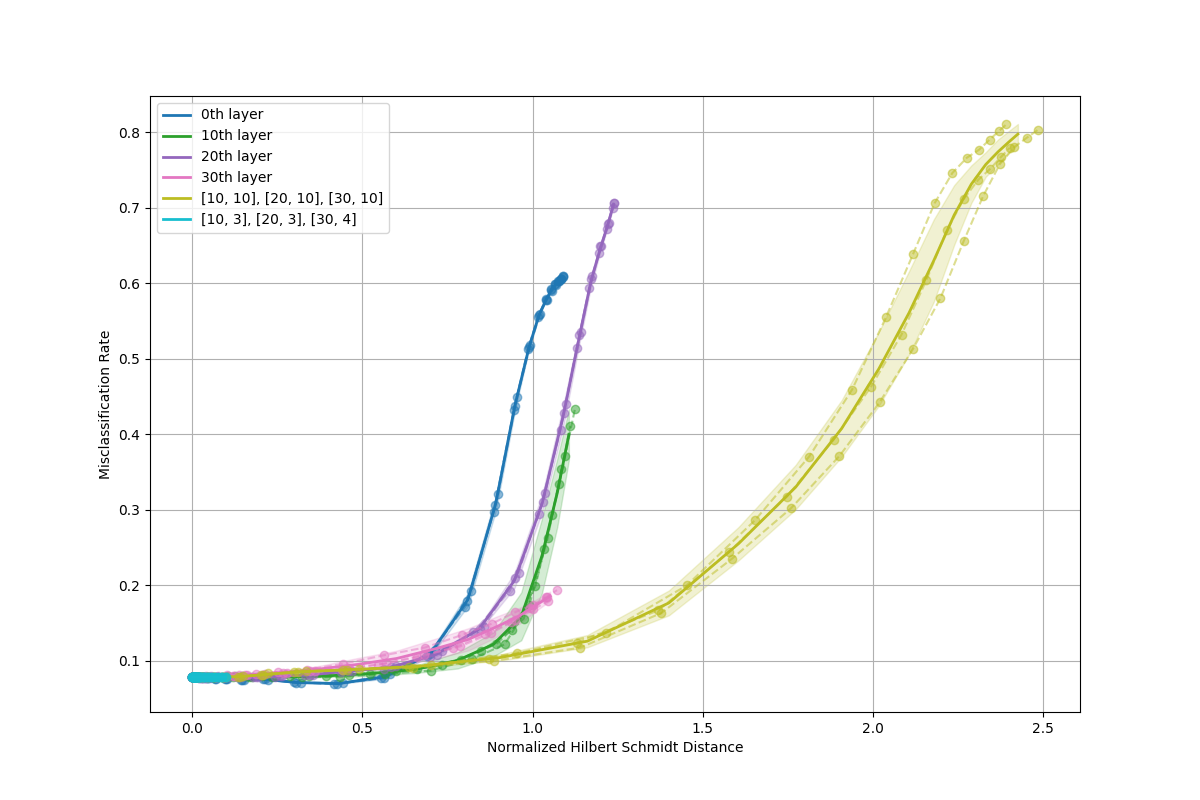}}  
        \caption*{(a)}
    \end{minipage} \hfill
    \begin{minipage}{0.47\textwidth}  
        \centering
        \scalebox{\scaleB}{
        \includegraphics[width=\textwidth]{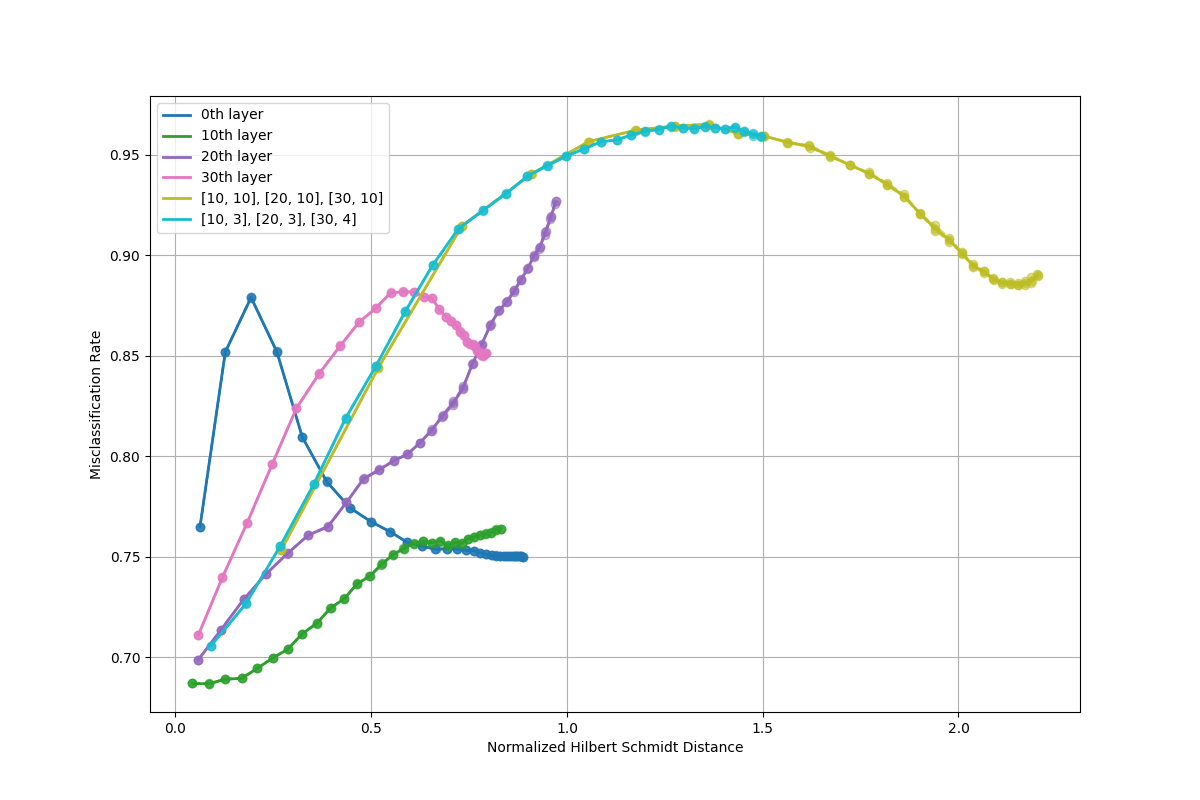}}  
        \caption*{(b)}
    \end{minipage}
    \caption{
Comparing
the effects of inserting an adversarial block consisting of 10 layers into a four-class classifier with 40 existing layers versus incorporating three adversarial blocks.  
    In contrast to Fig.  \ref{fig-exp-4}, where the adversarial layers act on all qubits,   here
the adversarial layers 
    act only on qubits number $5 ,6, 7$ and $8$. 
    }
    \label{fig-exp-12}
\end{figure}

\paragraph{MNIST and FMNIST:}
The results presented here correspond to the case where the adversarial gates act on a local set of qubits,  specifically qubits  $3 ,4$ and $5$ for Figures  \ref{fig-exp-5} to \ref{fig-exp-8} and qubits $5 ,6, 7$ and $8$ for Figures \ref{fig-exp-9} to \ref{fig-exp-12} (See Figure~\ref{fig:experimental-settup}'s caption for the qubit numbering scheme used in our experiments). 
In Figs.~\ref{fig-exp-5}–\ref{fig-exp-12},  the left-hand plots correspond to the MNIST dataset, and the right-hand plots correspond to the FMNIST dataset.

\begin{figure}[H]
    \centering
    \begin{minipage}{0.47\textwidth}  
        \centering
        \scalebox{\scaleA}{
        \includegraphics[width=\textwidth]{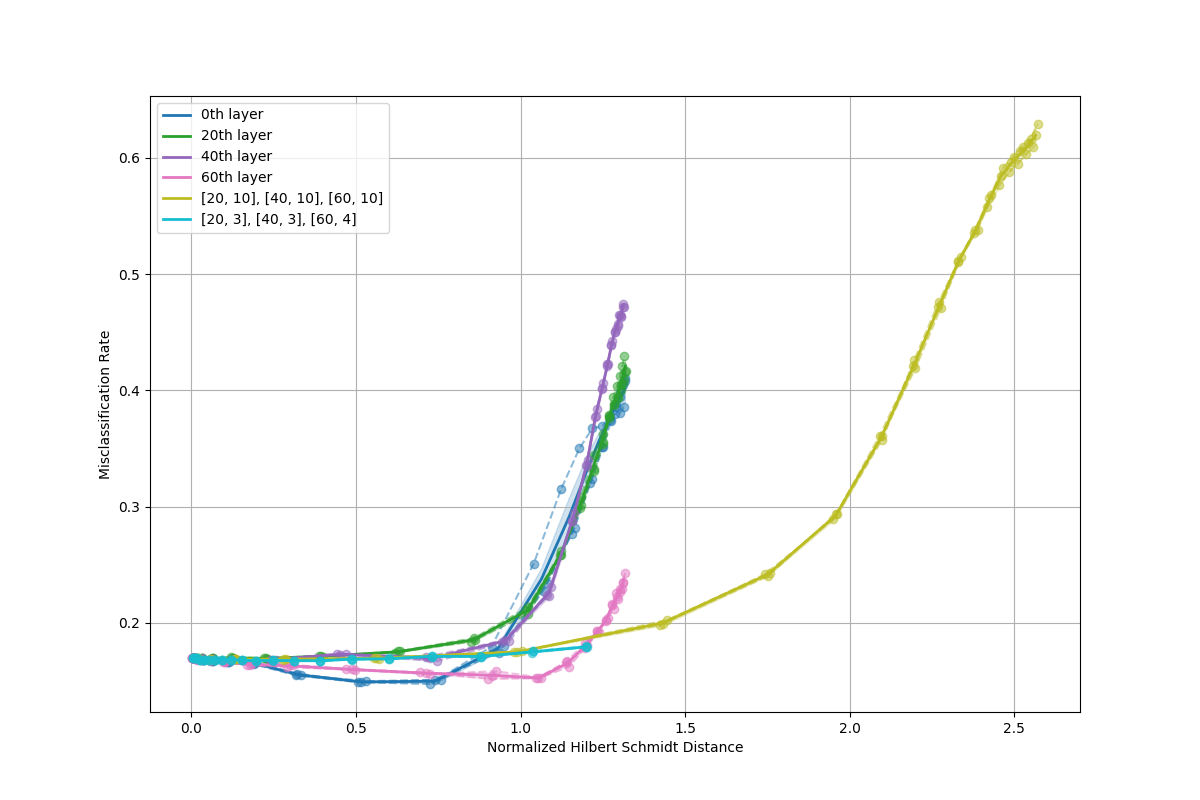}}  
        \caption*{(a)}
    \end{minipage} \hfill
    \begin{minipage}{0.47\textwidth}  
        \centering
        \scalebox{\scaleA}{
        \includegraphics[width=\textwidth]{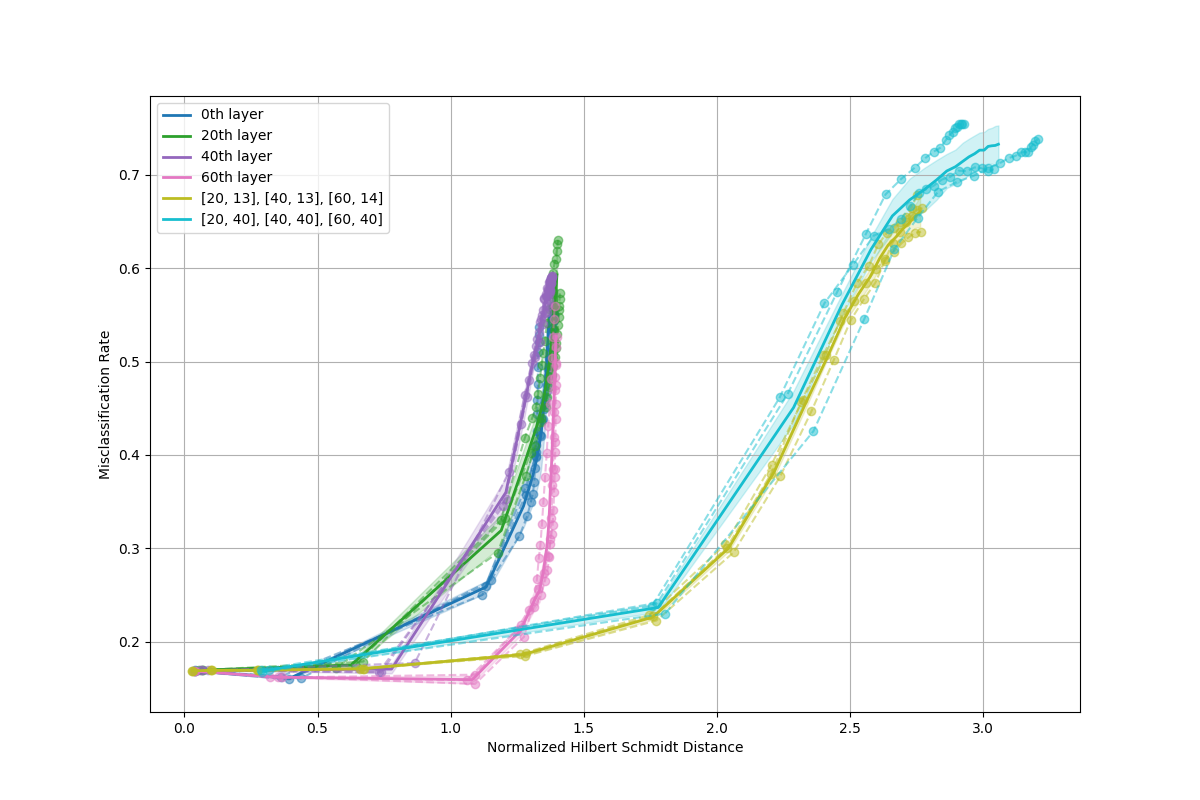}}  
        \caption*{(b)}
    \end{minipage}
    \caption{
Effect of inserting local adversarial blocks into an 80-layer classifier trained on CIFAR-2.  The left plot shows the effect of adding a 10-layer block, while the right plot shows the effect of adding a 40-layer block.  The adversarial layers 
    act only on qubits number $5,6,7$ and $8$. 
In contrast to Figures~\ref{fig-exp-5} to \ref{fig-exp-12},  where the left and right plots correspond to classifiers trained on MNIST and FMNIS,  both plots here present results for a classifier trained on CIFAR-2.    
    }
    \label{fig-experiments-cifar-3}
\end{figure}

\paragraph{CIFAR-2:}
Fig.~\ref{fig-experiments-cifar-2} below shows the effect of inserting local adversarial gates acting on qubits  $3 ,4$ and $5$ whereas Fig.~\ref{fig-experiments-cifar-3} corresponds to the case where the adversarial gates act on qubits $5 ,6, 7$ and $8$.

\begin{figure}[H]
    \centering
    \begin{minipage}{0.47\textwidth}  
        \centering
        \scalebox{\scaleA}{
        \includegraphics[width=\textwidth]{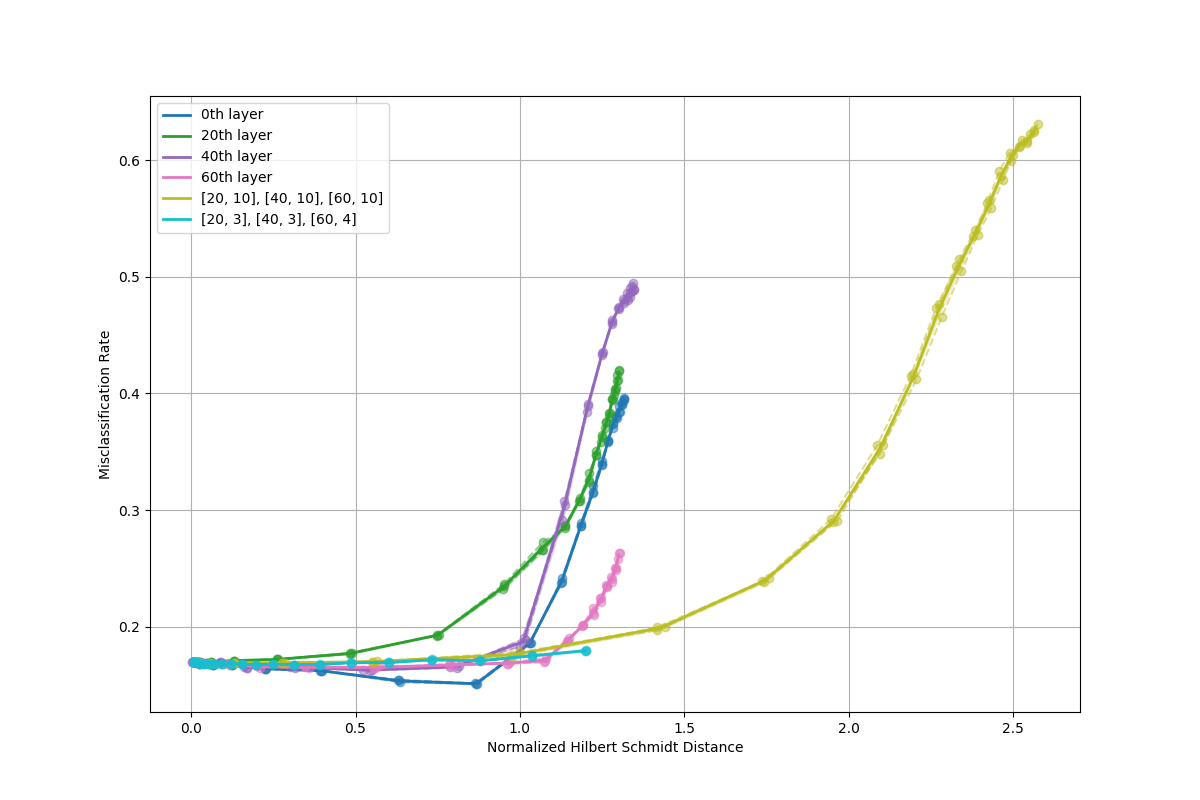}}  
        \caption*{(a)}
    \end{minipage} \hfill
    \begin{minipage}{0.47\textwidth}  
        \centering
        \scalebox{\scaleA}{
        \includegraphics[width=\textwidth]{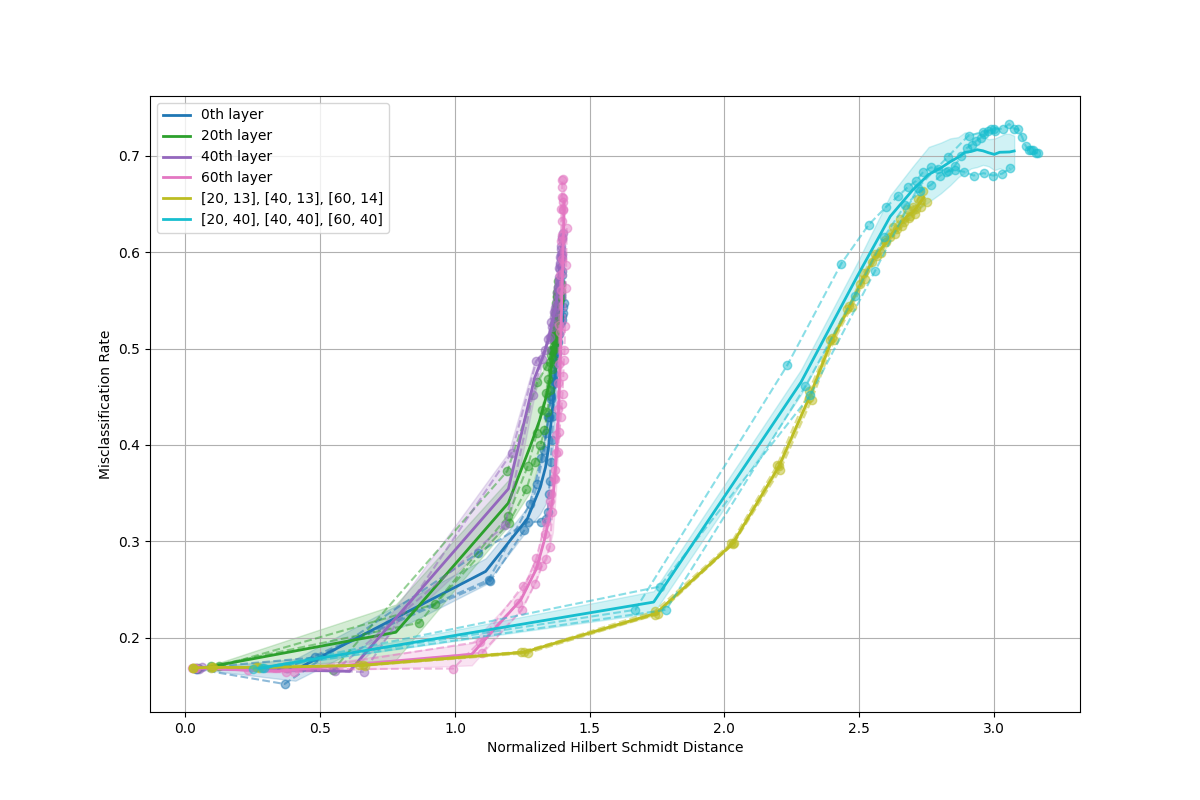}}  
        \caption*{(b)}
    \end{minipage}
    \caption{
Impact of inserting local adversarial blocks into an 80-layer classifier trained on CIFAR-2.  The left plot shows the effect of adding a 10-layer block, while the right plot shows the effect of adding a 40-layer block.  The adversarial layers 
    act only on qubits number $3, 4$ and $5$.  
    For the cases where multiple adversarial blocks are inserted, the legends specifies the insertion depth and the number of layers in each added block.
In contrast to Figures~\ref{fig-exp-5} to \ref{fig-exp-12}  where the left and right plots correspond to classifiers trained on MNIST and FMNIS,  both plots here present results for a classifier trained on CIFAR-2.    
    }
    \label{fig-experiments-cifar-2}
\end{figure}

\subsubsection{Rationale for Selected Hyperparameters}
\label{sec:appendix-hyperparameters}

This section details the empirical results underpinning our selection of hyperparameters, in particular the learning rate and $\gamma$.

\begin{figure}[H]
    \centering
    \begin{minipage}{0.47\textwidth}  
        \centering
        \scalebox{\myscale}{
        \includegraphics[width=\textwidth]{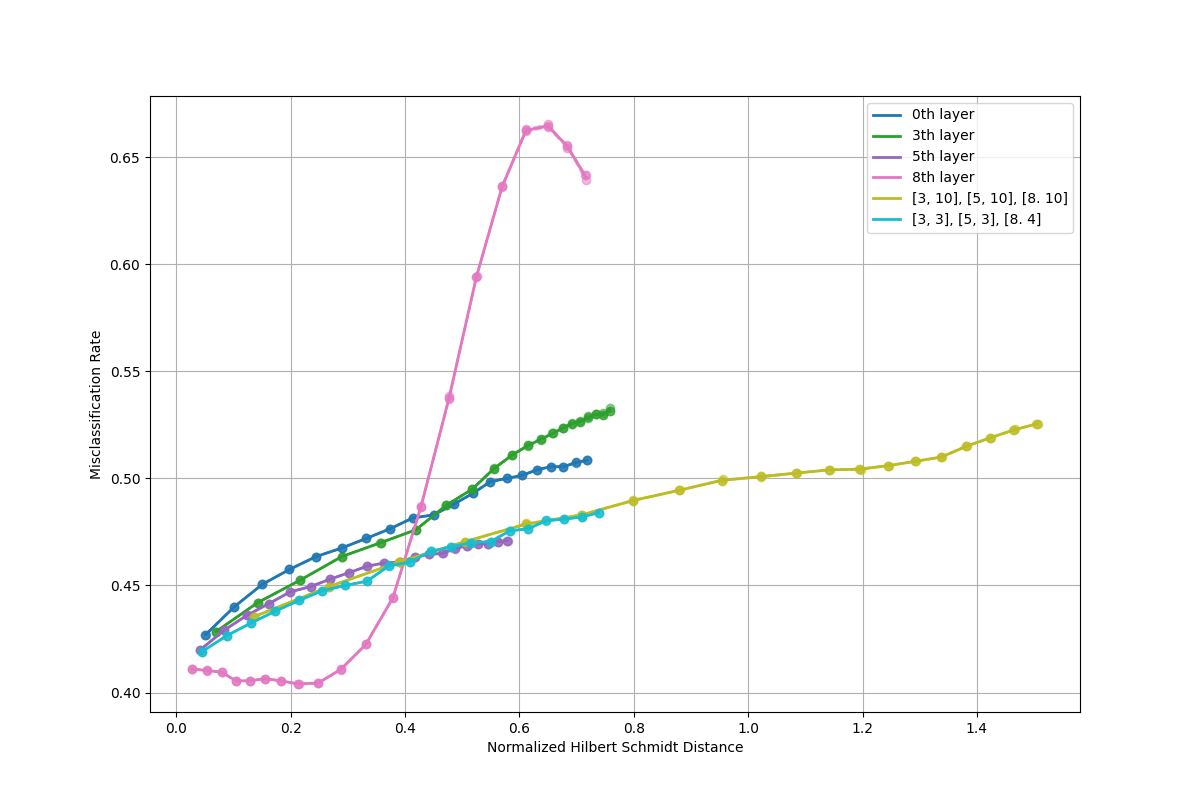}}  
    \end{minipage} \hfill
    \begin{minipage}{0.47\textwidth}  
        \centering
        \scalebox{\myscale}{
        \includegraphics[width=\textwidth]{Figures/plots-v3/mnist-zero-gamma/f-2-10-10.png}}  
    \end{minipage}
    \caption{
Effects of inserting an adversarial block consisting of 10 layers into a binary classifier with 10 existing layers trained on FMNIST versus incorporating three adversarial blocks.  
 The left plot shows the results when the adversarial layers are trained with a learning rate of 0.001,  whereas the right plot corresponds to a learning rate of 0.005, which yields better adversarial performance on FMNIST.  
    }
    \label{fig-lr-fmnist}
\end{figure}

\begin{figure}[H]
    \centering
    \begin{minipage}{0.47\textwidth}  
        \centering
        \scalebox{\scaleB}{
        \includegraphics[width=\textwidth]{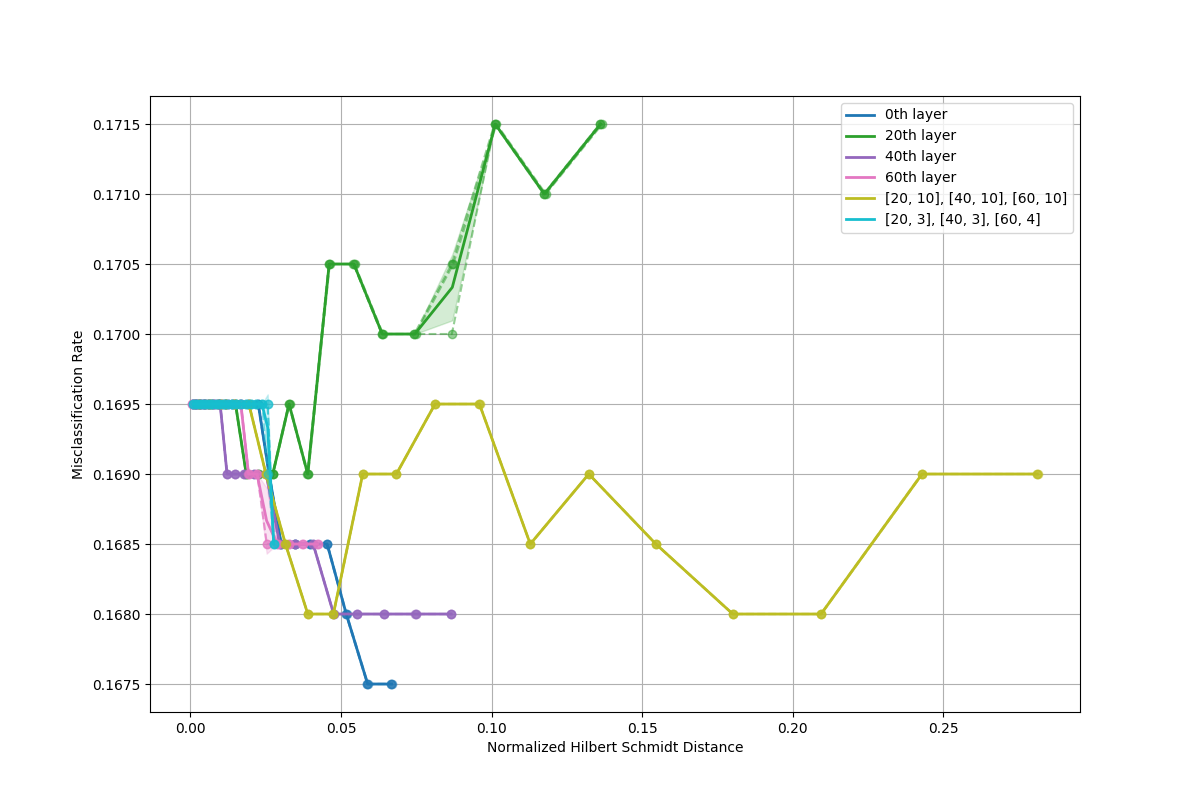}}  
    \end{minipage} \hfill
    \begin{minipage}{0.47\textwidth}  
        \centering
        \scalebox{\scaleB}{
        \includegraphics[width=\textwidth]{Figures/plots-v3/cifar-zero-gamma/10-adv.png}}  
    \end{minipage}
    \caption{
    Effects of inserting an adversarial block consisting of 10 layers into a binary classifier with 80 existing layers trained on CIFAR-2 versus incorporating three adversarial blocks.  
 The left plot shows the results when the adversarial layers are trained with a learning rate of 0.001,  whereas the right plot corresponds to a learning rate of 0.005, which yields better adversarial performance on CIFAR-2.  
    }
    \label{fig-lr-cifar}
\end{figure}

\paragraph{Learning Rates:} 
For MNIST,  we employ the same learning rate for training both the classifier and the adversarial layers.  As demonstrated in Figures \ref{fig-lr-fmnist} and \ref{fig-lr-cifar}, higher learning rates are adopted for FMNIST and CIFAR-2 because they result in improved adversarial performance.

\begin{figure}[H]
    \centering
    \begin{minipage}{0.33\textwidth}  
        \centering
        \scalebox{\scaleA}{
        \includegraphics[width=\textwidth]{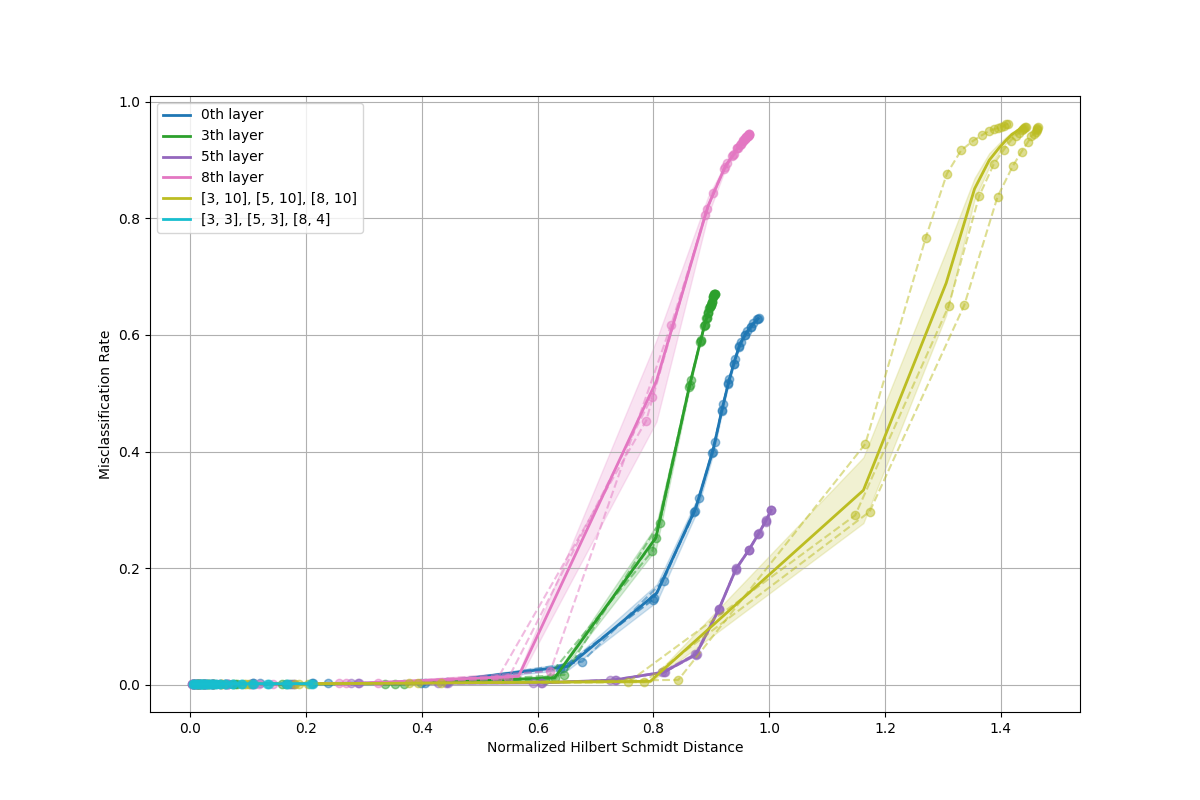}}  
    \end{minipage} \hfill
    \begin{minipage}{0.33\textwidth}  
        \centering
        \scalebox{\scaleA}{
        \includegraphics[width=\textwidth]{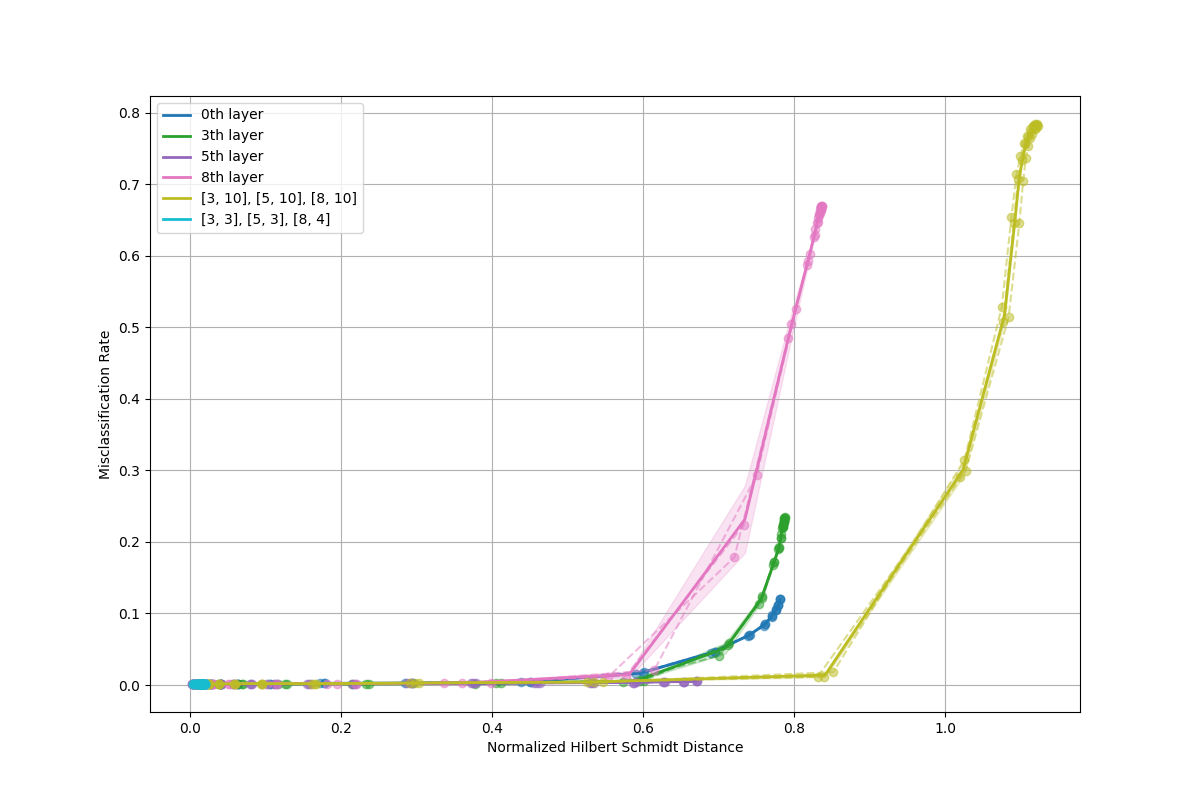}}  
    \end{minipage}
    \begin{minipage}{0.33\textwidth}  
        \centering
        \scalebox{\scaleA}{
        \includegraphics[width=\textwidth]{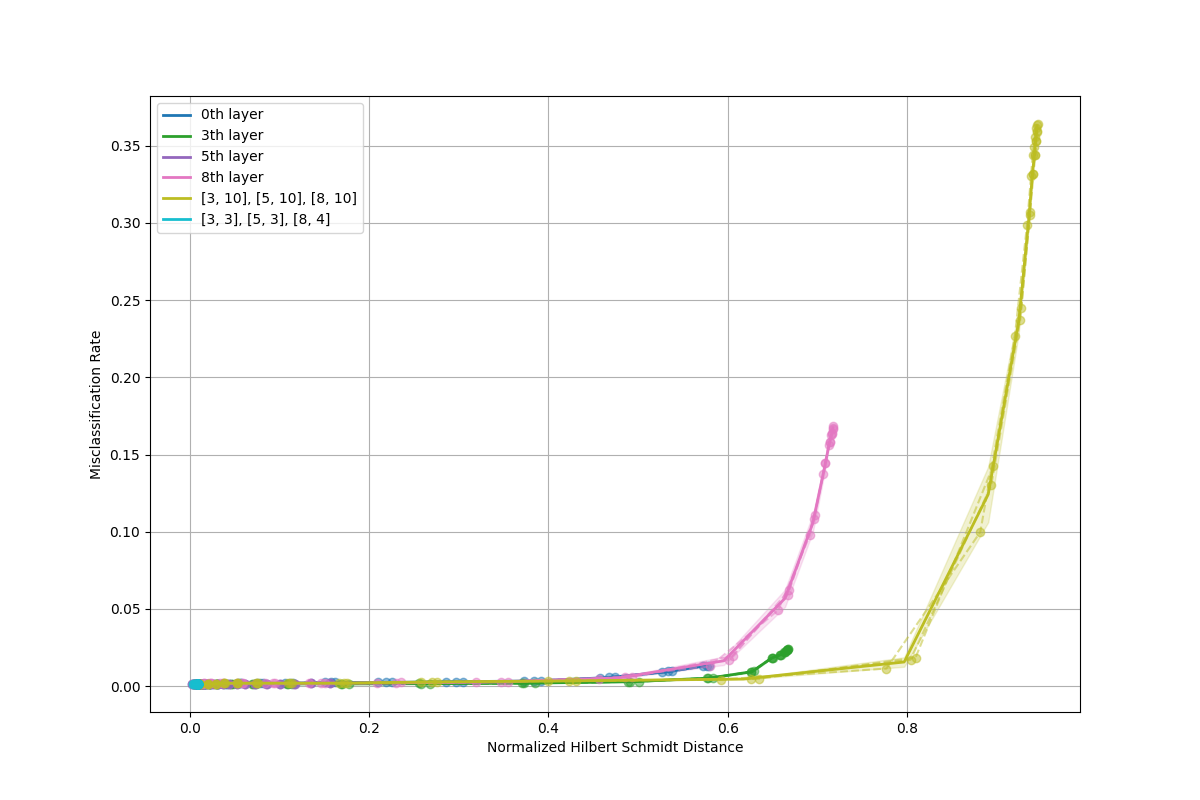}}  
    \end{minipage} \hfill

    \caption{
    Effect of constraining the attack strength when adversarial layers are incorporated into a model with 10 existing layers trained on MNIST  for binary classification.   The leftmost subfigure imposes the weakest constraint on the strength, which gradually increases as we move from left to right. The parameter $\gamma$ is set to $0.5,  1,$ and $1.5$ in the subfigures from left to right.  
    }
    \label{fig-gamma-m}
\end{figure}

\begin{figure}[H]
    \centering
    \begin{minipage}{0.33\textwidth}  
        \centering
        \scalebox{\scaleA}{
        \includegraphics[width=\textwidth]{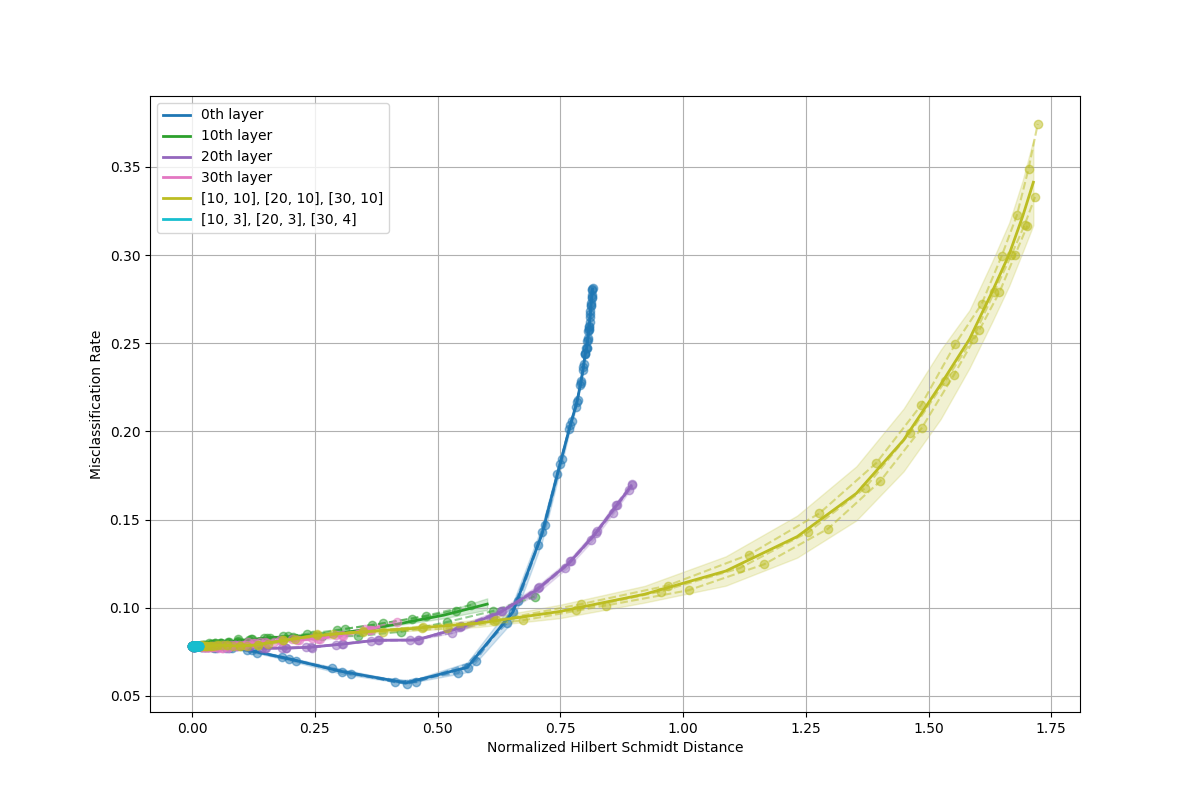}}  
    \end{minipage} \hfill
    \begin{minipage}{0.33\textwidth}  
        \centering
        \scalebox{\scaleA}{
        \includegraphics[width=\textwidth]{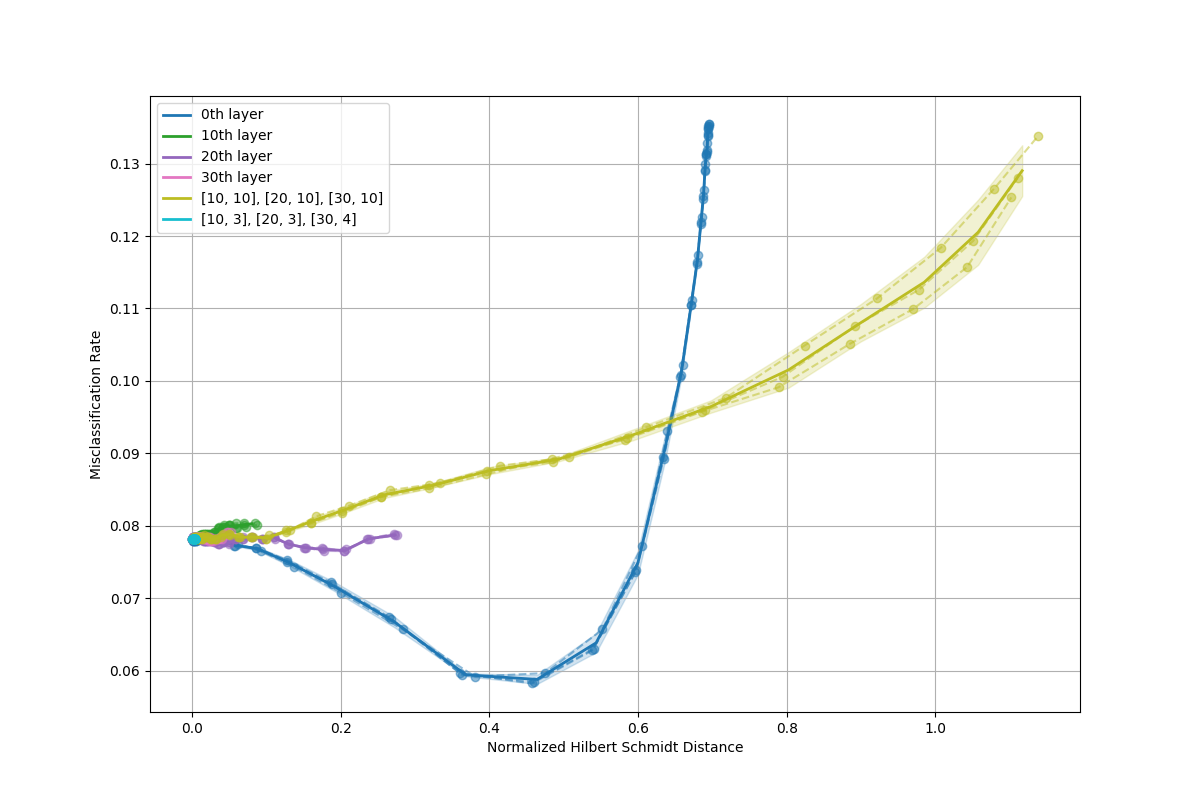}}  
    \end{minipage}
    \begin{minipage}{0.33\textwidth}  
        \centering
        \scalebox{\scaleA}{
        \includegraphics[width=\textwidth]{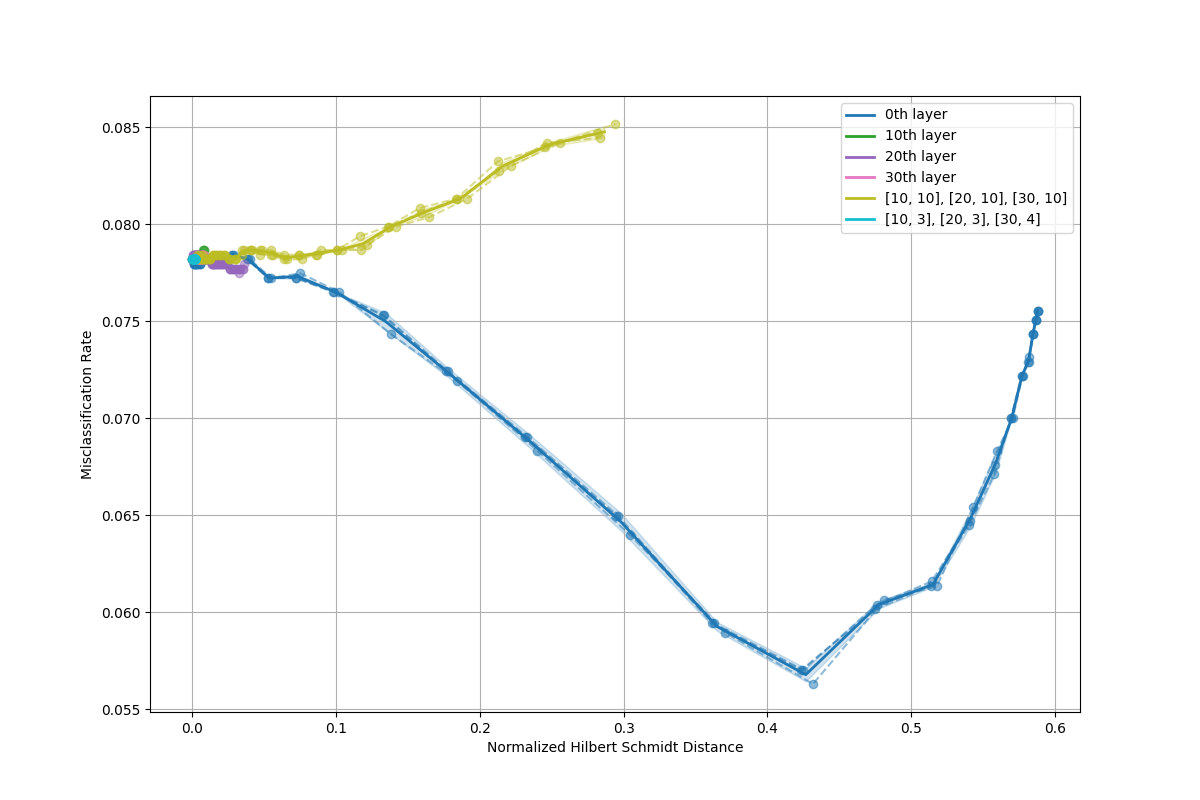}}  
    \end{minipage} \hfill

    \caption{
    Effect of constraining the attack strength when adversarial layers are incorporated into a model with 40 existing layers trained on MNIST  for four-class classification.   The parameter $\gamma$ is set to $0.5,  1,$ and $1.5$ in the subfigures from left to right.  
    }
    \label{fig-gamma-m4}
\end{figure}

\begin{figure}[H]
    \centering
    \begin{minipage}{0.33\textwidth}  
        \centering
        \scalebox{\scaleA}{
        \includegraphics[width=\textwidth]{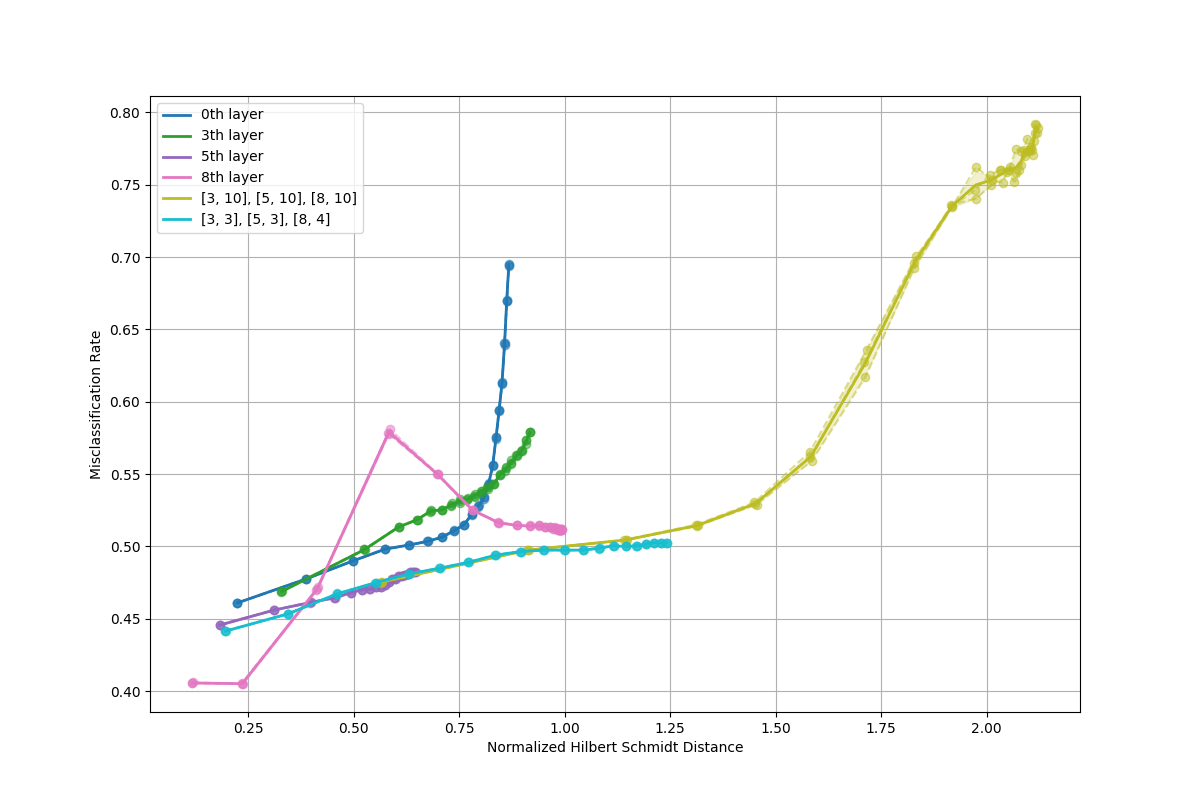}}  
    \end{minipage} \hfill
    \begin{minipage}{0.33\textwidth}  
        \centering
        \scalebox{\scaleA}{
        \includegraphics[width=\textwidth]{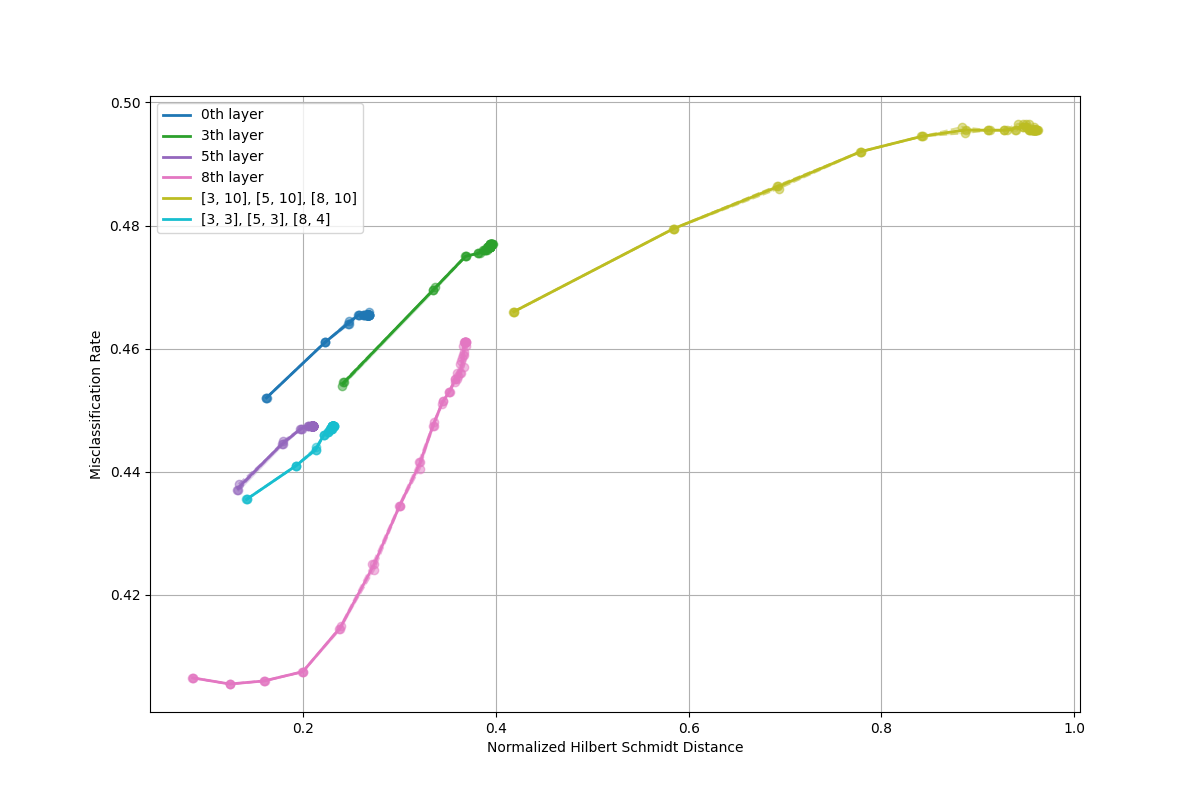}}  
    \end{minipage}
    \begin{minipage}{0.33\textwidth}  
        \centering
        \scalebox{\scaleA}{
        \includegraphics[width=\textwidth]{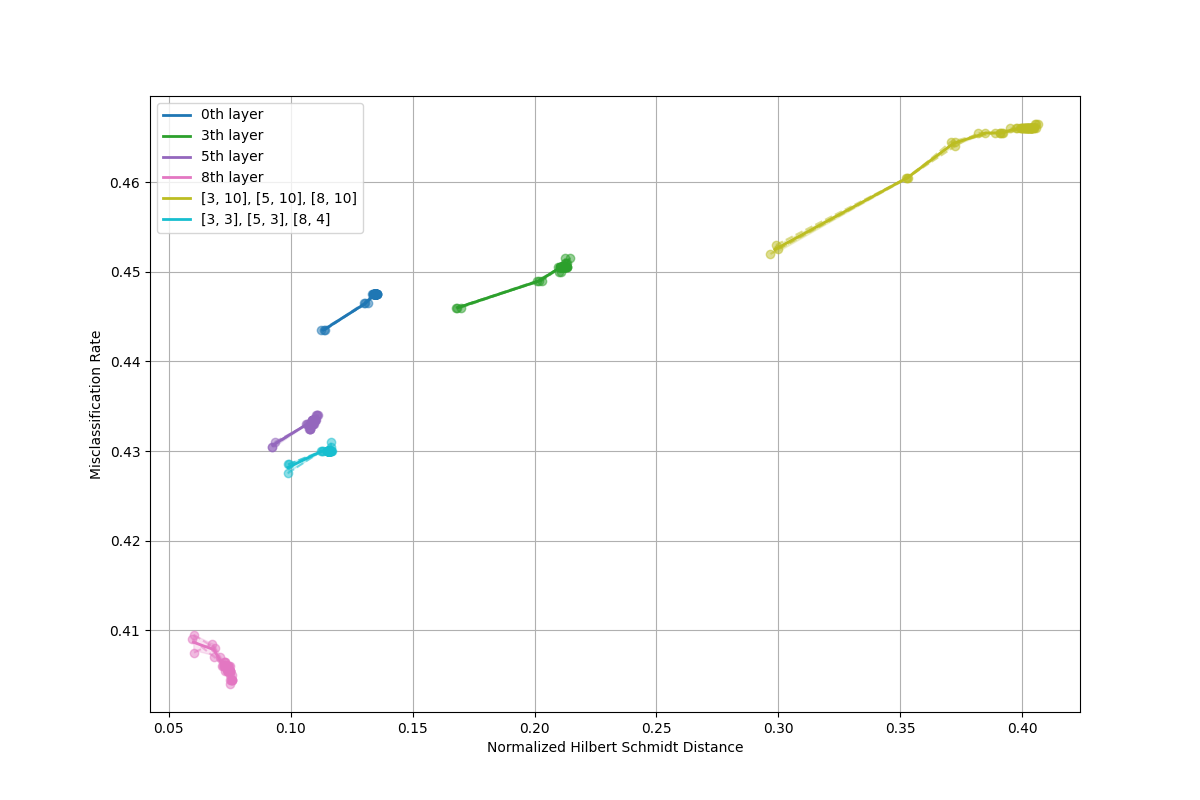}}  
    \end{minipage} \hfill

    \caption{
    Effect of constraining the attack strength when adversarial layers are incorporated into a model with 10 existing layers trained on FMNIST  for binary classification.   The parameter $\gamma$ is set to $0.1,  0.5$ and $1$ in the subfigures from left to right.  
    }
    \label{fig-gamma-f}
\end{figure}

\begin{figure}[H]
    \centering
    \begin{minipage}{0.33\textwidth}  
        \centering
        \scalebox{\scaleA}{
        \includegraphics[width=\textwidth]{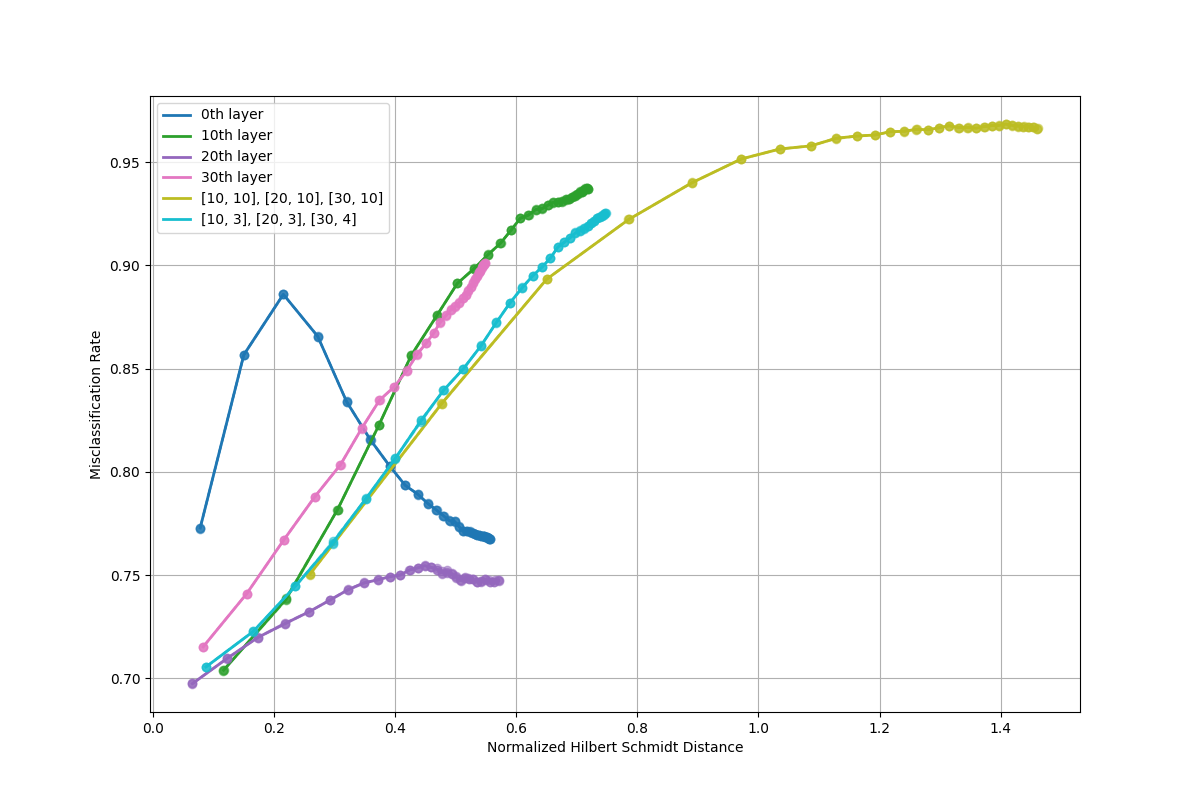}}  
    \end{minipage} \hfill
    \begin{minipage}{0.33\textwidth}  
        \centering
        \scalebox{\scaleA}{
        \includegraphics[width=\textwidth]{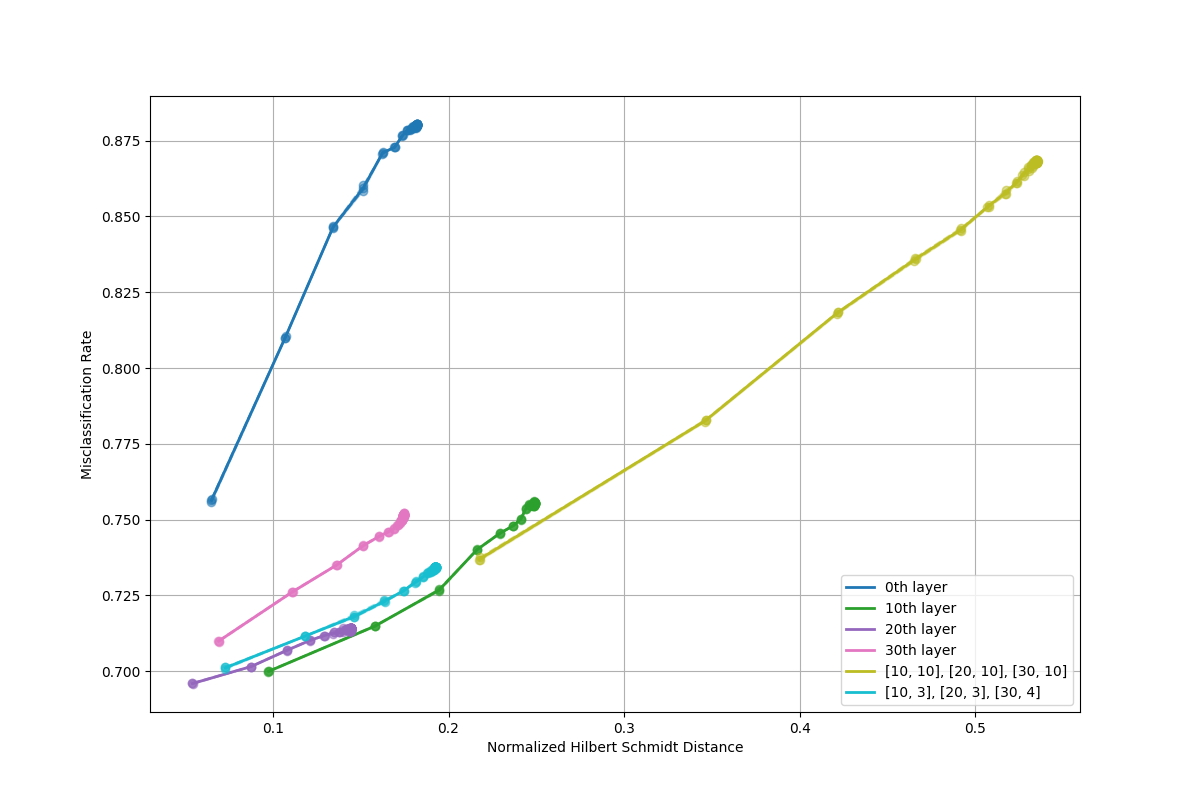}}  
    \end{minipage}
    \begin{minipage}{0.33\textwidth}  
        \centering
        \scalebox{\scaleA}{
        \includegraphics[width=\textwidth]{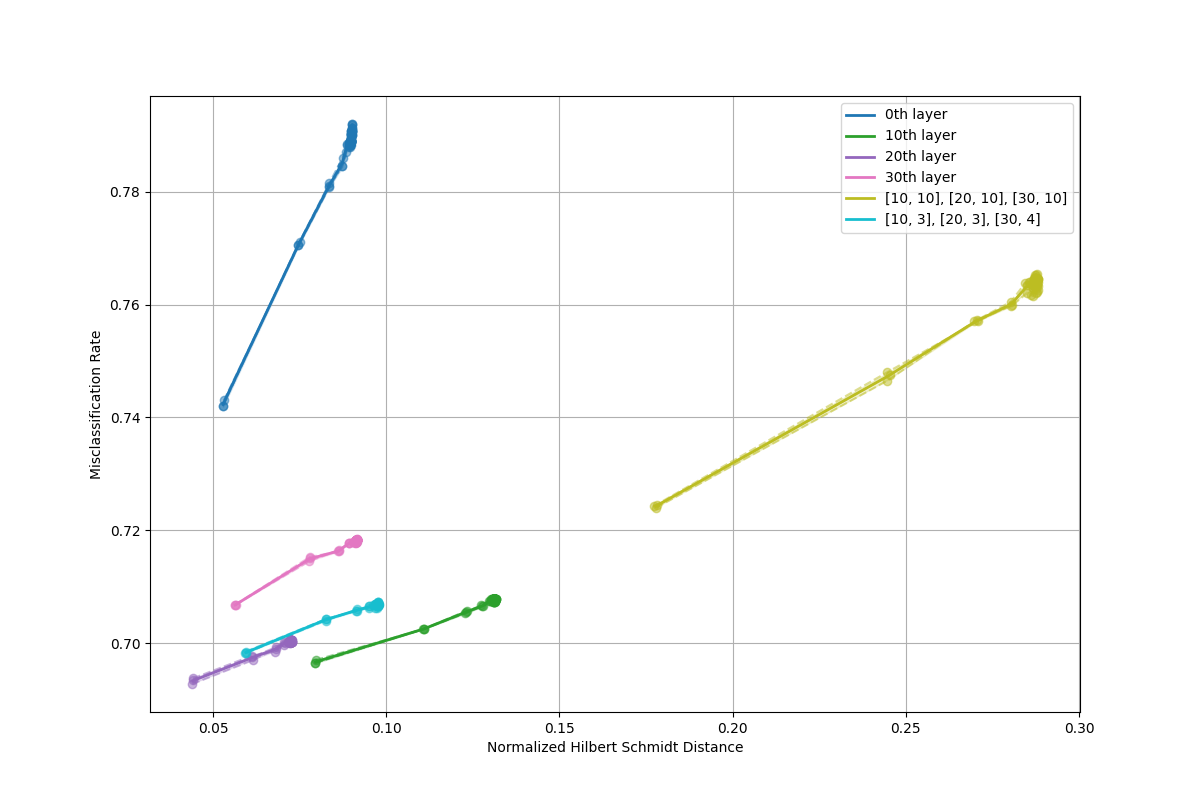}}  
    \end{minipage} \hfill

    \caption{
    Effect of constraining the attack strength when adversarial layers are incorporated into a model with 40 existing layers trained on FMNIST  for four-class classification.   The parameter $\gamma$ is set to $0.1,  0.5$ and $1$ in the subfigures from left to right.  
    }
    \label{fig-gamma-f4}
\end{figure}

\paragraph{Constraining the Attack Strength:} 
Figures \ref{fig-gamma-m} to \ref{fig-gamma-f4} demonstrate that when the attack strength is constrained by increasing the parameter $\gamma$ in (\ref{equ-adv-loss-l2}),  
the attack scenarios that are more successful than others before the constraint is applied remain more successful afterward.  For example, in Figure \ref{fig-gamma-m}, which illustrates the effect of adding adversarial blocks to a 10-layer classifier trained on MNIST for binary classification, the block inserted between the 8th and 9th layer of the classifier is more successful than the other adversarial perturbations in the leftmost subfigure,  and it continues to outperform them as we move from the leftmost to the rightmost subfigure,  increasing $\gamma$ and the attack constraint. Figures  \ref{fig-gamma-m} to \ref{fig-gamma-f4} further show that the plots corresponding to more constrained attacks resemble the early-epoch behavior of the plots obtained when the attacks are unconstrained. Therefore, it is sufficient to study the attacks in the unconstrained setting with $\gamma = 0$.

\subsection{Evaluating our Theoretical Bound}
 \label{sec:appendix-evaluating-theoretical} 

Theorem \ref{theorem-confidence-haar} provides a probabilistic bound on $\vert y_k(\sigma) - \hat{y}_k(\sigma) \vert$,  where $\hat{y}_k(\sigma)$ and  $y_k(\sigma)$ denote the probabilities of assigning label $k$ to  an input sample $\sigma$ before and after an adversarial attack, respectively.   
This result is particularly useful in the binary-classification setting. In that case, once we know the pre-attack probabilities assigned to the two labels, the bound in the theorem allows us to determine whether an adversarial attack can flip the classifier’s predicted label for $\sigma$. 
To assess whether Theorem \ref{theorem-confidence-haar}  is useful in practice, we must determine how tight the bound it provides actually is. The experiments below are designed to evaluate this.

\begin{figure}[H]
    \centering
    \begin{minipage}{0.33\textwidth}  
        \centering
        \scalebox{\scaleB}{
        \includegraphics[width=\textwidth]{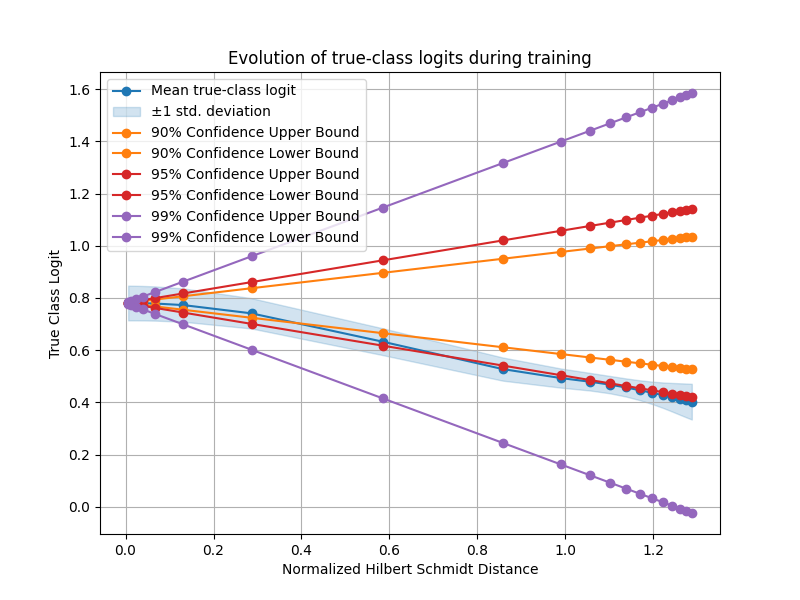}}  
		 \caption*{(a)}
    \end{minipage} \hfill
    \begin{minipage}{0.33\textwidth}  
        \centering
        \scalebox{\scaleB}{
        \includegraphics[width=\textwidth]{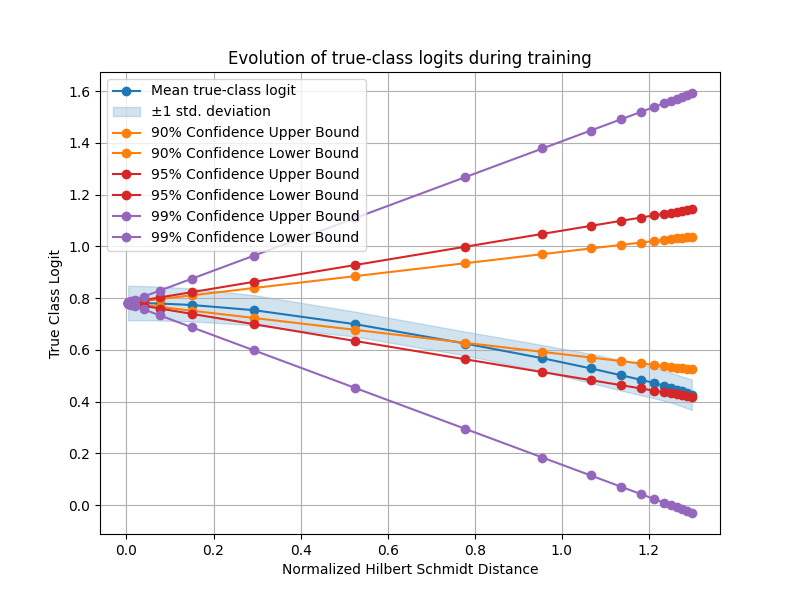}}  
        \caption*{(b)}
    \end{minipage}
    \begin{minipage}{0.33\textwidth}  
        \centering
        \scalebox{\scaleB}{
        \includegraphics[width=\textwidth]{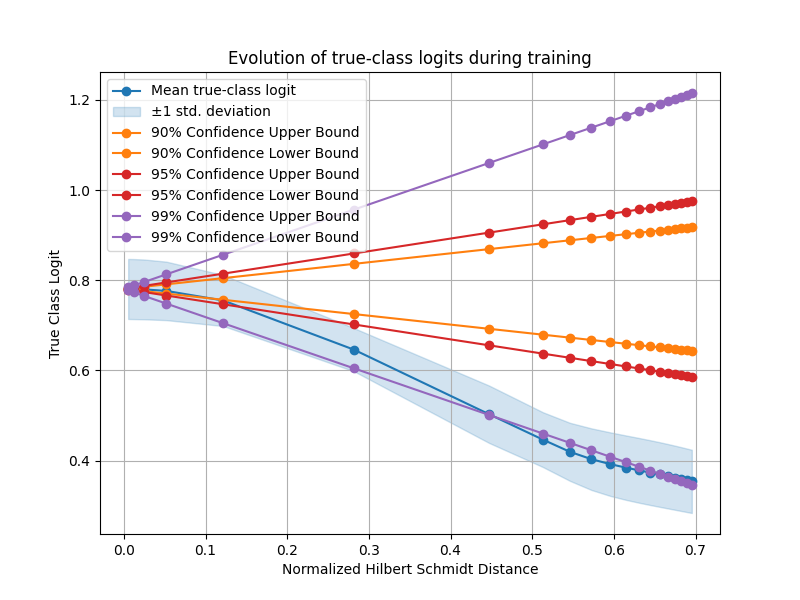}}  
        \caption*{(c)}
    \end{minipage} \hfill

    \caption{ Evolution of the average probability assigned to the true class for a 10-layer quantum classifier trained on MNIST as we increase the strength of the attack, where the strength is measured in terms of the normalized Hilbert Schmidt distance.  Subfigure (a) corresponds to a 10-layer adversarial block inserted before the 1st layer of the classifier, subfigure (b) to a 10-layer adversarial block inserted after the 5th layer and before the 6th layer, while subfigure (c) corresponds to inserting three 10-layer adversarial blocks after layers 3, 5, and 8 of the classifier.  In subfigures (a) and (b), we observe that the lower bounds provided by Theorem \ref{theorem-confidence-haar}, corresponding to confidence levels of $90\%$ and $95\%$, are close to the actual probabilities assigned to the true class and predict them well. However, in subfigure (c), the actual probabilities are lower than the bounds predicted at the $90\%$ and $95\%$ confidence levels and are closer to the bound predicted at the $99\%$ confidence level by Theorem  \ref{theorem-confidence-haar}.  }
    \label{fig-theoretical-m10}
\end{figure}

\begin{figure}[H]
    \centering
    \begin{minipage}{0.33\textwidth}  
        \centering
        \scalebox{\scaleB}{
        \includegraphics[width=\textwidth]{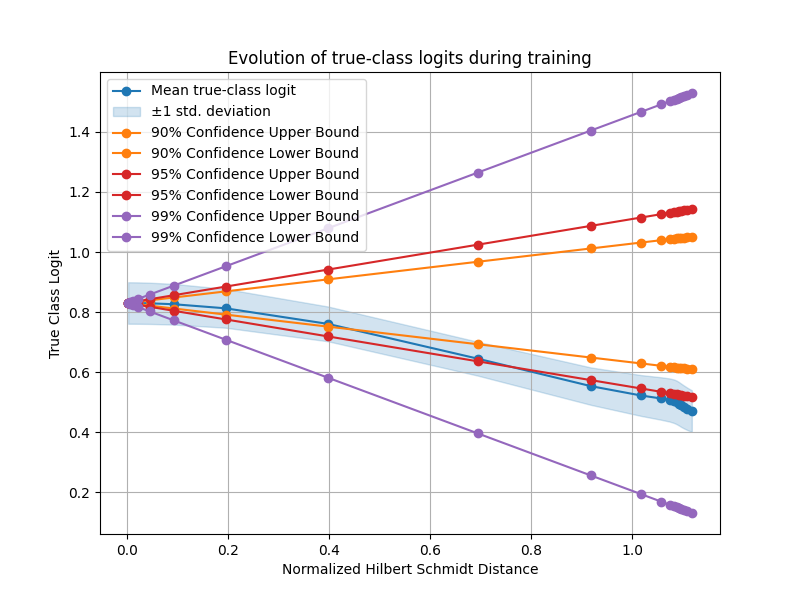}}  
		 \caption*{(a)}
    \end{minipage} \hfill
    \begin{minipage}{0.33\textwidth}  
        \centering
        \scalebox{\scaleB}{
        \includegraphics[width=\textwidth]{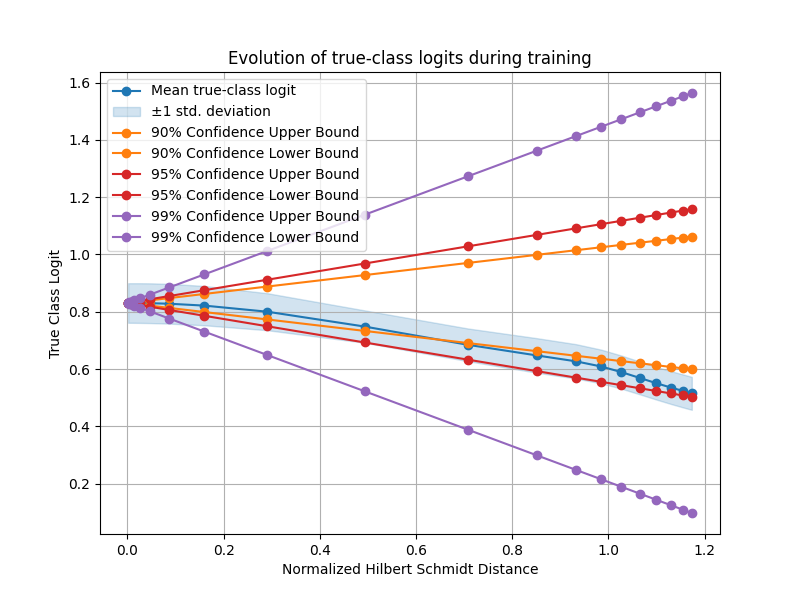}}  
        \caption*{(b)}
    \end{minipage}
    \begin{minipage}{0.33\textwidth}  
        \centering
        \scalebox{\scaleB}{
        \includegraphics[width=\textwidth]{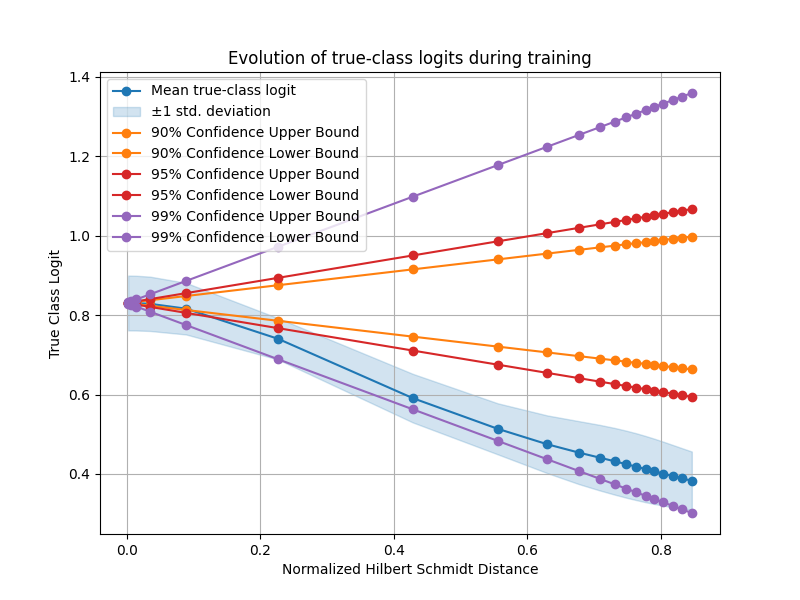}}  
        \caption*{(c)}
    \end{minipage} \hfill

    \caption{
Evolution of the average probability assigned to the true class for a 20-layer quantum classifier trained on MNIST as we increase the strength of the attack.    Subfigure (a) corresponds to a 10-layer adversarial block inserted before the 1st layer of the classifier, subfigure (b) to a 10-layer adversarial block inserted after the 10th layer and before the 11th layer, while subfigure (c) corresponds to inserting three 10-layer adversarial blocks after layers 5, 10, and 15.    }
    \label{fig-theoretical-m20}
\end{figure}

\begin{figure}[H]
    \centering
    \begin{minipage}{0.33\textwidth}  
        \centering
        \scalebox{\scaleB}{
        \includegraphics[width=\textwidth]{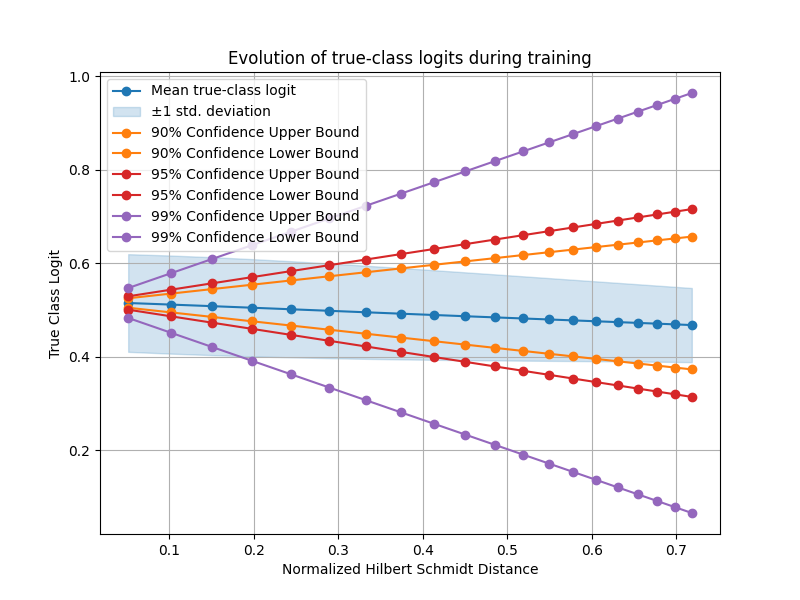}}  
		 \caption*{(a)}
    \end{minipage} \hfill
    \begin{minipage}{0.33\textwidth}  
        \centering
        \scalebox{\scaleB}{
        \includegraphics[width=\textwidth]{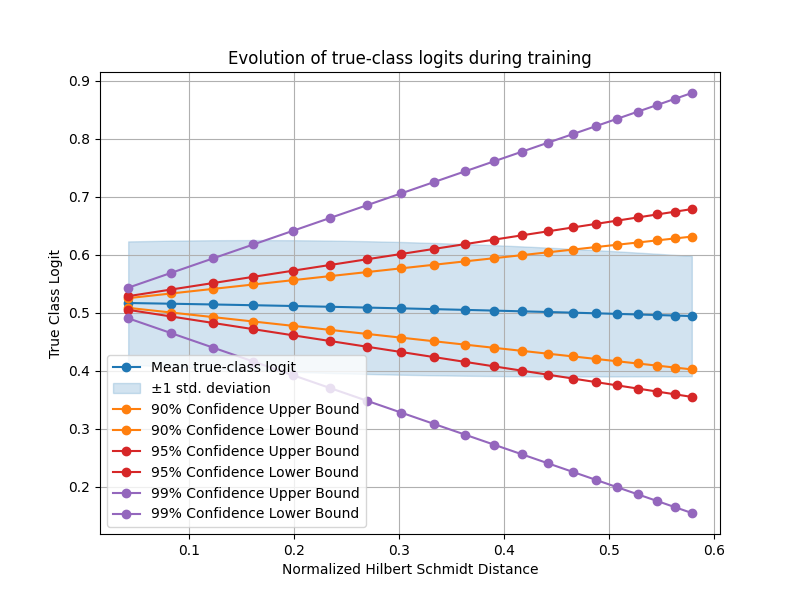}}  
        \caption*{(b)}
    \end{minipage}
    \begin{minipage}{0.33\textwidth}  
        \centering
        \scalebox{\scaleB}{
        \includegraphics[width=\textwidth]{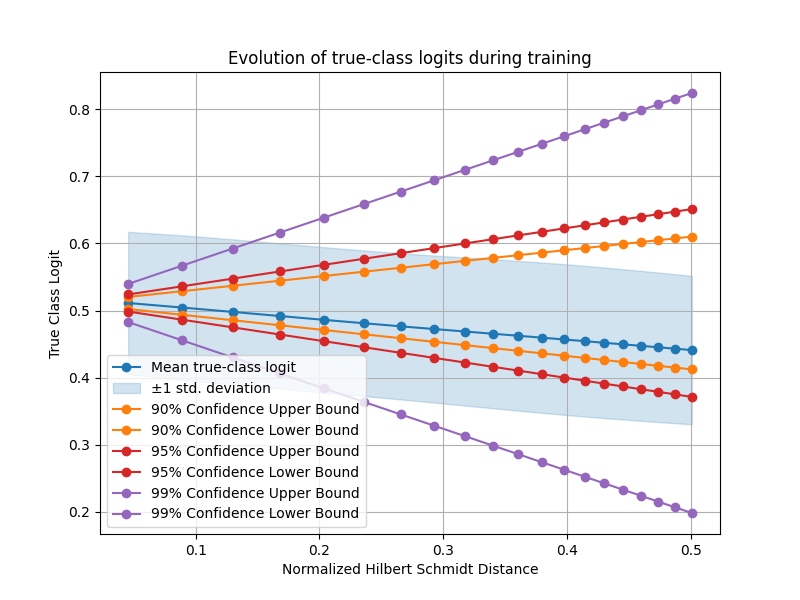}}  
        \caption*{(c)}
    \end{minipage} \hfill

    \caption{Evolution of the average probability assigned to the true class for a 10-layer quantum classifier trained on FMNIST as we increase the strength of the attack.  Subfigure (a) corresponds to a 10-layer adversarial block inserted before the 1st layer of the classifier, subfigure (b) to a 10-layer adversarial block inserted after the 5th layer and before the 6th layer, while subfigure (c) corresponds to inserting three 10-layer adversarial blocks after layers 3, 5, and 8.    }
    \label{fig-theoretical-f10}
\end{figure}

\begin{figure}[H]
    \centering
    \begin{minipage}{0.33\textwidth}  
        \centering
        \scalebox{\scaleB}{
        \includegraphics[width=\textwidth]{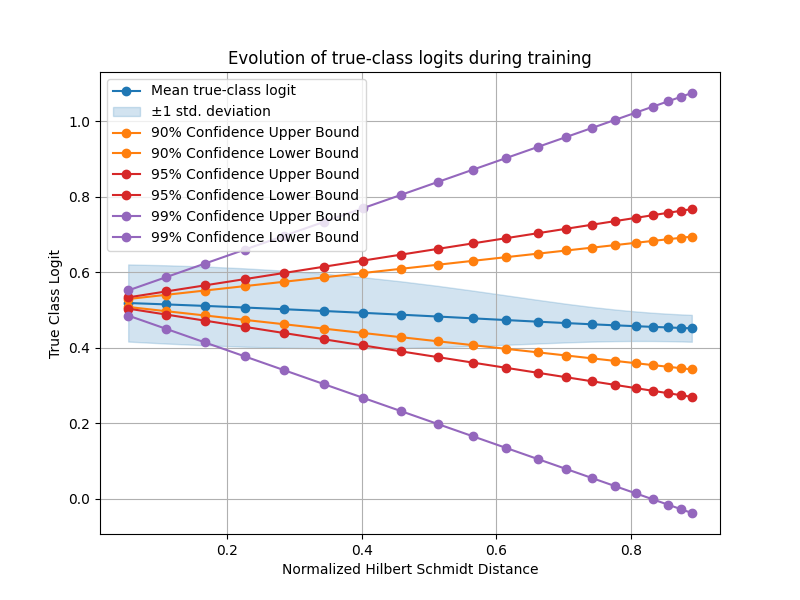}}  
		 \caption*{(a)}
    \end{minipage} \hfill
    \begin{minipage}{0.33\textwidth}  
        \centering
        \scalebox{\scaleB}{
        \includegraphics[width=\textwidth]{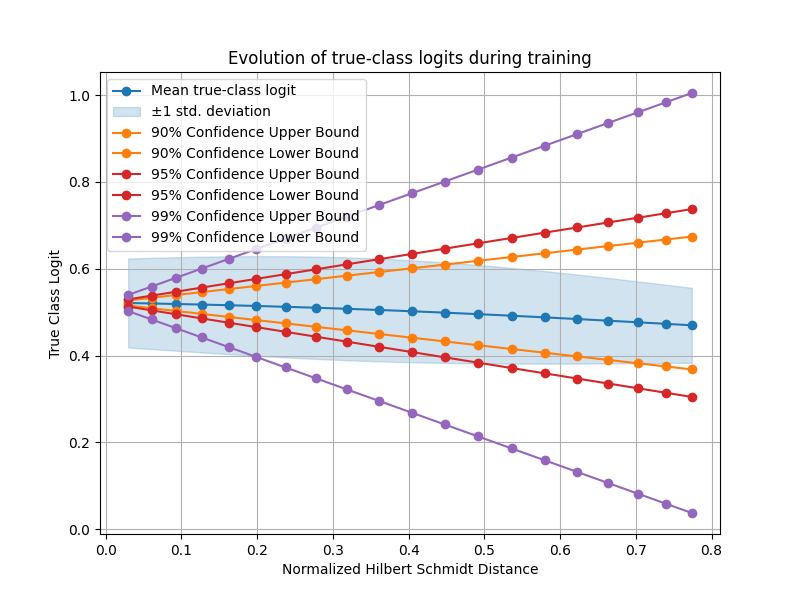}}  
        \caption*{(b)}
    \end{minipage}
    \begin{minipage}{0.33\textwidth}  
        \centering
        \scalebox{\scaleB}{
        \includegraphics[width=\textwidth]{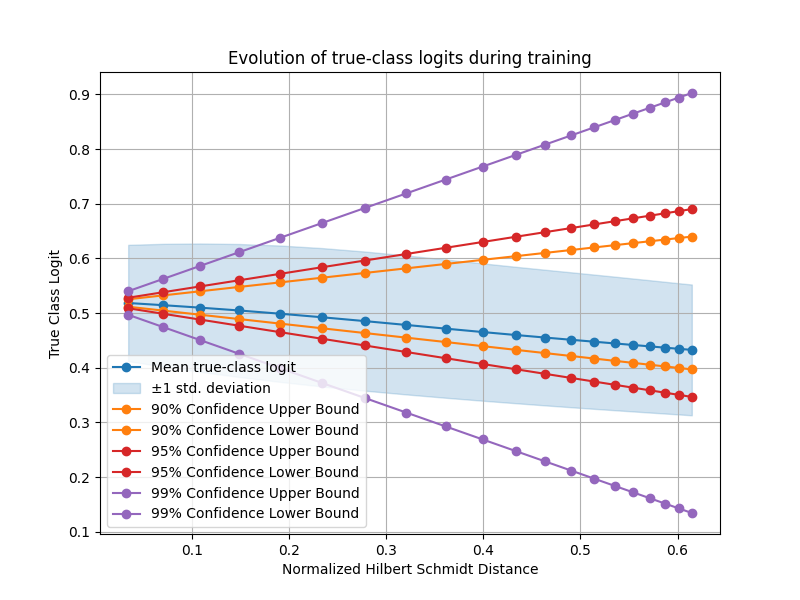}}  
        \caption*{(c)}
    \end{minipage} \hfill

    \caption{
Evolution of the average probability assigned to the true class for a 20-layer quantum classifier trained on FMNIST as we increase the strength of the attack.    Subfigure (a) corresponds to a 10-layer adversarial block inserted before the 1st layer of the classifier, subfigure (b) to a 10-layer adversarial block inserted after the 10th layer and before the 11th layer, while subfigure (c) corresponds to inserting three 10-layer adversarial blocks after layers 5, 10, and 15.      }
    \label{fig-theoretical-f20}
\end{figure}

\begin{figure}[H]
    \centering
    \begin{minipage}{0.33\textwidth}  
        \centering
        \scalebox{\scaleB}{
        \includegraphics[width=\textwidth]{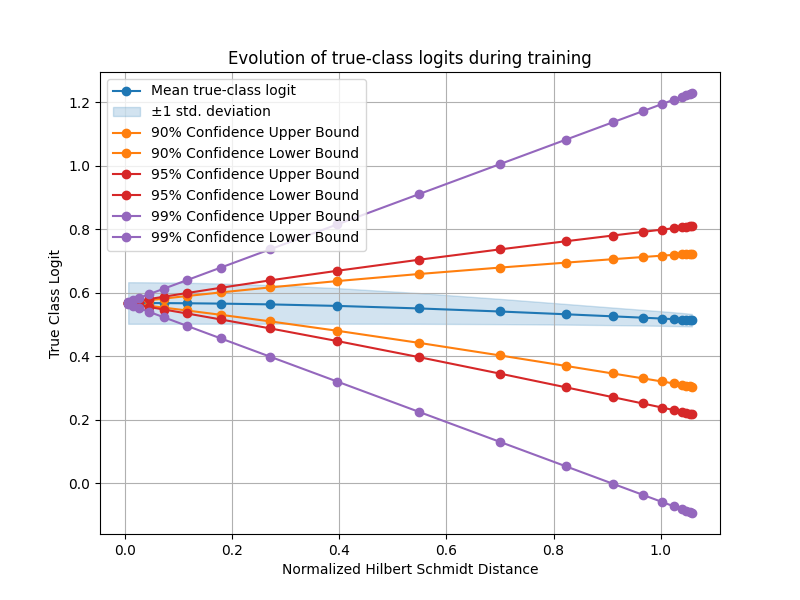}}  
		 \caption*{(a)}
    \end{minipage} \hfill
    \begin{minipage}{0.33\textwidth}  
        \centering
        \scalebox{\scaleB}{
        \includegraphics[width=\textwidth]{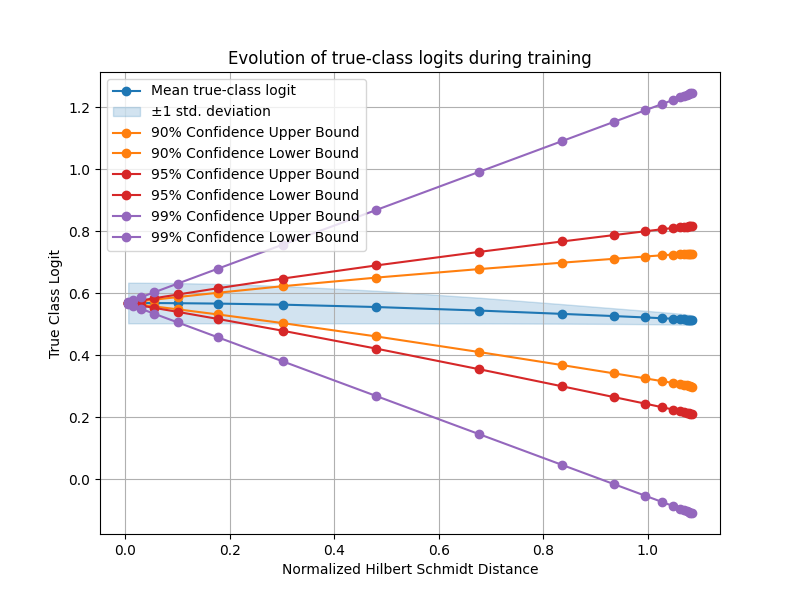}}  
        \caption*{(b)}
    \end{minipage}
    \begin{minipage}{0.33\textwidth}  
        \centering
        \scalebox{\scaleB}{
        \includegraphics[width=\textwidth]{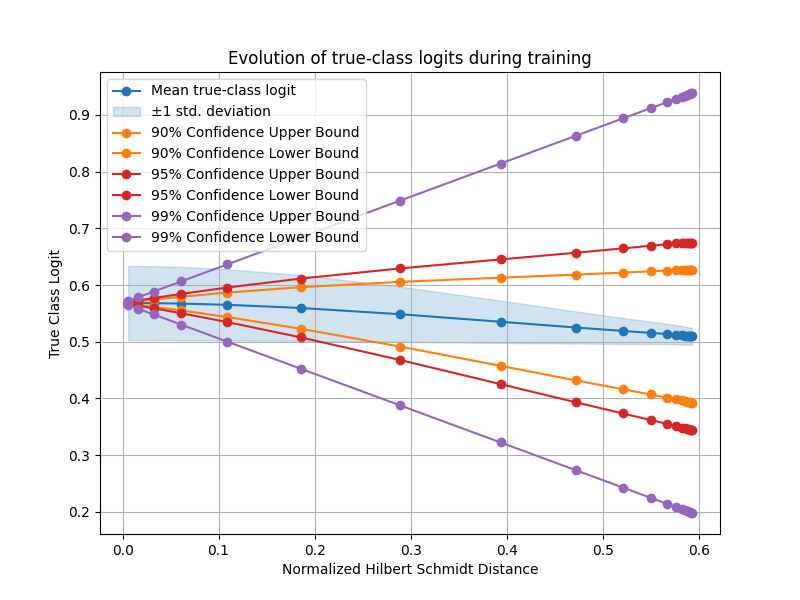}}  
        \caption*{(c)}
    \end{minipage} \hfill

    \caption{
Evolution of the average probability assigned to the true class for a 80-layer quantum classifier trained on CIFAR-2 as we increase the strength of the attack.    Subfigure (a) corresponds to a 10-layer adversarial block inserted before the 1st layer of the classifier, subfigure (b) to a 10-layer adversarial block inserted after the 40th layer and before the 41th layer, while subfigure (c) corresponds to inserting three 10-layer adversarial blocks after layers 20, 40, and 60.     }
    \label{fig-theoretical-cifar}
\end{figure}

In Figs.~\ref{fig-theoretical-m10}-\ref{fig-theoretical-cifar},  
the blue plots show how the average probability assigned to the true class evolves as we increase the strength of the adversarial attack.  Specifically, for each input sample $\sigma$,  we take $k$ to be its true label, and then compute the mean of $\hat{y}_k(\sigma)$ over all samples in the test dataset.  Each blue dot on the plots corresponds to a training epoch, for which both the average probability and the strength of the attack are computed at the end of the epoch. The shaded light-blue region shows $\pm 1$ standard deviation.
The orange, red, and purple plots show the upper and lower bounds predicted by Theorem \ref{theorem-confidence-haar} for 
$\hat{y}_k(\sigma)$ for a given strength of attack,  corresponding to confidence levels of $90\%, 95\%, $ and $99\%$, respectively.
Figs.~\ref{fig-theoretical-m10} and \ref{fig-theoretical-m20} showing 10- and 20-layer classifiers, depict results for MNIST,   while Figs.~\ref{fig-theoretical-f10} and \ref{fig-theoretical-f20} correspond to 10- and 20-layer classifiers trained on FMNIST, and Fig.~\ref{fig-theoretical-cifar} corresponds to an 80-layer classifier trained on CIFAR-2.  All classifiers are trained for binary classification.  
These figures show that for MNIST and FMNIST, the lower bounds provided by Theorem \ref{theorem-confidence-haar}, corresponding to confidence levels of $90\%$ and $95\%$, are in most cases close to the actual probabilities assigned to the true class and predict them well.  However, for CIFAR-2, 
the bounds provided by Theorem \ref{theorem-confidence-haar} can become loose as the strength of the attack increases. 
Nonetheless, this demonstrates that, while the bounds may occasionally loosen, Theorem \ref{theorem-confidence-haar} provides useful predictions in the majority of cases.  For CIFAR-2,  the lower bounds remain tight relative to the actual probabilities when the attack strength is low. Given that increasing attack strength typically reduces stealthiness,  these findings highlight the practical utility of our bound for stealthy adversarial attacks.

\end{document}